# SOME GEOMETRIC ASPECTS OF VARIATIONAL PROBLEMS IN FIBRED MANIFOLDS

**DEMETER KRUPKA**

**Department of Theoretical Physics**
**J. E. Purkyně University**
**Brno, Czechoslovakia**

# Contents

## 1. Introduction

**A few introductory remarks.** This work contains some notes on the geometric structure of the calculus of variations in fibered manifolds.

Let $(Y,\pi,X)$ be a fibered manifold, $(J^rY,\pi_r,X)$ its $r$-jet prolongation (Section 2), $n=\dim X$. Denote by $\Gamma(\pi)$ the set of all cross sections of the fibered manifold $(Y,\pi,X)$. If there is given an $n$-form $\lambda$ on $J^rY$, we can integrate the $n$-form $j^r\gamma^*\lambda$ on $X$ obtained by means of the $r$-jet prolongation $j^r\gamma$ of any cross section $\gamma\in\Gamma(\pi)$ and get, to each compact domain $c\subset X$, the number

$$\lambda_c(\gamma)=\int_c j^r\gamma^*\lambda.$$

This work takes notice of the behavior of the real function

(1.1) $$\Gamma(\pi)\ni\gamma\to\lambda_c(\gamma)\in\mathbf{R}.$$

We are interested especially in the most simple and frequently applied case $r=1$.

The work deals with the following questions:

1) what are the reasonable underlying *geometric structures* for the calculus of variations of functionals (1.1); what are the objects appearing in the standard classical calculus of variations, from the geometrical point of view;

2) how can one obtain conditions for a cross section $\gamma\in\Gamma(\pi)$ to be an extremal of the functional (1.1), in a completely invariant manner; how to obtain invariant *first variation formula*;

3) what is the geometrical structure of the *Euler equations*, known from the variational calculus; what $n$-forms $\lambda$ lead to "*identically satisfied*" Euler equations;

4) what are the so-called *canonical variational problems* in fibered manifolds, and how can be a given problem transformed into the canonical form;

5) what are the *symmetry transformations* of a variational problem, how can one formulate various problems concerning the behavior of (1.1) under some mappings of the fibered manifold $(Y,\pi,X)$;

6) what are the so-called *generally invariant* variational problems, and what are necessary and sufficient conditions for $\lambda$ to define such a problem.

In addition, some explicit calculations are performed, and the corresponding formulas are given.

For some other questions we do not touch, related to the functional (1.1), we refer to [10], [32].

**Fundamental structures and definitions.** Our approach to the calculus of variations is based on the theory of jets. As the main source we use for this [17]; see also [1], [3], [20], [23], [31], [35], [38], [39]. In many of these works the theory of jets is applied to the calculus of variations (e.g. [10], [20], [31], [35], [38]). We also use the bundle of infinite jets (see [39]). As an effect the first variation formula (6.22) may be written in a compact form also for *non-vertical* variations (compare with the formula (6.23)). As a serious difficulty in the use of the infinite jets in a more systematic way we feel that, in general, in our concept of the jet prolongation of projectable vector fields, the infinite jet prolongation has some unusual properties: its "horizontal" and "vertical" parts do not generate local one-parameter groups of transformations (at least in the sense we work with; compare with the exposition in Section 3).

In Section 3 we introduce the notion of a projectable vector field on $Y$, and prolong such a vector field to the jet spaces. This is motivated by *Trautman's* considerations concerning symmetries of variational problems [38] who has used them in a coordinate form (see also [35] and [20]). This time it is clear that such vector fields are very useful for the calculus.

Section 4 contains definitions of *horizontal* and *pseudovertical* forms on the jet spaces $J^r Y$. The first definition is obviously known (see e.g. [2]) while the second one, appearing in [20], is motivated by what is sometimes called the *Lepage's equivalence relation* (see [4], [27], [35]). The relation assigns $n$-forms, defining the same variational problem, and can be expressed in fibered manifolds by means of a linear mapping, $h$ (4.1), from the space of general forms to the space of horizontal forms. The mapping $h$ can be extended from $n$-forms to $(n+1)$-forms (see (4.15)); in this way we come to the notion of the so-called *Lepagian* $n$-forms. These are closely related to the *Euler form* (5.9) and, at the same time, to the Euler equations for extremals.

As for the notion of *variations* we note that the only type of variations used in this work is that one defined by vector fields. More precisely we shall work only with such "deformations" of cross sections that can be defined as local one-parameter groups of transformations of the underlying manifolds.

**The Euler mapping.** Roughly speaking, by the *Euler mapping* we mean the mapping assigning to each $n$-form $\lambda$ (or to each *Lagrange function*) the set of *Euler expressions* (left-hand sides of the Euler equations for extremals considered as functions on a jet space), or, as we say, the *Euler form* (see (5.13)). We give a complete characterization of the Euler mapping describing its kernel (Section 5). This is done in terms of exterior forms and their exterior differential. The result presents a generalization of a well known classical theorem [8] concerning the so-called *divergence expressions*, to the case of more "independent variables" or, which is the same, to the case of multiple integrals (1.1). An explicit formula for Lagrange functions leading to zero Euler expressions, is derived (see (5.16)).

**First variation formula.** Leaning on the theory of *Lepagian* $n$-forms $\lambda$ (Section 5) the basic formula of the calculus of variations of functionals (1.1), the *first variation formula*, is derived in a completely invariant way. This is the first variation formula of the so-called canonical variational problems. If the given $n$-form $\lambda$ is not Lepagian, one can apply the procedure of *Sniatycki* [35] generalizing *É. Cartan's* [6] and *Lepage's* approach to the variational problems, and thus obtain an equivalent variational problem in the *canonical* form (see also [4]). The procedure of *Sniatycki* is explained in a somewhat different way in Section 5 (4.17). Another "canonization" procedure is described by (6.27).

We note that our concept of canonical variational problems is essentially due to *Hermann* [13], [14].

**Invariant variational problems.** Considerations concerning invariant variational problems are also of geometrical nature. For instance the important formula (7.5) or (7.6), used for a classification of symmetry transformations, is derived in a completely intrinsic manner without references to a coordinate system (compare with [38]). The classification of symmetries of the canonical variational problems is studied, due to *Trautman*. The problem of critical points (extremals) with *prescribed* symmetry transformations is stated, and its solution is obtained in rather a simple form (see (7.12), (7.14)). It is characterized as the solution of a system of Euler equations depending on generators of the prescribed one-parameter groups of transformations. As an interesting special case that may appear in the variational calculus in tensor bundles, we discuss the so-called *generally covariant* variational theories and obtain conditions for the $n$-form $\lambda$ defining such theory (7.20).

(See especially [38], [35], [20], and also [11], [13], [15], [18], [19], [22], [25], [28], [29], [31], [34], [36], [37].)

**Some formulas.** In this paragraph there are collected some formulas frequently needed in calculations. We often use them without explicitly mentioning it.

Let $f: X \to Y$ be a morphism of manifolds $X$, $Y$, let $\eta$ be a differential $p$-form on $Y$; then the *pull-back* $f^*\eta$ of $\eta$ by $f$ is a $p$-form on $X$ defined as

$$f^*\eta = (\eta \circ f) \cdot (Tf)^p$$

(see e.g. [26]). Let $Z$ be another manifold and $g: Z \to X$ a morphism. Then

$$(f \circ g)^* \eta = g^* f^* \eta, \quad f^* d\eta = df^* \eta.$$

If $\lambda$ is another form on $Y$, then

$$f^*(\eta \wedge \lambda) = f^*\eta \wedge f^*\lambda.$$

In these formulas $Tf$ denotes the *tangent map* to $f$, $d$ and $\wedge$ denote the *exterior differential* and the *wedge product* of forms, respectively.

Let $\xi$ be a vector field on $Y$ and $\alpha_t^\xi$ its one-parameter local group of transformations of $Y$. The *Lie derivative* $\vartheta(\xi)\eta$ of $\eta$ by $\xi$ is defined as

$$\vartheta(\xi)\eta = \left\{ \frac{d}{dt} \alpha_t^{\xi *} \eta \right\}_0 .$$

Take a point $y \in Y$ and tangent vectors $\xi_1, \xi_2, \ldots, \xi_{p-1}$, to $Y$ at the point; then the *contraction* of $\eta$ by $\xi$ is a $(p-1)$-form on $Y$ defined as

$$\langle i(\xi)\eta, \xi_1 \times \xi_2 \times \ldots \times \xi_{p-1} \rangle = \langle \eta, \xi(x) \times \xi_1 \times \xi_2 \times \ldots \times \xi_{p-1} \rangle.$$

Here $\langle\ ,\ \rangle$ denotes the natural pairing of forms and vector fields. Let $\eta_1, \eta_2$ be two arbitrary forms on $Y$ and let $\eta_1$ be of degree $r$. Then the introduced operations obey the following rules (omitting the standard bilinearity and linearity conditions):

$$\vartheta(\xi)\eta = i(\xi)d\eta + di(\xi)\eta,$$

$$\vartheta(\xi)(\eta_1 \wedge \eta_2) = \vartheta(\xi)\eta_1 \wedge \eta_2 + \eta_1 \wedge \vartheta(\xi)\eta_2,$$

$$\vartheta(\xi)d\eta = d\vartheta(\xi)\eta,$$

$$i(\xi)(\eta_1 \wedge \eta_2) = i(\xi)\eta_1 \wedge \eta_2 + (-1)^r \eta_1 \wedge i(\xi)\eta_2.$$

If $\xi_0$ is another vector field on $Y$, then

$$\vartheta(\xi)i(\xi_0)\eta = i(\xi_0)\vartheta(\xi)\eta.$$

(For definitions and properties of these operations see e.g. [7], [2], [13], [26], [36].)

**Notation.** All throughout the work the following standard notation is used:

| | | |
|---|---|---|
| $\mathbf{R}^n$ | - | $n$-dimensional real Euclidean space |
| $\mathbf{R}^1 = \mathbf{R}$ | - | the real line |
| $\mathscr{A}(X)$ | - | an atlas on a manifold $X$ |
| $\alpha_t^\xi$ | - | the local one-parameter group generated by a vector field $\xi$ |
| $X$ | - | an $n$-dimensional real orientable paracompact manifold |
| $(TX, \tau_X, X)$ | - | the tangent bundle of $X$ |
| $(Y, \pi, X)$ | - | a fibered manifold |
| $\omega$ | - | a volume element form on $X$ |
| $\Gamma(\pi)$ | - | the set of all cross sections of $(Y, \pi, X)$ |
| $(J^r, \pi_r, X)$ | - | the $r$-jet prolongation of $(Y, \pi, X)$ |

All manifolds and differentiable mappings (morphisms) are supposed to be of class $C^\infty$. The *composition* of mappings $f$ and $g$ is denoted by $\circ$; we also write

$$g \circ f = gf.$$

The *r-jet prolongation* of a cross section $\gamma \in \Gamma(\pi)$ is defined by

$$(j^r\gamma)(x) = j_x^r\gamma.$$

Finally, we note that we use the usual summation convention throughout the work, but we show summation in detail if needed, especially in more complicated formulas. If the symbol of summation is omitted we suppose that the summation is obvious.

## 2. Fundamental structures

**Fibered manifolds.** By a *fibroid manifold* we mean each triple $(Y, \pi, X)$, where $Y$ and $X$ are differentiable manifolds and $\pi : Y \to X$ is a surjective submersion. A mor-phism $\gamma: X \to Y$ satisfying

$$\pi \circ \gamma = id_X$$

will be called a *cross section* of the fibered manifold $(Y, \pi, X)$. The set of all cross sections of $(Y, \pi, X)$ will be denoted by $\Gamma(\pi)$.

Throughout the work we suppose that we are given a fibered manifold $(Y, \pi, X)$ with finite-dimensional *Y* and *X*; we denote

$$n = \dim X, \quad n + m = \dim Y,$$

where *n* and *m* are some positive integers.

**Fibered charts.** By definition of $\pi$, to each $y_0 \in Y$ one can find a chart $(V, \psi)$ with center $y_0$, such that $\psi$ is of the form

(2.1) $$\psi(y) = (\varphi(\pi(y)), \psi_0(y))$$

for some chart on *X*, $(\pi(V), \varphi)$, with center $\pi(y_0)$. Charts of the described properties will be called *fibered charts* [25].

We usually write $(x_i, y_\mu)$, $1 \le i \le n$, $1 \le \mu \le m$, for the coordinate functions defined by (2.1), and $\psi = (\varphi, \psi_0)$.

**Jets.** In this paper we freely use the symbols $\partial/\partial x_k$, $D_k$ for partial derivatives, and *D* for derivatives, in the sense of [1], [9].

Let $x \in X$, $\gamma_1, \gamma_2 \in \Gamma(\pi)$, and let *r* be an arbitrary natural number. The cross sections $\gamma_1$, $\gamma_2$ are said to be *r-equivalent* at the point *x*, if $\gamma_1(x) = \gamma_2(x)$ and if there exists a fibered chart $(V, \psi)$ with center $\gamma_1(x) = \gamma_2(x)$ such that

$$D^k \psi_0 \gamma_1 \varphi^{-1}(\varphi(x)) = D^k \psi_0 \gamma_2 \varphi^{-1}(\varphi(x))$$

for all $k = 1, 2, \ldots, r$; here $\psi = (\varphi, \psi_0)$. The class of *r*-equivalence at *x* containing a cross section $\gamma$ is denoted $j_x^r \gamma$ and called the *r-jet* of $\gamma$ at the point *x*. The set of all $j_x^r \gamma$, where $x \in X$ and $\gamma \in \Gamma(\pi)$, is denoted by $J^r Y$ or just $J^r$ if there is no danger of confusion. We put $J^0 Y = Y$ and define mappings $\pi_{r,s}$, $r \le s$ and $\pi_r$ by the formulas

(2.2) $$\pi_{r,s}(j_x^r \gamma) = j_x^s \gamma, \quad \pi_r(j_x^r \gamma) = x.$$

If $\gamma \in \Gamma(\pi)$, we define

$$j^r \gamma(x) = j_x^r \gamma,$$

and call the mapping $j^r \gamma$ the *r-jet prolongation* of $\gamma$.

Denote by $L_k^s(\mathbf{R}^n, \mathbf{R}^m)$ the vector space of *k*-linear symmetric mappings from $\mathbf{R}^n$ to $\mathbf{R}^m$. Let $(V, \psi)$ be a fibered chart on *Y*, $\psi = (\varphi, \psi_0)$, and consider a pair $(Z_r, \zeta_r)$, where $Z_r = \pi_{r,0}^{-1}(V)$, and for $j_x^r \gamma \in Z_r$,

(2.3) $$\zeta_r(j_x^r \gamma) = (\varphi(x), \psi_0(\gamma(x)), D\psi_0 \gamma \varphi^{-1}(\varphi(x)), \ldots, D^r \psi_0 \gamma \varphi^{-1}(\varphi(x))).$$

Evidently, $\zeta_r$ maps $Z_r$ into $\mathbf{R}^n \times \mathbf{R}^m \times L(\mathbf{R}^n, \mathbf{R}^m) \times \ldots \times L_s^r(\mathbf{R}^n, \mathbf{R}^m)$. It is known that there exists a unique manifold structure on $J^rY$ such that, for any choice of $(V,\psi)$, $(Z_r,\zeta_r)$ is a chart. We shall always consider $J^rY$ as a differentiable manifold with this structure.

It can be shown then that the triples $(J^rY,\pi_{r,s},J^sY)$, $(J^rY,\pi_r,X)$ are fibered manifolds, and each $j^r\gamma$ defined by $\gamma\in\Gamma(\pi)$ a cross section of $(J^rY,\pi_r,X)$.

We call $(Z_r,\zeta_r)$ the *canonical chart* on $J^rY$ (*associated* with $(V,\psi)$). The fibered manifold $(J^rY,\pi_r,X)$ is called the *r-jet prolongation* of $(Y,\pi,X)$. We also speak about $J^rY$ as the *r*-jet prolongation of *Y*; no confusion can possibly arise from this.

If we write $(x_i,y_\mu)$ for the coordinates on *Y* defined by a fibered chart $(V,\psi)$, then for the sake of simplicity of notation, coordinates on $J^rY$ defined by $(Z_r,\zeta_r)$ will be denoted by

$$\zeta_r=(x_i,y_\mu,z_{i\mu},z_{i_1i_2\mu},\ldots,z_{i_1i_2\ldots i_r\mu}) \tag{2.4}$$

where

$$1\le i_1\le i_2\le\ldots\le i_s\le n,\quad s\le r.$$

(Compare with [1], [17], [22], [23].)

**Infinite jets.** Now we wish to discuss the case when $r=\infty$. Then the set $J^\infty Y$ is defined in the same manner as in the previous paragraph the set $J^rY$. $J^\infty Y$ will be given an infinite dimensional manifold structure, modeled on a locally convex Frechet space. Since there appear only a few infinite-dimensional manifolds in this paper it seems to be most convenient to introduce them without references to the general definitions (see [39], [3]).

Let us consider the finite dimensional vector space $L_s^k(\mathbf{R}^n,\mathbf{R}^m)$ of *k*-linear symmetric mappings from $\mathbf{R}^n$ to $\mathbf{R}^m$. Define

$$F=\mathbf{R}^n\times\mathbf{R}^m\times\prod_{k=1}^{\infty}L_s^k(\mathbf{R}^n,\mathbf{R}^m),$$

and consider in $F$ the structure of a locally convex space with the topology of projective limit of spaces $\mathbf{R}^n$, $\mathbf{R}^m$, $L(\mathbf{R}^n,\mathbf{R}^m)$, $L_s^2(\mathbf{R}^n,\mathbf{R}^m)$, …, by the natural projections. With the described structure $F$ is a Frechet space [33] (for details see [5], [33]).

Let $(Y,\pi,X)$ be our fibered manifold. For a fibered chart $(V,\psi)$ on *Y*, let us put $Z_\infty=\pi_{\infty,0}^{-1}(V)$, where $\pi_{\infty,0}\colon J^\infty Y\to Y$ denotes the natural projection. Put for $j_x^\infty\gamma\in Z_\infty$

$$\zeta_\infty(j_x^\infty\gamma)=(\varphi(x),\psi_0(\gamma(x)),D\psi_0\gamma\varphi^{-1}(\varphi(x)),D^2\psi_0\gamma\varphi^{-1}(\varphi(x)),\ldots). \tag{2.5}$$

The mapping $\zeta_\infty$ takes values in the Frechet space $F$. If $\mathscr{A}(Y)$ is an atlas on *Y* formed by fibered charts, we can define $(Z_{\infty,\iota},\zeta_{\infty,\iota})$ for each $(V_\iota,\psi_\iota)\in A(Y)$ in the same manner as above.

*Consider $J^\infty Y$ with the topology of the projective limit by the mappings $\pi_\infty,\pi_{\infty,r}$. Then the following conditions hold:*

1) $\bigcup_\iota Z_{\infty,\iota}=J^\infty Y$.
2) $\zeta_{\infty,\iota}\colon Z_{\infty,\iota}\to F$ *is homeomorphism of* $Z_{\infty,\iota}$ *onto the open set* $\zeta_{\infty,\iota}(Z_{\infty,\iota})\subset F$,
3) *If* $Z_{\infty,\iota}\cap Z_{\infty,\kappa}\neq 0$, *then the mapping*

$$\zeta_{\infty,\kappa}\zeta_{\infty,\iota}^{-1}:\zeta_{\infty,\iota}(Z_{\infty,\iota}\cap Z_{\infty,\kappa})\to\zeta_{\infty,\kappa}(Z_{\infty,\iota}\cap Z_{\infty,\kappa})$$

*is of the form*

$$(x',\gamma'(x'),D\gamma(x'),D^2\gamma'(x'),\dots)\to(\varphi_\kappa\varphi_\iota^{-1}(x'),\psi_{\kappa 0}\psi_\iota^{-1}(x',\gamma'(x)),$$
$$D\psi_{\kappa 0}\psi_\iota^{-1}(x',\gamma'(x'))\circ(id,D\gamma'(x'))\circ D\varphi_\iota\varphi_\kappa^{-1}(\varphi_\kappa\varphi_\iota^{-1}(x')),\dots),$$

*and is therefore differentiable in the sense of the differential calculus in locally convex spaces* ([39], III, 3.2).

In other words, the set $\mathscr{A}(J^\infty Y)$ of all pairs $(Z_{\infty,\iota},\zeta_{\infty,\iota})$ is a locally convex atlas of class $C^\infty$ on $J^\infty Y$, and can be used to define the structure of a locally convex differentiable manifold of class $C^\infty$ in $J^\infty Y$, modeled on the Frechet space $F$.

We always consider $J^\infty Y$ as this differentiable manifold and call each pair $(Z_{\infty,\iota},\zeta_{\infty,\iota})$ the *canonical chart* on $J^\infty Y$ (*associated* with $(V_\iota,\psi_\iota)$).

Consider the mappings $\pi_{\infty,r}$, $\pi_\infty$; it is easily seen that they are submersions in the sense of the theory of locally convex manifolds [39]. This is why we shall speak about the triples $(J^\infty Y,\pi_{\infty,r},J^rY)$, $(J^\infty Y,\pi_\infty,X)$, $r=1,2,\dots$, as about fibered manifolds. The fibered manifold $(J^\infty Y,\pi_\infty,X)$ will be called the *infinite jet prolongation* of $(Y,\pi,X)$. Similarly as in the finite dimensional case we shall also call $J^\infty Y$ the infinite jet prolongation off $Y$, and write $J^\infty Y=J^\infty$.

(Definitions and proofs can be consulted with [1], [3], [16], [23].)

**The tangent bundle of** $J^\infty Y$. Consider the locally convex manifold $J^\infty Y$ and the atlas $\mathscr{A}(J^\infty Y)$ on it defined before. Let $j_x^\infty\gamma\in J^\infty Y$ be an arbitrary point. Let us consider the set of the triples $(Z_\infty,\zeta_\infty,w)$, where $j_x^\infty\gamma\in Z_\infty$, $(Z_\infty,\zeta_\infty)\in A(J^\infty Y)$, and $w\in F$. We have an equivalence relation $\sim$ in the set, defined as follows:

$$(2.6)\qquad (Z_\infty,\zeta_\infty,\bar{w})\sim(\bar{Z}_\infty,\bar{\zeta}_\infty,w)\Leftrightarrow D\bar{\zeta}_\infty\zeta_\infty^{-1}(\zeta_\infty(j_x^\infty\gamma))\cdot w=\bar{w}.$$

In agreement with the usual terminology [26], the equivalence class with respect to the equivalence relation is called the *tangent vector* to the manifold $J^\infty Y$ at the point $j_x^\infty\gamma$. In this Section we shall write $\{Z_\infty,\zeta_\infty,w\}_{j_x^\infty\gamma}$ for the tangent vectors containing the triple $(Z_\infty,\zeta_\infty,w)$. The set of all tangent vectors at a point $j_x^\infty\gamma$ is called the *tangent space* to the manifold $J^\infty Y$ at $j_x^\infty\gamma$ , and is denoted $T_{j_x^\infty\gamma}J^\infty Y$.

In the following we shall work with the vector space $L_b(F,F)$ of *continuous* linear mappings from $F$ to $F$ with the topology of uniform convergence on all bounded subsets of $F$; it can be checked that $L_b(F,F)$ is a Frechet space [33] (see also [5]).

Put

$$TJ^\infty Y=\bigcup_{j_x^\infty\gamma}T_{j_x^\infty\gamma}J^\infty Y,\quad \tau(\{Z_\infty,\zeta_\infty,w\}_{j_x^\infty\gamma})=j_x^\infty\gamma,$$

and for an arbitrary fibered chart $(V_0,\psi_0)\in A(Y)$ with the corresponding canonical chart $(Z_{0\infty},\zeta_{0\infty})$ on $J^\infty Y$,

$$(2.7)\qquad \begin{aligned} &T_{j_x^\infty\gamma}\zeta_{0,\infty}\cdot\{Z_\infty,\zeta_\infty,w\}_{j_x^\infty\gamma}=D\zeta_{0,\infty}\zeta^{-1}(\zeta(j_x^\infty\gamma))\cdot w,\\ &T\zeta_{0,\infty}(\{Z_\infty,\zeta_\infty,w\}_{j_x^\infty\gamma})=(\zeta_{0,\infty}(j_\infty\gamma),T_{j_x^\infty\gamma}\zeta_{0,\infty}\cdot\{Z_\infty,\zeta_\infty,w\}_{j_x^\infty\gamma}).\end{aligned}$$

*Consider the set* $TJ^\infty Y$ *with the uniquely determined topology defined by the assumption that for each* $(Z,\zeta)\in A(J^\infty Y)$, $\tau^{-1}(Z)$ *is an open set and* $T\zeta$ *is a homeomorphism. Then the triple* $(TJ^\infty Y,\tau,J^\infty Y)$ *has the following properties:*

1) *If* $\mathscr{A}(J^\infty Y)=\{(Z_\iota,\zeta_\iota)\}$ *is an atlas on* $J^\infty Y$, *then* $\mathscr{A}(TJ^\infty Y)=\{(\tau^{-1}(Z_\iota),T\zeta_\iota)\}$ *is an atlas on* $TJ^\infty Y$ *(in the sense of* [39]*).*

2) *For each* $j_x^\infty\gamma\in J^\infty Y$, *the mapping* $T_{j_x^\infty\gamma}\zeta_0$ *defined by chart* $(Z_0,\zeta_0)$ *with center* $j_x^\infty\gamma$, *is a bijection and defines in* $T_{j_x^\infty\gamma}J^\infty Y$ *the structure of a locally convex space isomorphic to* $F$.

3) *Denote* $pr_\iota: Z_\iota\times F\to F$ *the natural projection. To each* $\iota$ *there is an isomorphism* $\rho_\iota:\tau^{-1}(Z_\iota)\to Z_\iota\times F$ *such that on* $\tau^{-1}(Z_\iota)$, $\tau=pr_\iota\circ\rho_\iota$, *and that the restriction* $\rho_\iota|_{j_x^\infty\gamma}$ *of* $\rho_\iota$ *to* $\tau^{-1}(j_x^\infty\gamma)$ *is an isomorphism of locally convex spaces. Furthermore, for* $Z_\iota\cap Z_\kappa\neq\varnothing$, *the mapping*

$$Z_\iota\cap Z_\varkappa\ni j_x^\infty\gamma\to(\rho_\iota\circ\rho_\varkappa^{-1})|_{j_x^\infty\gamma}\in L_b(F,F)$$

*is a morphism* (i.e. is differentiable).

The proof of all these assertions is straightforward, and is based on [39], [33], [5]. With the defined differentiable structure, the triple $(TJ^\infty Y,\tau,J^\infty Y)$ is said to be the *tangent bundle* of $J^\infty Y$.

**The total tangent bundle.** Let $j_x^\infty\gamma\in J^\infty Y$ be a point and let $(Z,\zeta)$ be a canonical chart with center $j_x^\infty\gamma$ associated with a fibered chart $(V,\psi)$ on $Y$. Let us write $\psi=(\varphi,\psi_0)$ as before, and consider a vector $w=(v,w_0,w_1,\ldots)\in F$ satisfying

$$w_0=D\psi_0\gamma\varphi^{-1}(\varphi(x))\cdot v,\quad w_1=D^2\psi_0\gamma\varphi^{-1}(\varphi(x))\cdot v,\ldots,$$

or, which is the same,

$$w=D\zeta J^\infty\gamma\varphi^{-1}(\varphi(x))\cdot v. \tag{2.8}$$

It follows from the chain rule for derivatives that the following assertion holds: the class of equivalence (with respect to the relation (2.6)) containing $(Z,\zeta,w)$ satisfying $w=D\zeta J^\infty\gamma\varphi^{-1}(\varphi(x))\cdot v$ is equal to the equivalence class containing $(\bar Z,\bar\zeta,\bar w)$ if and only if $w=D\bar\zeta\zeta^{-1}(\zeta(j_x^\infty\gamma))\cdot w$, i.e. if and only if $\bar v=D\bar\varphi\varphi^{-1}(\varphi(x))\cdot v$. This means that the condition (2.8) is invariant; each tangent vector $\{Z,\zeta,w\}_{j_x^\infty\gamma}$ satisfying (2.8) is called the *total tangent vector* on $J^\infty Y$. We shall write $tT_{j_x^\infty\gamma}J^\infty Y$ for the set of all total tangent vectors at a point $j_x^\infty\gamma$ and introduce the notation

$$tTJ^\infty Y=\bigcup_{j_x^\infty\gamma}tT_{j_x^\infty\gamma}J^\infty Y.$$

Let $\tau_t$ denotes the restriction of $\tau$ to the set $tTJ^\infty Y$. In this paragraph we shall briefly describe certain properties of the triple $(tTJ^\infty Y,\tau_t,J^\infty Y)$.

Consider an atlas $\mathscr{A}(J^\infty Y)$ on $J^\infty Y$ consisting of canonical charts.

*Let* $\mathscr{A}(tTJ^\infty Y)$ *be the set of all pairs* $(\tau_\iota^{-1}(Z_\iota),tT\zeta_\iota)$, *where* $tT\zeta_\iota$ *is a mapping from* $\tau_\iota^{-1}(Z_\iota)$ *to* $\zeta_\iota(Z_\iota)\times\mathbf{R}^n$ *defined as*

(2.9) $$tT\zeta_\iota(\{Z,\zeta,D\zeta J^\infty\gamma\varphi^{-1}(\varphi(x))\cdot v\}_{j_x^\infty\gamma})=(\zeta_\iota(j_x^\infty\gamma,D\varphi_\iota\varphi^{-1}(\varphi(x))\cdot v).$$

*Then:*

1) $\bigcup_\iota \tau_\iota^{-1}(Z_\iota)=tTJ^\infty Y$.
2) *For each* $\iota$ *the mapping* $tT\zeta_\iota$ *is bijective.*
3) *The mapping* $tT\zeta_\iota\circ(tT\zeta_\kappa)^{-1}$ *is of the form*

(2.10) $$\begin{aligned}&(\zeta_\kappa(j_x^\infty\gamma),D\varphi_\kappa\varphi^{-1}(\varphi(x))\cdot v)\to\\&\quad\to(\zeta_\iota\zeta_\kappa^{-1}(\zeta_\kappa(j_x^\infty\gamma)),D\varphi_\iota\varphi_\kappa^{-1}(\varphi_\kappa(x))\cdot D\varphi_\kappa\varphi^{-1}(\varphi(x))\cdot v),\end{aligned}$$

*and is therefore a morphism.*

To put it differently, the triple $(tTJ^\infty Y,\tau_t,J^\infty Y)$ has the properties of the vector bundle (compare with [26]). We call it the on $J^\infty Y$.

(See [39], [26], [16], and compare with approach in [3].)

**The bundle of vertical vectors on** $J^\infty Y$. With similar arguments as in the previous paragraph we call each tangent vector $\{Z,\zeta,w\}_{j_x^\infty\gamma}\in TJ^\infty Y$ satisfying

$$w=(0,w_0,w_1,\ldots)$$

a *vertical vector* on $J^\infty Y$. The set of all vertical vectors at a point $j_x^\infty\gamma$ is denoted $vT_{j_x^\infty\gamma}J^\infty Y$, and we put

$$vTJ^\infty Y=\bigcup_{j_x^\infty\gamma}vT_{j_x^\infty\gamma}J^\infty Y.$$

The triple $(vTJ^\infty Y,\tau_v,J^\infty Y)$ with $\tau_v$ denoting the restriction of $\tau$ to $vTJ^\infty Y$, has analogous properties as, say, $(TJ^\infty Y,\tau,J^\infty Y)$ ($vTJ^\infty Y$ is of course a manifold modeled on the locally convex space $F\times\mathbf{R}^m\times\prod L_s^k(\mathbf{R}^n,\mathbf{R}^m)$). We call the triple $(vTJ^\infty Y,\tau_v,J^\infty Y)$ the *bundle of vertical vectors* on $J^\infty Y$.

**Morphisms of locally convex manifolds.** A *morphism* $f:P\to Q$ of locally convex manifolds is defined formally in the same way as in the finite-dimensional case. If $f$ is a morphism, one can define the *tangent morphism* $Tf:TP\to TQ$. Let $p\in P$ be a point, $(U,\varphi)$ a chart with center $p$, $(V,\psi)$ a chart with center $f(p)$. Then

(2.11) $$Tf(\{U,\varphi,u\}_p)=\{V,\psi,D\psi f\varphi^{-1}(\varphi(p))\cdot u\}_{f(p)}.$$

Let us now define what we mean by a morphism of (finite dimensional) fibered manifolds. Let $(Y,\pi,X)$ and $(\overline{Y},\overline{\pi},\overline{X})$ be two fibered manifolds. A pair $(f,f_0)$ is called a *morphism* of these fibered manifolds, if $f:Y\to\overline{Y}$ and $f_0:X\to\overline{X}$ are morphisms of differentiable manifolds, and

$$\overline{\pi}\circ f=f_0\circ\pi.$$

(See [26], [39].)

**Additional remarks.** It can be shown that both $(tTJ^{\infty}Y, \tau_t, J^{\infty}Y)$ and $(vTJ^{\infty}Y, \tau_v, J^{\infty}Y)$ can be considered as subbundles of the tangent bundle $(TJ^{\infty}Y, \tau, J^{\infty}Y)$ (in the sense of definitions analogous to the case of manifolds modeled on Banach spaces [26]). The space $T_{j_x^{\infty}\gamma} J^{\infty}Y$ is the topological direct product of its subspaces $tT_{j_x^{\infty}\gamma} J^{\infty}Y$ and $vT_{j_x^{\infty}\gamma} J^{\infty}Y$; we write

$$T_{j_x^{\infty}\gamma} J^{\infty}Y = tT_{j_x^{\infty}\gamma} J^{\infty}Y \oplus vT_{j_x^{\infty}\gamma} J^{\infty}Y,$$

or

(2.12) $$TJ^{\infty}Y = tTJ^{\infty}Y \oplus vTJ^{\infty}Y.$$

Let us consider the pull-back $(\pi_{\infty}^{*}TX, \pi_{\infty}^{*}\tau_X, J^{\infty}Y)$ [26]. It is immediately clear that the mapping

(2.13) $$\begin{aligned} &\pi_{\infty}^{*}TX \ni (j_x^{\infty}\gamma, \{U,\varphi,u\}_x) \to T_x j^{\infty}\gamma \cdot \{U,\varphi,u\}_x \\ &= \{Z,\zeta, D\zeta j^{\infty}\gamma\varphi^{-1}(\varphi(p))\cdot u\}_{j_x^{\infty}\gamma} \in tTJ^{\infty}Y \end{aligned}$$

(together with the identity of $J^{\infty}Y$) is an isomorphism of vector bundles $(\pi_{\infty}^{*}TX, \pi_{\infty}^{*}\tau_X, J^{\infty}Y)$, $(tTJ^{\infty}Y, \tau_t, J^{\infty}Y)$.

**The bundle of covariant antisymmetric tensors on $J^{\infty}Y$.** We are now going to consider another type of bundles, represented by the bundle of multilinear forms on $J^{\infty}Y$. Since we use only some differential forms of a special type on $J^{\infty}Y$, particularly those which can be obtained as the pull-back by the mappings $\pi_{\infty,r}$,, we are not in need to give quite general definitions. We define what could be called the bundle of $p$-linear *continuous* antisymmetric forms on $J^{\infty}Y$ (also separately continuous forms could be treated).

Let $F$ be our locally convex space. It can be proved that the vector space $L_a^p(F,\mathbf{R})$ of all *continuous* antisymmetric $p$-linear forms on $F$ with the topology of uniform convergence on all bounded subsets of $F$ is a Frechet space [5], [33].

Let $j_x^{\infty}\gamma \in J^{\infty}Y$, and consider the set of all triples $(Z,\zeta,u)$, where $(Z,\zeta)$ is a chart on $J^{\infty}Y$ with center $j_x^{\infty}\gamma$ and $u \in L_a^p(F,\mathbf{R})$. In this set there is an equivalence relation $\sim$ defined as follows:

(2.14) $$(Z,\zeta,u) \sim (\bar{Z},\bar{\zeta},\bar{u}) \Leftrightarrow \bar{u} \circ (D\bar{\zeta}\zeta^{-1}(\zeta(j^{\infty}Y)),\ldots D\bar{\zeta}\zeta^{-1}(\zeta(j^{\infty}Y))) = u.$$

The equivalence class containing $(Z,\zeta,u)$ is denoted by $\{Z,\zeta,u\}_{j_x^{\infty}\gamma}$. Further, we set

(2.15) $$\begin{aligned} &\left\langle \{Z,\zeta,u\}_{j_x^{\infty}\gamma}, \{Z,\zeta,w_1\}_{j_x^{\infty}\gamma} \times \{Z,\zeta,w_2\}_{j_x^{\infty}\gamma} \times \ldots \times \{Z,\zeta,w_p\}_{j_x^{\infty}\gamma} \right\rangle \\ &= \langle u, w_1 \times w_2 \times \ldots \times w_p \rangle, \end{aligned}$$

where the right-hand side is equal to $u(w_1, w_2, \ldots w_p)$. Thus, the equivalence class $\{Z,\zeta,u\}_{j_x^{\infty}\gamma}$ defines a $p$-linear continuous antisymmetric form on $T_{j_x^{\infty}\gamma} J^{\infty}Y$. The set of all such classes is denoted $L_a^p T_{j_x^{\infty}\gamma} J^{\infty}Y$, and we write

$$L_a^p TJ^{\infty}Y = \bigcup_{j_x^{\infty}\gamma} L_a^p T_{j_x^{\infty}\gamma} J^{\infty}Y, \quad \rho_{\infty}(\{Z,\zeta,u\}_{j_x^{\infty}\gamma}) = j_x^{\infty}\gamma.$$

If $(Z_0,\zeta_0)$ is a chart on $J^\infty Y$ with center $j^\infty_x\gamma$, define

(2.16) $$\zeta^p_0(\{Z,\zeta,u\}_{j^\infty_x\gamma})=(\zeta_0(j^\infty_x\gamma),u_0(D\zeta\zeta_0^{-1}(\zeta_0(j^\infty_x\gamma)),\ldots,D\zeta\zeta_0^{-1}(\zeta_0(j^\infty_x\gamma))));$$

the mapping $\zeta^p_0$ takes values in $\zeta_0(Z_0)\times L^p_a(F,\mathbf{R})$. If $\mathscr{A}(J^\infty Y)=\{(Z_\iota,\zeta_\iota)\}$ is an atlas on $J^\infty Y$, denote $\mathscr{A}(L^p_a TJ^\infty Y)$ the set of all pairs $(\rho^{-1}_\infty(Z_\iota),\zeta^p_\iota)$.

*The following conditions take place:*

1) $\bigcup_\iota \rho^{-1}(Z_\iota)=L^p_a TJ^\infty Y$.
2) $\zeta^p_\iota$ *is a bijection, for all* $\iota$.
3) T*he mappings* $\zeta^p_\iota\circ(\zeta^p_\kappa)^{-1}$ *are morphisms.*

In other words, $\mathscr{A}(L^p_a TJ^\infty Y)$ is a *locally convex atlas* on $L^p_a TJ^\infty Y$*;* with the structure defined by this atlas and with the obvious property

4) *the restriction of* $\zeta^p_\iota$ *to* $\rho^{-1}_\infty(j^\infty_x\gamma)$, $j^\infty_x\gamma\in Z_\iota$, *is a linear isomorphism of the Frechet spaces* $L^p_a T_{j^\infty_x\gamma} J^\infty Y$, $L^p_a(F,\mathbf{R})$, we call the triple $(L^p_a TJ^\infty Y,\rho_\infty,J^\infty Y)$ the *bundle of covariant antisymmetric tensors* on $J^\infty Y$.

## 3. Jet prolongations of vector fields

**Projectable vector fields.** Consider a fibered manifold $(Y,\pi,X)$. Let $\Xi$ be a vector field on $Y$ and $\xi$ a vector field on $X$. The pair $(\Xi,\xi)$ is said to be $\pi$*-related,* if for any $y\in Y$

(3.1) $$T\pi\cdot\Xi(y)=\xi(\pi(y)).$$

The equality is also written as $T\pi\circ\Xi=\xi\circ\pi$. Denote $\alpha^\Xi_t$ (res. $\alpha^\xi_t$) the local one-parameter group of transformations generated by $\Xi$ (res. $\xi$). The pair $(\Xi,\xi)$ is $\pi$-related if and only if

(3.2) $$\pi\circ\alpha^\Xi_t=\alpha^\xi_t\circ\pi.$$

Let $(V,\psi)$, $\psi=(\varphi,\psi_0)$ be a fibered chart on $Y$ (2.1). Then the pair $(\Xi,\xi)$ is $\pi$-related if and only if $\Xi$ and $\xi$ are expressed by means of the fibered chart as

(3.3) $$\xi=\xi_k\frac{\partial}{\partial x_k},\quad \Xi=\xi_k\frac{\partial}{\partial x_k}+\Xi_\mu\frac{\partial}{\partial y_\mu}.$$

It is clear that if to a given vector field $\Xi$ on $Y$ there exists a vector field $\xi$ on $X$ such that the pair $(\Xi,\xi)$ is $\pi$-related, then $\xi$ is uniquely determined by this condition. Each vector field $\Xi$ on $Y$ with the property that such the $\xi$ does exist, is called *projectable.*

**Jet prolongations of local automorphisms.** Let $(\alpha,\alpha_0)$ be a local automorphism of the fibered manifold $(Y,\pi,X)$. This means that $\alpha$ is defined on an open set $V\subset Y$, $\alpha_0$ is defined on $\pi(V)\subset X$, $\alpha$ and $\alpha_0$ are isomorphisms, and, according to the defi-nition of morphisms of fibered manifolds,

$$\pi \circ \alpha = \alpha_0 \circ \pi$$

on $V$. Let $r$ be a positive integer. $(\alpha, \alpha_0)$ defines a local automorphism $(j^r\alpha, \alpha_0)$ of the fibered manifold $(J^r, \pi_r, X)$ by the formula

$$j^r\alpha(j^r_x\gamma) = j^r_{\alpha_0(x)}\alpha\gamma\alpha_0^{-1}. \tag{3.4}$$

Let $(V, \psi)$ be a fibered chart on $Y$ with center $\gamma(x)$ and $(\overline{V}, \overline{\psi})$ a fibered chart with center $\alpha(\gamma(x))$. Then we can write by means of the corresponding canonical charts on $J^r$ (2.3)

$$\begin{aligned} &\overline{\zeta}_r(j^r_{\alpha_0(x)}\alpha\gamma\alpha_0^{-1}) = (\overline{\varphi}\alpha_0\varphi^{-1}(\varphi(x)), \overline{\psi}_0\alpha\psi^{-1}(\psi(\gamma(x))), \\ &\quad D\overline{\psi}_0\alpha\psi^{-1} \circ \psi\gamma\varphi^{-1} \circ \varphi\alpha_0\overline{\varphi}^{-1})(\overline{\varphi}\varphi^{-1}(\varphi(x))), \dots, \\ &\quad D^r\overline{\psi}_0\alpha\psi^{-1} \circ \psi\gamma\varphi^{-1} \circ \varphi\alpha_0\overline{\varphi}^{-1})(\overline{\varphi}\varphi^{-1}(\varphi(x))). \end{aligned} \tag{3.5}$$

It follows that 1) the right-hand side of (3.4) really depends only on $j^r_x\gamma$ (by the chain rule [9]), 2) $j^r\alpha$ depends differentiably on $j^r_x\gamma$, 3) $j^r\alpha$ is an isomorphism on $\pi^{-1}_{r,0}(V)$. Furthermore, the relations

$$\pi_r \circ j^r\alpha = \alpha_0 \circ \pi_r, \tag{3.6}$$

$$\pi_{r,s} \circ j^r\alpha = j^s\alpha \circ \pi_{r,s} \tag{3.7}$$

hold on $\pi^{-1}_{r,0}(V)$. The pair $(j^r\alpha, \alpha_0)$ is thus a local automorphism of the fibered manifold $(J^r, \pi_r, X)$; we call it the *r-jet prolongation* of the local automorphism $(\alpha, \alpha_0)$ of $(Y, \pi, X)$.

In the previous paragraph there were introduced the so called projectable vector fields on $Y$. They are characterized by the property that their local one-parameter groups are just local automorphisms of $(Y, \pi, X)$. We can prolong the local automorphisms by the described procedure and thus obtain some local one-parameter groups of local automorphisms of $(J^r, \pi_r, X)$. In turn we will be led to certain vector fields defined on the space $J^r$.

**Jet prolongations of vector fields.** Let $\Xi$ be a projectable vector field on $Y$. Its *r-jet prolongation* is a vector field on $J^r$ defined by

$$j^r\Xi(j^r_x\gamma) = \left\{\frac{d}{dt} j^r_{\alpha^\xi_t(x)}\alpha^\Xi_t\gamma\alpha^\xi_{-t}\right\}_0. \tag{3.8}$$

Let $(V, \psi)$ be a fibered chart on $Y$, $(x_i, y_\mu)$ the coordinates defined by this chart, and consider the canonical chart on $J^r$ associated with $(V, \psi)$. Assume that $\Xi$ is represented by (3.3). By definition,

$$\begin{aligned} &x_i(j^r_{\alpha^\xi_t(x)}\alpha^\Xi_t\gamma\alpha^\xi_{-t}) = x_i(\alpha^\xi_t(x)), \\ &y_\mu(j^r_{\alpha^\xi_t(x)}\alpha^\Xi_t\gamma\alpha^\xi_{-t}) = y_\mu(\alpha^\Xi_t(\gamma(x))), \end{aligned} \tag{3.9}$$

$$z_{i\mu}(j^r_{\alpha_t^\xi(x)}\alpha_t^\Xi\,\gamma\alpha_{-t}^\xi)=D_i(y_\mu\circ\alpha_t^\Xi\,\gamma\alpha_{-t}^\xi\varphi^{-1})(\varphi(\alpha_t^\xi(x))),$$
$$\ldots$$
$$z_{i_1i_2\ldots i_r\mu}(j^r_{\alpha_t^\xi(x)}\alpha_t^\Xi\,\gamma\alpha_{-t}^\xi)=D_{i_1}D_{i_2}\ldots D_{i_r}(y_\mu\circ\alpha_t^\Xi\,\gamma\alpha_{-t}^\xi\varphi^{-1})(\varphi(\alpha_t^\xi(x))),$$

where $D_i$ denotes the *i*-th partial derivative. Evidently the identity

$$\begin{aligned}&D_qD_{i_1}D_{i_2}\ldots D_{i_s}(y_\mu\circ\alpha_t^\Xi\gamma\alpha_{-t}^\xi\varphi^{-1})(\varphi(\alpha_t^\xi(x)))\\&\quad=D_l(D_{i_1}D_{i_2}\ldots D_{i_s}(y_\mu\circ\alpha_t^\Xi\gamma\alpha_{-t}^\xi\varphi^{-1})\circ\varphi\alpha_t^\xi\varphi^{-1})(x_1,x_2,\ldots,x_n)\\&\quad\cdot D_q(x_l\circ\alpha_{-t}^\xi\varphi^{-1})(\varphi(\alpha_t^\xi(x)))\end{aligned}$$

holds so that

$$\begin{aligned}&\left\{\frac{d}{dt}D_{i_1}D_{i_2}\ldots D_{i_s}(y_\mu\circ\alpha_t^\Xi\,\gamma\alpha_{-t}^\xi\varphi^{-1})(\varphi(\alpha_t^\xi(x)))\right\}_0\\&\quad=\left\{\frac{d}{dt}D_l(D_{i_1}D_{i_2}\ldots D_{i_{s-1}}(y_\mu\circ\alpha_t^\Xi\,\gamma\alpha_{-t}^\xi\varphi^{-1})\circ\varphi\alpha_t^\xi\,\varphi^{-1})(x_1,x_2,\ldots,x_n)\right\}_0\delta_{i_sl}\\&\quad+z_{li_1i_2\ldots i_{s-1}\mu}\cdot\left\{\frac{d}{dt}D_{i_s}(x_l\circ\alpha_{-t}^\xi\varphi^{-1})(\varphi(\alpha_t^\xi(x)))\right\}_0.\end{aligned}$$

Here $\delta_{il}$ stands for the Kronecker symbol. But

$$\left\{\frac{d}{dt}D_{i_s}(x_l\circ\alpha_{-t}^\xi\varphi^{-1})(\varphi(\alpha_t^\xi(x)))\right\}_0+D_{i_s}\zeta_l=0, \tag{3.10}$$

and we can write

$$\begin{aligned}&\left\{\frac{d}{dt}D_{i_1}D_{i_2}\ldots D_{i_s}(y_\mu\circ\alpha_t^\Xi\gamma\alpha_{-t}^\xi\varphi^{-1})(\varphi(\alpha_t^\xi(x)))\right\}_0\\&\quad=D_{i_s}\left\{\frac{d}{dt}D_{i_1}D_{i_2}\ldots D_{i_{s-1}}(y_\mu\circ\alpha_t^\Xi\gamma\alpha_{-t}^\xi\varphi^{-1})\circ\varphi\alpha_t^\xi\varphi^{-1})\right\}_0(x_1,x_2,\ldots,x_n)\\&\quad-z_{li_1i_2\ldots i_{s-1}\mu}\cdot D_{i_s}\zeta_l.\end{aligned} \tag{3.11}$$

If we now write

$$j^r\Xi=\xi_k\frac{\partial}{\partial x_k}+\Xi_\mu\frac{\partial}{\partial y_\mu}+\Xi_{i\mu}\frac{\partial}{\partial z_{i\mu}}+\ldots+\sum_{i_1\le i_2\le\ldots\le i_r}\Xi_{i_1i_2\ldots i_r\mu}\frac{\partial}{\partial z_{i_1i_2\ldots i_r\mu}} \tag{3.12}$$

for the coordinate expression of $j^r\Xi$ we see that (3.11) represents a recurrent formula for the components $\Xi_\mu$, $\Xi_{i\mu}$, … of the vector field. For the sake of simplicity of notation, introduce the abbreviation

$$\frac{df}{dx_i}=\frac{\partial f}{\partial x_k}+\frac{\partial f}{\partial y_\mu}z_{i\mu}+\ldots+\sum_{i_1\le i_2\le\ldots\le i_r}\frac{\partial f}{\partial z_{i_1i_2\ldots i_r\mu}}z_{ii_1i_2\ldots i_r\mu} \tag{3.13}$$

( $f$ - arbitrary function on $\pi_{r,0}^{-1}(V)$). According to [23], we shall call the expression

(3.13) the *formal derivative* of $f$ by $x_i$. Some useful properties of the formal derivative are derived in [23]. Returning to our coordinate expression (3.12), we see that

$$\Xi_{i_0 i_1 i_2 \dots i_s \mu} = \frac{d\Xi_{i_1 i_2 \dots i_s \mu}}{dx_{i_0}} - z_{l i_1 i_2 \dots i_r \mu} \frac{\partial \xi_l}{\partial x_{i_0}}. \tag{3.14}$$

Applying the formula we immediately have

$$\begin{aligned} \Xi_{i\mu} &= \frac{d\Xi_{\mu}}{dx_i} - z_{i\mu} \frac{\partial \xi_l}{\partial x_i}, \\ \Xi_{ij\mu} &= \frac{d}{dx_i}\left(\frac{d\Xi_{\mu}}{dx_j}\right) - z_{il\mu}\frac{\partial \xi_l}{\partial x_j} - z_{l\mu}\frac{\partial^2 \xi_l}{\partial x_i \partial x_j} - z_{lj\mu}\frac{\partial \xi_l}{\partial x_i}. \end{aligned} \tag{3.15}$$

(Compare with [20], [38].)

**Infinite jet prolongation of local automorphisms and vector fields.** Let $(\alpha,\alpha_0)$ be a local automorphism of the fibered manifold $(Y,\pi,X)$. It has been shown that for each positive integer $r$ one can construct a local automorphism $(j^r\alpha,\alpha_0)$ of the fibered manifold $(J^r,\pi_r,X)$ such that for $s \le r$ (3.7)

$$\pi_{r,s} \circ j^r\alpha = j^s\alpha \circ \pi_{r,s}.$$

Consider the open sets $V$ and $\pi(V) = U$ on which $\alpha$ and $\alpha_0$ are defined, respectively. Each mapping $j^r\alpha$ can be composed with the projection $\pi_{\infty,r}$. Thus, for each $r$ we get a morphism $j^r\alpha \circ \pi_{\infty,r} : \pi_{\infty,0}^{-1}(V) \to j^r\alpha(\pi_{r,0}^{-1}(V))$; for $s \le r$

$$j^s\alpha \circ \pi_{\infty,s} = j^s\alpha \circ \pi_{r,s} \circ \pi_{\infty,r} = \pi_{r,s} \circ (j^r\alpha \circ \pi_{\infty,r}).$$

This shows that there is a unique morphism $j^\infty\alpha : \pi_{\infty,0}^{-1}(V) \to J^\infty$ satisfying the condition

$$\pi_{\infty,r} \circ j^\infty\alpha = j^r\alpha \circ \pi_{\infty,r} \tag{3.16}$$

for each $r$ (see [39], III. 3.2). The same argument may be applied to $\alpha^{-1}$. The considerations show that $j^\infty\alpha$ is an isomorphism, commuting with $\pi_\infty$. Analogously as in in the finite case, the pair $(j^\infty\alpha,\alpha_0)$ will be called the *infinite jet prolongation* of the local automorphism $(\alpha,\alpha_0)$.

Let $\Xi$ be a projectable vector field on $Y$ with $\xi$ defined by the condition $T\pi \circ \Xi = \xi \circ \pi$. Consider the curve $t \to j^\infty_{\alpha_t^\xi(x)} \alpha_t^\Xi \gamma \alpha_{-t}^\xi$ in $J^\infty$. The curve is a morphism so that there exists its tangent vector

$$j^\infty\Xi(j^\infty_x\gamma) = \left\{\frac{d}{dt} j^\infty_{\alpha_t^\xi(x)} \alpha_t^\Xi \gamma \alpha_{-t}^\xi\right\}_0. \tag{3.17}$$

This gives rise to a vector field $j^\infty_x\gamma \to j^\infty\Xi(j^\infty_x\gamma)$ on $J^\infty$ which is called the *infinite jet prolongation* of the projectable vector field $\Xi$. It is clear that

$$T\pi_{\infty,r} \circ j^\infty\Xi = j^r\Xi \circ \pi_{\infty,r}. \tag{3.18}$$

**Total and vertical vector fields on** $J^\infty Y$**.** *Total vector fields* on $J^\infty$ are cross sections of the fibered manifold $(tTJ^\infty Y, \tau_t, J^\infty Y)$. *Vertical vector fields* on $J^\infty$ are cross sections of the fibered manifold $(vTJ^\infty Y, \tau_v, J^\infty Y)$.

Let $\Xi$ be a projectable vector field on $Y$ and define $\xi$ by the relation $T\pi \circ \Xi = \xi \circ \pi$. We have, at any point $j_x^\infty \gamma \in J^\infty$, an identity

$$j^\infty \Xi(j_x^\infty \gamma) = T_x j^\infty \gamma \cdot \xi(x) + (j^\infty \Xi(j_x^\infty \gamma) - T_x j^\infty \gamma \cdot \xi(x)).$$

Let us denote

$$(3.19)\qquad \begin{aligned} t(j^\infty \Xi)(j_x^\infty \gamma) &= T_x j^\infty \gamma \cdot \xi(x), \\ v(j^\infty \Xi)(j_x^\infty \gamma) &= j^\infty \Xi(j_x^\infty \gamma) - T_x j^\infty \gamma \cdot \xi(x). \end{aligned}$$

Clearly, these vectors are well-defined, i.e. do not depend on the representatives used for their definitions. We also use the notation

$$(3.20)\qquad \xi^\infty (j_x^\infty \gamma) = t(j^\infty \Xi)(j_x^\infty \gamma).$$

It can be proved that the mappings $j_x^\infty \gamma \to \xi^\infty(j_x^\infty \gamma)$, $j_x^\infty \gamma = v(j^\infty \Xi)(j_x^\infty \gamma)$ are morphisms. Thus, $\xi^\infty$ is a total vector field on $J^\infty$ and $v(j^\infty \Xi)$ is a vertical vector field on $J^\infty$. One must be careful, however, when working with these vectorsfields. In general (when $\xi \neq 0$) the notion of the local one-parameter group cannot be joined with them, at least not in the usual sense. We shall only need the formula

$$(3.21)\qquad j^\infty \Xi = \xi^\infty + v(j^\infty \Xi)$$

for the invariant decomposition of $j^\infty \Xi$ into the total and vertical parts.

**Example.** It is known how the local one-parameter group of a vector field on a differentiable manifold can be used to construct a vector field on each tensor bundle over this manifold ([36], Chap. II, § 8). Since the vector fields obtained by this procedure are projectable (with respect to the natural projection on the differentiable manifold), they can be prolonged to the corresponding jet spaces. Such vector fields are of great interest in the calculus of variations in fibered manifolds. We wish to write up explicitly the coordinate expressions for their 1-jet prolongation.

It is not so difficult to perform the calculations in full generality. Let $\sigma$ denote the *type* of tensors which will be considered; in detail, $\sigma$ may be regarded as the set $\{1,2,\dots,r;r+1,r+2,\dots,r+s\}$ of indices, in which some of them, say, the first *r*, are regarded as *contravariant.* The corresponding tensor bundle of a manifold *X* is denoted $(T_\sigma X, \tau_\sigma, X)$. To simplify our notiation,, we write $\otimes\{k_1,k_2,\dots,k_r;j_1,j_2,\dots,j_s\}_\sigma$ for the base vectors defined by a chart on *X* with center *x*; $\otimes\{k_1,k_2,\dots,k_r;j_1,j_2,\dots,j_s\}_\sigma$ is simply some *tensor product* of the base vectors $\partial/\partial x_k$, $dx_j$, with the order prescribed by $\sigma$. Notice that the symbols $k_i$ are used for the contravariant indices.

Let $\xi$ be a vector field on *X*. By means of a chart $(U,\varphi)$, $\varphi = (x_1, x_2, \dots, x_n)$, its local one-parameter group $\alpha_t^\xi$ is expressed as

$$(x_i \circ \alpha_t^\xi)(x) = f_i(t, x_1, x_2, \dots, x_n).$$

It is known that the induced one-parameter group acting on each space of tensors on

$X$ (the action being written multiplicatively) satisfies the rules

1) $\alpha_t^\xi \cdot (u \otimes v) = (\alpha_t^\xi \cdot u) \otimes (\alpha_t^\xi \cdot v)$,

2) $\alpha_t^\xi \cdot \dfrac{\partial}{\partial x_i} = \dfrac{\partial f_k}{\partial x_i} \dfrac{\partial}{\partial x_k}$,

3) $\alpha_t^\xi \cdot dx_j = \dfrac{\partial f_j}{\partial x_l} dx_l$,

taking place for any tensors $u$ and $v$ (of course we must have in mind that the tensors $\partial/\partial x_k$, $dx_j$, are defined at the point $x$, while the same symbols $\partial/\partial x_k$, $dx_j$ on the right-hand side are used for tensors at the point $\alpha_t^\xi(x)$). It follows that

$$\alpha_t^\xi \cdot \otimes \{k_1, k_2, \ldots, k_r; j_1, j_2, \ldots, j_s\}_\sigma$$
$$= \frac{\partial f_{i_1}}{\partial x_{k_1}} \frac{\partial f_{i_2}}{\partial x_{k_2}} \cdots \frac{\partial f_{i_r}}{\partial x_{k_r}} \cdot \frac{\partial f_{j_1}}{\partial x_{l_1}} \frac{\partial f_{j_2}}{\partial x_{l_2}} \cdots \frac{\partial f_{j_s}}{\partial x_{l_s}} \cdot \otimes_{\{i_1, i_2, \ldots, i_r; l_1, l_2, \ldots, l_s\}_\sigma}.$$

If $y_{\{i_1, i_2, \ldots, i_r; l_1, l_2, \ldots, l_s\}_\sigma}$ stands for the coordinate functions on $T_\sigma X$ defined by the base vectors $\{i_1, i_2, \ldots, i_r; l_1, l_2, \ldots, l_s\}_\sigma$, the induced mapping, denoted now by $\overline{\alpha}_t^\xi$, has the following coordinate expression:

$$x_i \circ \overline{\alpha}_t^\xi = f_i,$$
$$y_{\{i_1, i_2, \ldots, i_r; l_1, l_2, \ldots, l_s\}_\sigma} \circ \overline{\alpha}_t^\xi$$
$$= y_{\{k_1, k_2, \ldots, k_r; j_1, j_2, \ldots, j_s\}_\sigma} \frac{\partial f_{i_1}}{\partial x_{k_1}} \frac{\partial f_{i_2}}{\partial x_{k_2}} \cdots \frac{\partial f_{i_r}}{\partial x_{k_r}} \cdot \frac{\partial f_{j_1}}{\partial x_{l_1}} \frac{\partial f_{j_2}}{\partial x_{l_2}} \cdots \frac{\partial f_{j_s}}{\partial x_{l_s}}.$$

Differentiating with respect to $t$ at $t = 0$ we get

$$\left\{ \frac{d}{dt} x_i \circ \overline{\alpha}_t^\xi \right\}_0 = \xi_i,$$
$$\left\{ \frac{d}{dt} y_{\{i_1, i_2, \ldots, i_r; l_1, l_2, \ldots, l_s\}_\sigma} \circ \overline{\alpha}_t^\xi \right\}_0$$
$$= (\delta_{i_1 p} \delta_{k_1 q} \delta_{i_2 k_2} \ldots \delta_{i_r k_r} \delta_{j_1 l_1} \delta_{j_2 l_2} \ldots \delta_{j_s l_s} + \delta_{i_1 k_1} \delta_{i_2 p} \delta_{k_2 q} \delta_{i_3 k_3} \ldots \delta_{i_r k_r} \delta_{j_1 l_1} \delta_{j_2 l_2} \ldots \delta_{j_s l_s}$$
$$+ \ldots + \delta_{i_1 k_1} \delta_{i_2 k_2} \ldots \delta_{i_r k_r} \delta_{j_1 l_1} \delta_{j_2 l_2} \ldots \delta_{j_{s-1} l_{s-1}} \delta_{j_s p} \delta_{l_s q}) \cdot \frac{\partial \xi_p}{\partial x_q} \cdot y_{\{k_1, k_2, \ldots, k_r; j_1, j_2, \ldots, j_s\}_\sigma}.$$

If we write $\sigma_{\{k_1, k_2, \ldots, k_r; j_1, j_2, \ldots, j_s\}_\sigma, \{i_1, i_2, \ldots, i_r; l_1, l_2, \ldots, l_s\}_\sigma, p, q}$ for the bracket in the last expression and $\xi_\sigma$ for the vector field generated by the local transformations of $T_\sigma X$, the calculation leads to the expression

$$\xi_\sigma = \xi_i \frac{\partial}{\partial x_i} + \sigma_{\{k_1, k_2, \ldots, k_r; j_1, j_2, \ldots, j_s\}_\sigma, \{i_1, i_2, \ldots, i_r; l_1, l_2, \ldots, l_s\}_\sigma, p, q} \cdot \frac{\partial \xi_p}{\partial x_q}$$
$$\cdot y_{\{k_1, k_2, \ldots, k_r; j_1, j_2, \ldots, j_s\}_\sigma} \cdot \frac{\partial}{\partial y_{\{i_1, i_2, \ldots, i_r; l_1, l_2, \ldots, l_s\}_\sigma}}.$$

If we further write $A$, $B$, $\ldots$ for the sets $\{i_1, i_2, \ldots, i_r; l_1, l_2, \ldots, l_s\}_\sigma$ of "$\sigma$-admissible in-

dices" and

$$\sigma_{Ap}^{Bq} = \sigma_{\{k_1,k_2,\ldots,k_r;j_1,j_2,\ldots,j_s\}_\sigma,\{i_1,i_2,\ldots,i_r;l_1,l_2,\ldots,l_s\}_\sigma,p,q},$$

the formula becomes

$$\xi_\sigma = \xi_i \frac{\partial}{\partial x_i} + \sigma_{Ap}^{Bq} \cdot \frac{\partial \xi_p}{\partial x_q} \cdot y_B \cdot \frac{\partial}{\partial y_A}, \tag{3.22}$$

where the constants $\sigma_{Ap}^{Bq}$ are completely determined by the tensor character of $T_\sigma X$.

As mentioned before, the vector field $\xi_\sigma$ plays an important role in the variational calculus on fibered manifolds. It is used for the definition of so called *generally invariant* (or *covariant*) variational problems (see [38]). We shall return to this vector field in Section 7.

Notice that independently of the type $\sigma$ of the tensor bundle, $\xi_\sigma$ is always constructed in the same way from $\xi$ (particularly by means of the derivatives of $\xi$). This means that the procedure of the $r$-jet prolongation of such vector fields will be the same for all types of $\sigma$. From the general rule (3.15) we obtain

$$\begin{aligned} j^1\xi_\sigma &= \xi_k \frac{\partial}{\partial x_k} + \sigma_{Ap}^{Bq} \frac{\partial \xi_p}{\partial x_q} y_B \frac{\partial}{\partial y_A} \\ &+ \left( \sigma_{Ap}^{Bq} \frac{\partial^2 \xi_p}{\partial x_i \partial x_q} y_B + \sigma_{Ap}^{Bq} \frac{\partial \xi_p}{\partial x_q} z_{iB} - z_{lA} \frac{\partial \xi_l}{\partial x_i} \right) \frac{\partial}{\partial z_{iA}}; \end{aligned}$$

explicit expressions for *arbitrary* $r$-jet prolongations may now easily be derived.

As an example one can take the so called *"variational vector field"* on the tangent space $TX$; it is of the form

$$\xi_\sigma = \xi_k \frac{\partial}{\partial x_k} + \frac{\partial \xi_i}{\partial x_j} \cdot y_j \frac{\partial}{\partial y_i}.$$

$\xi_\sigma$ is used in the variational calculus of one independent variable [36].

**Jet prolongations of the Lie bracket.** The following proposition holds:

*Let $(Y,\pi,X)$ be a fibered manifold and let $\Xi_1$, $\Xi_2$ be two vector fields on Y. If both $\Xi_1$ and $\Xi_2$ are projectable then so is $[\Xi_1,\Xi_2]$ and for any r*

$$j^r[\Xi_1,\Xi_2] = [j^r\Xi_1, j^r\Xi_2].$$

*The mapping $\Xi \to j^r\Xi$, defined on projectable vector fields on Y, is an $\mathbf{R}$-linear isomorphism.*

(This can be proved by a straightforward calculation in local coordinates.)

## 4. Horizontal and pseudovertical differential forms on jet prolongations of fibered manifolds

**Horizontal and pseudovertical forms.** Let $(J^r,\pi_r,X)$ be the $r$-jet prolongation of a fibered manifold $(Y, \pi, X)$. Consider an exterior differential $p$-form $\rho$ on $J^r$. Choose $j_x^{r+1}\gamma \in J^{r+1}$ and define $h(\rho)$ by the relation

$$
\begin{aligned}
&\langle h(\rho)(j_x^{r+1}\gamma), \xi_1 \times \xi_2 \times \ldots \times \xi_p \rangle \\
&= \langle \rho(j_x^r\gamma), T_x j^r\gamma \cdot T\pi_{r+1} \cdot \xi_1 \times T_x j^r\gamma \cdot T\pi_{r+1} \cdot \xi_2 \times \ldots \times T_x j^r\gamma \cdot T\pi_{r+1} \cdot \xi_p \rangle,
\end{aligned} \tag{4.1}
$$

where $\xi_1$, $\xi_2$, …, $\xi_p$ are arbitrary vectors from $T_{j_x^{r+1}\gamma} J^{r+1}$. It is clear that $h(\rho)$ is a $p$-form on $J^{r+1}$ with the property that $\langle h(\rho)(j_x^{r+1}\gamma), \xi_1 \times \xi_2 \times \ldots \times \xi_p \rangle$ is vanishing whenever one of the vectors $\xi_1$, $\xi_2$, …, $\xi_p$ is vertical (i.e. $T\pi_{r+1} \cdot \xi_i = 0$). Such forms are usually called *horizontal*. Put

$$p(\rho) = \pi_{r+1,r}^* \rho - h(\rho); \tag{4.2}$$

the $p$-form $p(\rho)$ on $J^{r+1}$ fulfills

$$j^{r+1}\gamma^* p(\rho) = 0 \tag{4.3}$$

for any cross section $\gamma \in \Gamma(\pi)$. Accordingly, we define: A differential $p$-form $\eta$ on $J^r$ is said to be *pseudovertical,* if for any $\gamma \in \Gamma(\pi)$

$$j^r\gamma^* \eta = 0 \tag{4.4}$$

(compare with [20]).

The expression of $h(\rho)$ and $p(\rho)$ in canonical coordinates on $J^{r+1}$ may be directly derived from the definition. Notice that $h(\rho)$ is uniquely determined by the condition (4.1). Thus, if we give a coordinate form satisfying (4.1), we give at the same time the coordinate expression for $h(\rho)$. Write $\zeta_\iota$ for some canonical coordinates on $J^r$ ($\iota$ is running over a set of admissible indices) and

$$\rho = \frac{1}{p!} f_{\iota_1\iota_2\ldots\iota_p} d\zeta_{\iota_1} \wedge d\zeta_{\iota_2} \wedge \ldots \wedge d\zeta_{\iota_p} \tag{4.5}$$

for a $p$-form in this coordinates. Then if we denote by $\zeta_{q\iota}$, $q = 0,1,\ldots,n$, $\zeta_{0\iota} = \zeta_\iota$, the corresponding coordinates on $J^{r+1}$, we have

$$h(\rho) = \frac{1}{p!} f_{\iota_1\iota_2\ldots\iota_p} \zeta_{k_1\iota_1}\zeta_{k_2\iota_2}\ldots\zeta_{k_p\iota_p} dx_{k_1} \wedge dx_{k_2} \wedge \ldots \wedge dx_{k_p}. \tag{4.6}$$

For the form $p(\rho)$ we obtain

$$
\begin{aligned}
p(\rho) = \frac{1}{p!} f_{\iota_1\iota_2\ldots\iota_p} (&d\zeta_{\iota_1} \wedge d\zeta_{\iota_2} \wedge \ldots \wedge d\zeta_{\iota_p} \\
&- \zeta_{k_1\iota_1}\zeta_{k_2\iota_2}\ldots\zeta_{k_p\iota_p} dx_{k_1} \wedge dx_{k_2} \wedge \ldots \wedge dx_{k_p}).
\end{aligned} \tag{4.7}
$$

A useful example of $h(\rho)$ and $p(\rho)$ will be discussed in one of the next paragraphs.

**Main properties of horizontal and pseudovertical forms.** The following proposition describes some useful properties of horizontal and pseudovertical $p$-forms.

*If $\rho$ is a p-form on $J^r$, then there exist uniquely determined p-forms $h(\rho)$ and $p(\rho)$ on $J^{r+1}$ such that 1) $\pi^*_{r+1,r}\rho = h(\rho) + p(\rho)$,, 2) $h(\rho)$ is horizontal and $p(\rho)$ is pseudovertical. The mapping $\rho \to p(\rho)$ is linear (over the ring of functions) and its kernel is formed by all horizontal p-forms on $J^r$. For any p-form $\rho_1$ and q-form $\rho_2$ on $J^r$*

(4.8) $$p(\rho_1 \wedge \rho_2) = p(\rho_1) \wedge p(\rho_2) + h(\rho_1) \wedge p(\rho_2) + p(\rho_1) \wedge h(\rho_2).$$

*If $p > n = \dim X$, then $h(\rho) = 0$. Both $\rho$ and $h(\rho)$ are defined on $J^r$ if and only if $\rho$ is horizontal with respect to the projection $\pi_{r,r-1}$. Finally, let $(\alpha, \alpha_0)$ be an automorphism of the fibered manifold $(Y, \pi, X)$; then*

(4.9) $$p(j^r\alpha^*\rho) = j^{r+1}\alpha^* p(\rho).$$

(All these assertions immediately follow from the definitions.)

**Example: Horizontal and pseudovertical *n*-forms on $J^2Y$.** We shall illustrate the above proposition by giving explicit formulas for *n*-forms $\rho$ on $J^2$ ($n = \dim X$), which are horizontal with respect to the projection $\pi_{2,1}$ (one should remember that this is a necessary and sufficient condition for both $\rho$ and $p(\rho)$ to be defined on $J^2$).

Let $\rho$ be an *n*-form on $J^2$, horizontal with respect to $\pi_{2,1}$. Since any coordinate expression of $\rho$ does not contain $dz_{ij\mu}$ ($(x_i, y_\mu, z_{i\mu}, z_{ij\mu})$ being the coordinates on $J^2$ defined by a canonical chart), it will be useful to introduce the notation

$$\zeta_{0k\mu} = z_{k\mu}, \quad \zeta_{00\mu} = y_\mu, \quad \zeta_{kl\mu} = z_{kl\mu}, \quad \zeta_{0\mu} = y_\mu, \quad \zeta_{k\mu} = z_{k\mu}.$$

Writing

(4.10) $$\begin{aligned} \rho = {} & f_0 dx_1 \wedge dx_2 \wedge \ldots \wedge dx_n \\ & + \sum_{r=1}^{n} \sum_{s_1 < s_2 < \ldots < s_r} \sum_{\sigma_1, \sigma_2, \ldots, \sigma_r} \sum_{q_1, q_2, \ldots, q_r} \frac{1}{r!} f^{\,s_1\, s_2 \ldots s_r}_{\substack{q_1 q_2 \ldots q_r \\ \sigma_1 \sigma_2 \ldots \sigma_r}} \\ & \cdot dx_1 \wedge dx_2 \wedge \ldots \wedge dx_{s_1 - 1} \wedge d\zeta_{q_1 \sigma_1} \wedge dx_{s_1 + 1} \wedge \ldots \wedge dx_{s_r - 1} \\ & \wedge d\zeta_{q_r \sigma_r} \wedge dx_{s_r + 1} \wedge \ldots \wedge dx_n, \end{aligned}$$

where $0 \le q_1, q_2, \ldots, q_r \le n$ and the summation is obvious, we see at once that the expression on the right-hand side is the most general *n*-form on $J^2$, horizontal with respect to $\pi_{2,1}$. Clearly, the functions $f^{\,s_1\, s_2 \ldots s_r}_{\substack{q_1 q_2 \ldots q_r \\ \sigma_1 \sigma_2 \ldots \sigma_r}}$ can be supposed antisymmetric in the lower couples of the indices $\substack{q_i \\ \sigma_i}$.

By definition

(4.11) $$h(\rho) = \left( f_0 + \sum f^{\,s_1\, s_2 \ldots s_r}_{\substack{q_1 q_2 \ldots q_r \\ \sigma_1 \sigma_2 \ldots \sigma_r}} \zeta_{s_1 q_1 \sigma_1} \zeta_{s_2 q_2 \sigma_2} \ldots \zeta_{s_r q_r \sigma_r} \right) dx_1 \wedge dx_2 \wedge \ldots \wedge dx_n,$$

where the summation is the same as in (4.10). The decomposition of $\rho$ into horizontal and pseudovertical forms is thus given explicitly as

(4.12) $$\rho = \left( f_0 + \sum f^{\,s_1\, s_2 \ldots s_r}_{\substack{q_1 q_2 \ldots q_r \\ \sigma_1 \sigma_2 \ldots \sigma_r}} \zeta_{s_1 q_1 \sigma_1} \zeta_{s_2 q_2 \sigma_2} \ldots \zeta_{s_r q_r \sigma_r} \right) dx_1 \wedge dx_2 \wedge \ldots \wedge dx_n$$

$$+\sum f^{s_1 s_2 \ldots s_r}_{\substack{q_1 q_2 \ldots q_r \\ \sigma_1\sigma_2 \ldots \sigma_r}} \Big(\frac{1}{r!} dx_1 \wedge \ldots \wedge d\zeta_{q_1\sigma_1} \wedge \ldots \wedge d\zeta_{q_r\sigma_r} \wedge \ldots \wedge dx_n$$

$$-\zeta_{s_1 q_1 \sigma_1} \zeta_{s_2 q_2 \sigma_2} \ldots \zeta_{s_r q_r \sigma_r} dx_1 \wedge dx_2 \wedge \ldots \wedge dx_n\Big).$$

**An extension of $h$ to $(n+1)$-forms.** We shall extend the mapping $\rho \to h(\rho)$ defined on $n$-forms on the $r$-jet prolongation of a fibered manifold, to $(n+1)$-forms defined on higher jet prolongation of the same fibered manifold. This is based on the following proposition:

*Let $\rho$ be an $(n+1)$-form on $J^r$. There exists one and only one $(n+1)$-form $\bar{\rho}$ on $J^{r+1}$ satisfying*

$$h(i(\xi)\pi^*_{r+1,r}\rho) = i(\xi)\bar{\rho}$$

*for all vertical vector fields $\xi$ (with respect to $\pi_{r+1}$) on $J^{r+1}$.*

To prove it, let us work with the coordinates $\zeta_{q\iota}$ introduced before (see (4.6)). Let us examine the condition

$$i(\xi)\bar{\rho}_0 = 0,$$

taking place for an $(n+1)$-form $\bar{\rho}_0$ on $J^{r+1}$ and for all vertical vector fields $\xi$ (with respect to $\pi_{r+1}$). Suppose

$$\bar{\rho}_0 = \frac{1}{(n+1)!} f_{\iota_0\iota_1\iota_2\ldots\iota_n} d\zeta_{\iota_0} \wedge d\zeta_{\iota_1} \wedge d\zeta_{\iota_2} \wedge \ldots \wedge d\zeta_{\iota_n};$$

then (compare with a formula in [7])

$$i(\zeta)\bar{\rho}_0 = \frac{1}{n!} \zeta_{\iota_0} f_{\iota_0\iota_1\iota_2\ldots\iota_n} d\zeta_{\iota_1} \wedge d\zeta_{\iota_2} \wedge \ldots \wedge d\zeta_{\iota_n},$$

and the condition leads to the equalities $f_{\iota_0\iota_1\iota_2\ldots\iota_n} = 0$ taking place for all $\iota_0 \neq 1, 2, \ldots, n$ (1, 2, …, $n$ index the coordinates on $X$). Since $f_{\iota_0\iota_1\iota_2\ldots\iota_n}$ are antisymmetric this means that $f_{\iota_0\iota_1\iota_2\ldots\iota_n} = 0$ whenever one of the indices $\iota_0$, $\iota_1$, $\iota_2$, …, $\iota_n$ is different from 1, 2, …, $n$. But $f_{k_0 k_1 k_2\ldots k_n}$, where $1 \leq k_0, k_1, k_2, \ldots, k_n \leq n$, vanishes identically, which shows that $\bar{\rho}_0 = 0$. In other words, if the $(n+1)$-form $\bar{\rho}$ from the proposition exists, it is unique. As for the existence, it suffices to find one such form on each coordinate neighborhood.

Let us now work with some canonical coordinates $\zeta_\iota$ on $J^r$. Put

$$\rho = \frac{1}{(n+1)!} f_{\iota_0\iota_1\iota_2\ldots\iota_n} d\zeta_{\iota_0} \wedge d\zeta_{\iota_1} \wedge d\zeta_{\iota_2} \wedge \ldots \wedge d\zeta_{\iota_n},$$

and define, with $\zeta_{ik} = \delta_{ik}$,

(4.13) $$\bar{\rho} = f_{\iota_0\iota_1\iota_2\ldots\iota_n} \zeta_{1\iota_1} \zeta_{2\iota_2} \ldots \zeta_{n\iota_n} d\zeta_{\iota_0} \wedge dx_1 \wedge dx_2 \wedge \ldots \wedge dx_n.$$

The equality $h(i(\xi)\pi^*_{r+1,r}\rho = i(\xi)\bar{\rho}$ follows by a direct calculation (see [7], p. 149,

formulas of Section 1, and (4.6)):

$$i(\xi)\pi^*_{r+1,r}\rho = \frac{1}{n!} f_{\iota_0\iota_1\iota_2\ldots\iota_n}\xi_{\iota_0} d\zeta_{\iota_1}\wedge d\zeta_{\iota_2}\wedge\ldots\wedge d\zeta_{\iota_n},$$

$$h(i(\xi)\pi^*_{r+1,r}\rho) = \frac{1}{n!} f_{\iota_0\iota_1\iota_2\ldots\iota_n}\xi_{\iota_0}\zeta_{k_1\iota_1}\zeta_{k_2\iota_2}\ldots\zeta_{k_n\iota_n} dx_1\wedge dx_2\wedge\ldots\wedge dx_n$$

$$= f_{\iota_0\iota_1\iota_2\ldots\iota_n}\xi_{\iota_0}\zeta_{1\iota_1}\zeta_{2\iota_2}\ldots\zeta_{n\iota_n} dx_1\wedge dx_2\wedge\ldots\wedge dx_n = i(\xi)\bar{\rho}.$$

This completes the proof.

We note that the form $\bar{\rho}$ can be invariantly defined by means of the relation

$$\begin{aligned}&\langle\bar{\rho}(j^{r+1}_x\gamma),\xi_0\times\xi_1\times\ldots\times\xi_n\rangle\\ &=\sum_{k=0}^{n}\langle\rho(j^r_x\gamma),T_x j^r\gamma\circ T\pi_{r+1}\cdot\xi_0\times T_x j^r\gamma\circ T\pi_{r+1}\cdot\xi_1\\ &\times\ldots\times T_x j^r\gamma\circ T\pi_{r+1}\cdot\xi_{k-1}\times T\pi_{r+1,r}\cdot\xi_k\times T_x j^r\gamma\circ T\pi_{r+1}\cdot\xi_{k+1}\\ &\times\ldots\times T_x j^r\gamma\cdot T\pi_{r+1}\cdot\xi_n\rangle\end{aligned}$$

in which $\xi_0$, $\xi_1$, …, $\xi_n$ are arbitrary vectors from the tangent space $T_{j^{r+1}_x\gamma}J^{r+1}$. Put

(4.14) $\qquad \tilde{h}(\rho) = \bar{\rho};$

then relation (4.13) takes the form

(4.15) $\qquad h(i(\xi)\pi^*_{r+1,r}\rho) = i(\xi)\tilde{h}(\rho).$

**Differential forms on $J^\infty Y$.** The only forms we shall work with on the infinite jet prolongation of a fibered manifold will be the pull-backs by the projections $\pi_{\infty,r}$. The pull-backs will be defined in a full analogy to the case of manifolds modeled on Banach spaces [26].

Let $\rho$ be a $p$-form on $J^r$ and let $\xi_1$, $\xi_2$, …, $\xi_p$ be tangent vectors at the point $j^\infty_x\gamma\in J^\infty$; define $(\pi^*_{\infty,r}\rho)(j^\infty_x\gamma)$:

(4.16)
$$\begin{aligned}&\langle(\pi^*_{\infty,r}\rho)(j^\infty_x\gamma),\xi_1\times\xi_2\times\ldots\times\xi_p\rangle\\ &=\langle\rho(j^r_x\gamma),T\pi_{\infty,r}\cdot\xi_1\times T\pi_{\infty,r}\cdot\xi_2\times\ldots\times T\pi_{\infty,r}\cdot\xi_p\rangle,\end{aligned}$$

or, symbolically,

$$\pi^*_{\infty,r}\rho = (\rho\circ\pi_{\infty,r})\cdot(T\pi_{\infty,r})^p.$$

The mapping $j^\infty_x\gamma\to(\pi^*_{\infty,r}\rho)(j^\infty_x\gamma)$ thus arising is a $p$-form on $J^\infty$, i.e. a cross section of the fibered manifold $(L^p_a TJ^\infty Y,\rho_\infty,J^\infty Y)$ defined in Section 2.

**Example.** In this paragraph we wish to illustrate the described definitions and methods, and take notice of one of their relations to the calculus of variations. We pose the following problem:

Suppose we have a horizontal n-form $\lambda$ on $J^2$ (with respect to $\pi_2$). Find an n-form $\Theta$ on $J^r$ with suitable r, satisfying the following conditions:

1) *In some canonical coordinates* (2.4) *on* $J^r$, $\Theta$ *is of the form*

$$\Theta = \lambda + f^i_\sigma \omega^i_\sigma + \frac{1}{2} f^i_{j\sigma} \omega^i_{j\sigma}, \tag{4.17}$$

*where*

$$f^i_{j\sigma} = f^j_{i\sigma},$$

$$\omega^i_\sigma = dx_1 \wedge dx_2 \wedge \ldots \wedge dx_{i-1} \wedge (dy_\sigma - z_{k\sigma} dx_k) \wedge dx_{i+1} \wedge dx_{i+2} \wedge \ldots \wedge dx_n,$$

$$\omega^i_{j\sigma} = dx_1 \wedge dx_2 \wedge \ldots \wedge dx_{i-1} \wedge (dz_{j\sigma} - z_{kj\sigma} dx_k) \wedge dx_{i+1} \wedge dx_{i+2} \wedge \ldots \wedge dx_n;$$

*in particular,*

$$h(\Theta) = \lambda.$$

2) $\tilde{h}(d\Theta)$ *is horizontal with respect to* $\pi_{r+1,0}$.

It will be proved in the next Sections that the problem is, in fact, motivated by the calculus of variations. We just note that because of the assumption 2), the (n+1)-form $\tilde{h}(d\Theta)$ well corresponds to the Euler equations of the calculus of variations, determining the extremals. Notice that if the answer is positive, i.e. if a form satisfying 1) and 2) does exist, we have proved, as a partial result, that the mapping $\rho \to h(\rho)$ maps onto the space of all horizontal n-forms on $J^2$.

Let us now consider an n-form $\Theta$ of the needed form (4.17). By (4.15), we shall examine the expression $h(i(\xi)\pi^*_{r+1,r} d\Theta)$ instead of $i(\xi)\tilde{h}(d\Theta)$, for an arbitrary vertical vector field $\xi$ on $J^{r+1}$. In the canonical coordinates

$$i(\xi)\omega^i_\sigma = (-1)^{i-1} \xi_\sigma \, dx_1 \wedge dx_2 \wedge \ldots \wedge \hat{dx_i} \wedge \ldots \wedge dx_n,$$

$$i(\xi)\omega^i_{j\sigma} = (-1)^{i-1} \xi_{j\sigma} \, dx_1 \wedge dx_2 \wedge \ldots \wedge \hat{dx_i} \wedge \ldots \wedge dx_n,$$

$$i(\xi) d\omega^i_\sigma = -\xi_{i\sigma} \, dx_1 \wedge dx_2 \wedge \ldots \wedge dx_n,$$

$$i(\xi)\omega^i_{j\sigma} = -\xi_{ij\sigma} \, dx_1 \wedge dx_2 \wedge \ldots \wedge dx_n.$$

Obviously, in these formulas $\hat{dx_i}$ means that $dx_i$ is missing, and $\xi_\sigma$, $\xi_{i\sigma}$, … denote the components of the vector field $\xi$ given by

$$\xi = \xi_\sigma \frac{\partial}{\partial y_\sigma} + \xi_{i\sigma} \frac{\partial}{\partial z_{i\sigma}} + \sum_{i_0 \le i_1 \le \ldots \le i_r} \xi_{i_0 i_1 \ldots i_r \sigma} \frac{\partial}{\partial z_{i_0 i_1 \ldots i_r \sigma}}. \tag{4.18}$$

Let us write ($n = \dim X$)

$$\lambda = \mathscr{L} dx_1 \wedge dx_2 \wedge \ldots \wedge dx_n.$$

Now

$$d\Theta = d\mathscr{L} \wedge dx_1 \wedge dx_2 \wedge \ldots \wedge dx_n + df^i_\sigma \wedge \omega^i_\sigma + f^i_\sigma d\omega^i_\sigma$$
$$+\frac{1}{2} df^i_{j\sigma} \wedge \omega^i_{j\sigma} + \frac{1}{2} f^i_{j\sigma} d\omega^i_{j\sigma}.$$

But the forms $\omega^i_\sigma$ and $\omega^i_{j\sigma}$ are pseudovertical so that, according to the rule $h(\rho \wedge \eta) = h(\rho) \wedge h(\eta)$ following from (4.8),

$$h(i(\xi)\pi^*_{r+1,r} d\Theta) = \left( \left( \frac{\partial \mathscr{L}}{\partial y_\sigma} - \frac{df^i_\sigma}{dx_i} \right) \xi_\sigma + \left( \frac{\partial \mathscr{L}}{\partial z_{i\sigma}} - f^i_\sigma - \frac{1}{2} \frac{df^i_{j\sigma}}{dx_j} \right) \xi_{i\sigma} \right.$$
$$\left. + \sum_i \left( \frac{\partial \mathscr{L}}{\partial z_{ii\sigma}} - \frac{1}{2} f^i_{i\sigma} \right) \xi_{ii\sigma} + \sum_{i<j} \left( \frac{\partial \mathscr{L}}{\partial z_{ij\sigma}} - f^i_{j\sigma} \right) \xi_{ij\sigma} \right) dx_1 \wedge dx_2 \wedge \ldots \wedge dx_n.$$

Remember that $d/dx_i$ denotes the formal derivative, introduced in Section 3 (3.14). Now condition 2) directly leads to the equalities defining $f^i_\sigma$ and $f^i_{j\sigma}$. Since $f^i_{j\sigma} = f^j_{i\sigma}$, we get

$$f^i_{j\sigma} = \frac{\partial \mathscr{L}}{\partial z_{ij\sigma}}, \quad i \neq j,$$

$$f^i_{i\sigma} = 2 \cdot \frac{\partial \mathscr{L}}{\partial z_{ii\sigma}},$$

$$f^i_\sigma = \frac{\partial \mathscr{L}}{\partial z_{i\sigma}} - \frac{d}{dx_i}\left( \frac{\partial \mathscr{L}}{\partial z_{ii\sigma}} \right) - \frac{1}{2} \sum_{j \neq i} \frac{d}{dx_j} \left( \frac{\partial \mathscr{L}}{\partial z_{ij\sigma}} \right),$$

the summation being obvious. We see that $\Theta$ is defined on $J^3$.

It might be of interest to obtain a coordinate expression for $\tilde{h}(d\Theta)$; we get

$$\tilde{h}(d\Theta) = \left( \frac{\partial \mathscr{L}}{\partial y_\sigma} - \frac{d}{dx_i}\left( \frac{\partial \mathscr{L}}{\partial z_{i\sigma}} \right) + \sum_{i \leq j} \frac{d}{dx_i} \left( \frac{d}{dx_j} \left( \frac{\partial \mathscr{L}}{\partial z_{ij\sigma}} \right) \right) \right) \tag{4.19}$$
$$\cdot\, dy_\sigma \wedge dx_1 \wedge dx_2 \wedge \ldots \wedge dx_n.$$

The function

$$E_\sigma = \frac{\partial \mathscr{L}}{\partial y_\sigma} - \frac{d}{dx_i}\left( \frac{\partial \mathscr{L}}{\partial z_{i\sigma}} \right) + \sum_{i \leq j} \frac{d}{dx_i} \left( \frac{d}{dx_j} \left( \frac{\partial \mathscr{L}}{\partial z_{ij\sigma}} \right) \right) \tag{4.20}$$

defined locally on $J^4$, is usually termed the *Euler expression* associated with the *Lagrange function* $\mathscr{L}$. (The Euler expressions are nothing but left-hand sides of the Euler equations of the calculus of variations. The Euler expressions will be frequently used in next Sections.)

If $\lambda$ is defined on $J^1$, then the corresponding Euler expressions are

$$E_\sigma = \frac{\partial \mathscr{L}}{\partial y_\sigma} - \frac{d}{dx_i}\left( \frac{\partial \mathscr{L}}{\partial z_{i\sigma}} \right).$$

In this case the $n$-form $\Theta$ has been obtained in [35] for a special case. Our approach was motivated, in part, by [4].

Let us summarize the results.

*There is one and only one n-form* $\Theta$ *on* $J^3$ *satisfying conditions* 1) *and* 2). *If in some canonical coordinates* $(x_i, y_\mu, z_{i\mu}, z_{ij\mu}, z_{ijk\mu})$ *the form* $\lambda$ *is expressed as*

(4.21) $$\lambda = \mathscr{L} dx_1 \wedge dx_2 \wedge \ldots \wedge dx_n,$$

*then*

(4.22) $$\begin{aligned}\Theta = \mathscr{L} dx_1 \wedge dx_2 \wedge \ldots \wedge dx_n &+ \left( \frac{\partial \mathscr{L}}{\partial z_{i\sigma}} - \frac{d}{dx_i}\left( \frac{\partial \mathscr{L}}{\partial z_{ii\sigma}} \right) - \frac{1}{2} \sum_{j \neq i} \frac{d}{dx_j} \left( \frac{\partial \mathscr{L}}{\partial z_{ji\sigma}} \right) \right) \omega^i_\sigma \\ &+ \frac{\partial \mathscr{L}}{\partial z_{ii\sigma}} \omega^i_{i\sigma} + \frac{1}{2} \sum_{j \neq i} \frac{\partial \mathscr{L}}{\partial z_{ij\sigma}} \omega^i_{j\sigma} .\end{aligned}$$

**Two formulas for the infinite pull-back of forms.** We shall need the following formulas:

1) *For any p-form* $\rho$ *on* $J^r$,

(4.23) $$\pi^*_{\infty,r+1} h(\rho) = h(\pi^*_{\infty,r} \rho).$$

2) *For any* $(n+1)$*-form* $\eta$ *on* $J^r$ *and any projectable vector field* $\Xi$ *on Y*

(4.24) $$i(v(j^\infty \Xi)) \tilde{h}(\pi^*_{\infty,r} \eta) = h(i(j^\infty \Xi) \pi^*_{\infty,r} \eta).$$

The first one is a direct consequence of definitions. To prove the second formula we write (3.21) $j^\infty \Xi = \xi^\infty + v(j^\infty \Xi)$, and

$$\begin{aligned}&\langle h(i(\xi^\infty)\pi^*_{\infty,r}\eta)(j^\infty_x \gamma), \xi_1 \times \xi_2 \times \ldots \times \xi_n \rangle \\ &= \langle (\pi^*_{\infty,r}\eta)(j^\infty_x \gamma), T_x j^\infty \gamma \cdot \xi(x) \times T_x j^\infty \gamma \cdot T\pi_\infty \cdot \xi_1 \times \ldots \times T_x j^\infty \gamma \cdot T\pi_\infty \cdot \xi_n \rangle \\ &= \langle (j^r \gamma^* \eta)(x), \xi(x) \times T\pi_\infty \cdot \xi_1 \times T\pi_\infty \cdot \xi_2 \times \ldots \times T\pi_\infty \cdot \xi_n \rangle = 0.\end{aligned}$$

for any vectors $\xi_1$, $\xi_2$, …, $\xi_n$ at a point $j^\infty_x \gamma \in J^\infty$. In these formulas the contraction by a vector field on $J^\infty$ and the operation $h$, are applied to $n$-forms on $J^\infty$ in an evident way. Now, by the invariant definition of $\tilde{h}(\pi^*_{\infty,r}\eta)$,

$$\begin{aligned}&\langle \tilde{h}(\pi^*_{\infty,r}\eta)(j^\infty_x \gamma), v(j^\infty \Xi)(j^\infty_x \gamma) \times \xi_1 \times \xi_2 \times \ldots \times \xi_n \rangle \\ &= \langle (\pi^*_{\infty,r}\eta)(j^\infty_x \gamma), v(j^\infty \Xi)(j^\infty_x \gamma) \times T_x j^\infty \gamma \circ T\pi_\infty \cdot \xi_1 \times \ldots \times T_x j^\infty \gamma \circ T\pi_\infty \cdot \xi_n \rangle \\ &= \langle h(i(v(j^\infty \Xi))\pi^*_{\infty,r}\eta)(j^\infty_x \gamma), \xi_1 \times \xi_2 \times \ldots \times \xi_n \rangle \\ &= \langle h(i(j^\infty \Xi)\pi^*_{\infty,r}\eta)(j^\infty_x \gamma), \xi_1 \times \xi_2 \times \ldots \times \xi_n \rangle.\end{aligned}$$

This finishes the proof.

## 5. The Euler mapping

**Explicit formula for $d\rho$, where $\rho$ is an *n*-form horizontal with respect to $\pi_{1,0}$.** At the beginning of this Section we give two coordinate formulas concerning *n*-forms defined on $J^1$.

Let us consider a fibered manifold $(Y,\pi,X)$ and denote by $J^r$ its *r*-jets prolongation. Let $\rho$ be an *n*-form on $J^1$, where $n=\dim X$. In the canonical coordinates (2.4) on $J^1$

$$\rho = g_0 dx_1\wedge dx_2\wedge\ldots\wedge dx_n$$

$$+\sum_{r=1}^{n}\sum_{s_1<s_2<\ldots<s_r}\sum_{\sigma_1,\sigma_2,\ldots,\sigma_r}\frac{1}{r!}g^{s_1s_2\ldots s_r}_{\sigma_1\sigma_2\ldots\sigma_r}dx_1\wedge dx_2\wedge\ldots\wedge dx_{s_1-1} \tag{5.1}$$

$$\wedge dy_{\sigma_1}\wedge dx_{s_1+1}\wedge\ldots\wedge dx_{s_r-1}\wedge dy_{\sigma_r}\wedge dx_{s_r+1}\wedge\ldots\wedge dx_n .$$

We take into account in this formula that $\rho$ is horizontal with respect to $\pi_{1,0}$, and we assume that the functions $g^{s_1s_2\ldots s_r}_{\sigma_1\sigma_2\ldots\sigma_r}$ are antisymmetric with respect to the subscripts (compare with the example given in Section 4).

Then one can derive by a direct but tedious calculation the following formula for the differential $d\rho$:

*If $\rho$ is expressed by* (5.1), *then*

$$d\rho=\left(\frac{\partial g_0}{\partial y_\mu}-\frac{\partial g^s_\mu}{\partial x_s}\right)dy_\mu\wedge dx_1\wedge dx_2\wedge\ldots\wedge dx_n$$

$$+\sum_{r=1}^{n+1}\sum_{s_1<s_2<\ldots<s_r}\sum_{\sigma_1,\sigma_2,\ldots,\sigma_r}\frac{1}{r!}\left(\frac{\partial g^{s_1s_2\ldots s_r}_{\sigma_1\sigma_2\ldots\sigma_r}}{\partial y_\mu}-\frac{1}{r+1}\left(\sum_{k<s_1}\frac{\partial g^{ks_1s_2\ldots s_r}_{\mu\sigma_1\sigma_2\ldots\sigma_r}}{\partial x_k}\right.\right.$$

$$\left.\left.+\sum_{s_1<k<s_2}\frac{\partial g^{s_1ks_2\ldots s_r}_{\sigma_1\mu\sigma_2\ldots\sigma_r}}{\partial x_k}+\ldots+\sum_{k>s_r}\frac{\partial g^{s_1s_2\ldots s_rk}_{\sigma_1\sigma_2\ldots\sigma_r\mu}}{\partial x_k}\right)\right)dy_\mu\wedge dx_1\wedge dx_2\wedge\ldots$$

$$\wedge dx_{s_1-1}\wedge dy_{\sigma_1}\wedge dx_{s_1+1}\wedge\ldots\wedge dx_{s_r-1}\wedge dy_{\sigma_r}\wedge dx_{s_r+1}\wedge\ldots\wedge dx_n$$

$$+\sum_{\sigma_1,\sigma_2,\ldots,\sigma_n}\frac{1}{n!}\frac{\partial g^{12\ldots n}_{\sigma_1\sigma_2\ldots\sigma_n}}{\partial y_\mu}\cdot dy_\mu\wedge dy_{\sigma_1}\wedge dy_{\sigma_2}\wedge\ldots\wedge dy_{\sigma_n}$$

$$+\frac{\partial g_0}{\partial z_{k\mu}}dz_{k\mu}\wedge dx_1\wedge dx_2\wedge\ldots\wedge dx_n \tag{5.2}$$

$$+\sum_{r=1}^{n}\sum_{s_1<s_2<\ldots<s_r}\sum_{\sigma_1,\sigma_2,\ldots,\sigma_r}\frac{1}{r!}\frac{\partial g^{s_1s_2\ldots s_r}_{\sigma_1\sigma_2\ldots\sigma_r}}{\partial z_{k\mu}}dz_{k\mu}\wedge dx_1\wedge dx_2\wedge\ldots$$

$$\wedge dx_{s_1-1}\wedge dy_{\sigma_1}\wedge dx_{s_1+1}\wedge\ldots\wedge dx_{s_r-1}\wedge dy_{\sigma_r}\wedge dx_{s_r+1}\wedge\ldots\wedge dx_n .$$

As for summation, notice that we often do not explicitly designate the summation over the indices $k$ and $\mu$.

**Explicit formula for $\tilde{h}(d\rho)$.** Coming out from the previous paragraph, we prove the following important proposition:

*Let $\rho$ be an n-form on $J^1$ and suppose that $\rho$ is horizontal with respect to $\pi_{1,0}$. Let $\omega$ be a volume element on X. Write in the canonical coordinates* (2.4) *on $J^2$*

(5.3) $$\omega_0 = dx_1 \wedge dx_2 \wedge \ldots \wedge dx_n$$

(5.4) $$A_{k\nu} = \frac{\partial g_0}{\partial z_{k\nu}} + \sum_{r=1}^{n} \sum_{s_1 < s_2 < \ldots < s_r} \sum_{\sigma_1, \sigma_2, \ldots, \sigma_r} \frac{\partial g^{s_1 s_2 \ldots s_r}_{\sigma_1 \sigma_2 \ldots \sigma_r}}{\partial z_{k\nu}} \cdot z_{s_1\sigma_1} z_{s_2\sigma_2} \ldots z_{s_r\sigma_r},$$

(5.5) $$h(\rho) = G\omega_0.$$

*Then*

(5.6) $$\tilde{h}(d\rho) = \left( \frac{\partial G}{\partial y_\nu} - \frac{d}{dx_k}\left( \frac{\partial G}{\partial z_{k\nu}} \right) \right) dy_\nu \wedge \omega_0 + \left( \frac{dA_{k\nu}}{dx_k} dy_\nu + A_{k\nu} dz_{k\nu} \right) \wedge \omega_0.$$

For the proof let us start by the formula (4.13); we have

$$\tilde{h}(dy_\nu \wedge \omega_0) = dy_\nu \wedge \omega_0,$$

$$\tilde{h}(dy_\mu \wedge dx_1 \wedge dx_2 \wedge \ldots \wedge dy_{\sigma_1} \wedge \ldots \wedge dy_{\sigma_r} \wedge \ldots \wedge dx_n)$$
$$= \tilde{h}\left( \frac{1}{(n+1)!} \epsilon^{\iota_0 \iota_1 \ldots \iota_{s_1} \ldots \iota_{s_r} \ldots \iota_n}_{\mu 1 \ldots \sigma_1 \ldots \sigma_r \ldots n} d\zeta_{\iota_0} \wedge d\zeta_{\iota_1} \wedge \ldots \wedge d\zeta_{\iota_n} \right)$$
$$= \zeta^{\iota_0 \iota_1 \ldots \iota_{s_1} \ldots \iota_{s_r} \ldots \iota_n}_{\mu\, 1 \ldots \sigma_1 \ldots \sigma_r \ldots n} \zeta_{1\iota_1} \zeta_{2\iota_2} \ldots \zeta_{n\iota_n} d\zeta_{\iota_0} \wedge \omega_0.$$

In the last formula $\epsilon^{ijk\ldots}_{lmn\ldots}$ is the completely antisymmetric symbol (see e.g. [7], p. 36). Clearly, all terms with $\iota_0 \neq \mu, \sigma_1, \sigma_2, \ldots, \sigma_r$ vanish so that

$$\tilde{h}(dy_\mu \wedge dx_1 \wedge dx_2 \wedge \ldots \wedge dy_{\sigma_1} \wedge \ldots \wedge dy_{\sigma_r} \wedge \ldots \wedge dx_n)$$
$$= \epsilon^{\iota_1 \ldots \iota_{s_1} \ldots \iota_{s_r} \ldots \iota_n}_{1 \ldots \sigma_1 \ldots \sigma_r \ldots n} \zeta_{1\iota_1} \zeta_{2\iota_2} \ldots \zeta_{n\iota_n} dy_\mu \wedge \omega_0$$
$$- \epsilon^{\iota_1 \ldots \iota_{s_1} \ldots \iota_{s_r} \ldots \iota_n}_{1 \ldots \mu \ldots \sigma_r \ldots n} \zeta_{1\iota_1} \zeta_{2\iota_2} \ldots \zeta_{n\iota_n} dy_{\sigma_1} \wedge \omega_0 - \ldots$$
$$- \epsilon^{\iota_1 \ldots \iota_{s_1} \ldots \iota_{s_r} \ldots \iota_n}_{1 \ldots \sigma_1 \ldots \mu \ldots n} \zeta_{1\iota_1} \zeta_{2\iota_2} \ldots \zeta_{n\iota_n} dy_{\sigma_r} \wedge \omega_0$$
$$= (\epsilon^{\iota_{s_1} \iota_{s_2} \ldots \iota_{s_r}}_{\sigma_1 \sigma_2 \ldots \sigma_r} \zeta_{s_1 \iota_{s_1}} \zeta_{s_2 \iota_{s_2}} \ldots \zeta_{s_r \iota_{s_r}} \delta_{\mu\nu} - \epsilon^{\iota_{s_1} \iota_{s_2} \ldots \iota_{s_r}}_{\mu \sigma_2 \ldots \sigma_r} \zeta_{s_1 \iota_{s_1}} \zeta_{s_2 \iota_{s_2}} \ldots \zeta_{s_r \iota_{s_r}} \delta_{\sigma_1 \nu}$$
$$- \ldots - \epsilon^{\iota_{s_1} \iota_{s_2} \ldots \iota_{s_{r-1}} \iota_{s_r}}_{\sigma_1 \sigma_2 \ldots \sigma_{r-1} \mu} \zeta_{s_1 \iota_{s_1}} \zeta_{s_2 \iota_{s_2}} \ldots \zeta_{s_r \iota_{s_r}} \delta_{\sigma_r \nu}) dy_\nu \wedge \omega_0$$
$$= \epsilon^{s_1 s_2 \ldots s_r}_{k_1 k_2 \ldots k_r} (\zeta_{k_1 \sigma_1} \zeta_{k_2 \sigma_2} \ldots \zeta_{k_r \sigma_r} \delta_{\mu\nu} - \zeta_{k_1 \mu} \zeta_{k_2 \sigma_2} \ldots \zeta_{k_r \sigma_r} \delta_{\sigma_1 \nu}$$
$$- \ldots - \zeta_{k_1 \sigma_1} \zeta_{k_2 \sigma_2} \ldots \zeta_{k_{r-1} \sigma_{r-1}} \zeta_{k_r \mu} \delta_{\sigma_r \nu}) dy_\nu \wedge \omega_0,$$

where we passed to the summation over $k_1, k_2, \ldots, k_r$ instead of $\iota_{s_1}, \iota_{s_2}, \ldots, \iota_{s_r}$. Writing $\zeta_{k\sigma} = z_{k\sigma}$, we obtain the desired formula. Analogous expressions for $d\rho$ given by (5.2), can be obtained in the same manner. One gets together

$$\tilde{h}(dy_\mu \wedge dx_1 \wedge dx_2 \wedge \ldots \wedge dy_{\sigma_1} \wedge \ldots \wedge dy_{\sigma_r} \wedge \ldots \wedge dx_n)$$

$$= \epsilon^{s_1 s_2 \ldots s_r}_{k_1 k_2 \ldots k_r} (\delta_{\mu\nu} z_{k_1\sigma_1} z_{k_2\sigma_2} \ldots z_{k_r\sigma_r} - \delta_{\sigma_1\nu} z_{k_1\mu} z_{k_2\sigma_2} \ldots z_{k_r\sigma_r} - \ldots$$
$$- \delta_{\sigma_r\nu} z_{k_1\sigma_1} z_{k_2\sigma_2} \ldots z_{k_{r-1}\sigma_{r-1}} z_{k_r\mu}) dy_\nu \wedge \omega_0,$$
$$\tilde{h}(dy_\mu \wedge dy_{\sigma_1} \wedge dy_{\sigma_2} \wedge \ldots \wedge dy_{\sigma_n})$$
$$= \epsilon^{12\ldots n}_{k_1 k_2 \ldots k_n} (\delta_{\mu\nu} z_{k_1\sigma_1} z_{k_2\sigma_2} \ldots z_{k_n\sigma_n} - \delta_{\sigma_1\nu} z_{k_1\mu} z_{k_2\sigma_2} \ldots z_{k_n\sigma_n} - \ldots$$
$$- \delta_{\sigma_n\nu} z_{k_1\sigma_1} z_{k_2\sigma_2} \ldots z_{k_{n-1}\sigma_{n-1}} z_{k_n\mu}) dy_\nu \wedge \omega_0,$$
$$\tilde{h}(dz_{k\mu} \wedge \omega_0) = dz_{k\mu} \wedge \omega_0,$$
$$\tilde{h}(dz_{k\mu} \wedge dx_1 \wedge dx_2 \wedge \ldots \wedge dy_{\sigma_1} \wedge \ldots \wedge dy_{\sigma_r} \wedge \ldots \wedge dx_n)$$
$$= \epsilon^{s_1 s_2 \ldots s_r}_{k_1 k_2 \ldots k_r} (z_{k_1\sigma_1} z_{k_2\sigma_2} \ldots z_{k_r\sigma_r} dz_{k\mu} - (z_{k_1 k\mu} z_{k_2\sigma_2} \ldots z_{k_r\sigma_r} \delta_{\sigma_1\nu} + \ldots$$
$$+ z_{k_1\sigma_1} z_{k_2\sigma_2} \ldots z_{k_{r-1}\sigma_{r-1}} z_{k_r k\mu} \delta_{\sigma_r\nu}) dy_\nu) \wedge \omega_0.$$

Thus $\tilde{h}(d\rho)$ becomes

$$\tilde{h}(d\rho) = \left( \left( \frac{\partial g_0}{\partial y_\mu} - \frac{\partial g^s_\mu}{\partial x_s} \right) dy_\mu \right.$$
$$+ \sum_{r=1}^{n} \sum_{s_1 < s_2 < \ldots < s_r} \sum_{\sigma_1, \sigma_2, \ldots, \sigma_r} \frac{1}{r!} \cdot \frac{\partial g^{s_1 s_2 \ldots s_r}_{\sigma_1 \sigma_2 \ldots \sigma_r}}{\partial y_\mu} \epsilon^{s_1 s_2 \ldots s_r}_{k_1 k_2 \ldots k_r} (\delta_{\mu\nu} z_{k_1\sigma_1} z_{k_2\sigma_2} \ldots z_{k_r\sigma_r}$$
$$- \delta_{\sigma_1\nu} z_{k_1\mu} z_{k_2\sigma_2} \ldots z_{k_r\sigma_r} - \ldots - \delta_{\sigma_r\nu} z_{k_1\sigma_1} z_{k_2\sigma_2} \ldots z_{k_{r-1}\sigma_{r-1}} z_{k_r\mu}) dy_\nu$$
$$- \sum_{r=1}^{n-1} \sum_{s_1 < s_2 < \ldots < s_r} \sum_{\sigma_1, \sigma_2, \ldots, \sigma_r} \frac{1}{(r+1)!} \left( \sum_{k < s_1} \frac{\partial g^{k s_1 s_2 \ldots s_r}_{\mu \sigma_1 \sigma_2 \ldots \sigma_r}}{\partial x_k} + \sum_{s_1 < k < s_2} \frac{\partial g^{s_1 k s_2 \ldots s_r}_{\sigma_1 \mu \sigma_2 \ldots \sigma_r}}{\partial x_k} \right.$$
$$\left. + \ldots + \sum_{k > s_r} \frac{\partial g^{s_1 s_2 \ldots s_r k}_{\sigma_1 \sigma_2 \ldots \sigma_r \mu}}{\partial x_k} \right) \epsilon^{s_1 s_2 \ldots s_r}_{k_1 k_2 \ldots k_r} (\delta_{\mu\nu} z_{k_1\sigma_1} z_{k_2\sigma_2} \ldots z_{k_r\sigma_r}$$
$$\text{(5.7)} \qquad - \delta_{\sigma_1\nu} z_{k_1\mu} z_{k_2\sigma_2} \ldots z_{k_r\sigma_r} - \ldots - \delta_{\sigma_r\nu} z_{k_1\sigma_1} z_{k_2\sigma_2} \ldots z_{k_{r-1}\sigma_{r-1}} z_{k_r\mu}) dy_\nu$$
$$+ \frac{\partial g_0}{\partial z_{k\mu}} dz_{k\mu} + \sum_{r=1}^{n} \sum_{s_1 < s_2 < \ldots < s_r} \sum_{\sigma_1, \sigma_2, \ldots, \sigma_r} \frac{1}{r!} \frac{\partial g^{s_1 s_2 \ldots s_r}_{\sigma_1 \sigma_2 \ldots \sigma_r}}{\partial z_{k\mu}} \epsilon^{s_1 s_2 \ldots s_r}_{k_1 k_2 \ldots k_r}$$
$$\cdot (z_{k_1 k\mu} z_{k_2\sigma_2} \ldots z_{k_r\sigma_r} \delta_{\sigma_1\nu} + \ldots + z_{k_1\sigma_1} z_{k_2\sigma_2} \ldots z_{k_{r-1}\sigma_{r-1}} z_{k_r k\mu} \delta_{\sigma_r\nu}) dy_\nu) \wedge \omega_0.$$

It is seen from this expression that the terms containing $dz_{k\mu}$ are just equal to $A_{k\mu} dz_{k\mu} \wedge \omega_0$. Further,

$$\frac{\partial G}{\partial y_\mu} = \frac{\partial g_0}{\partial y_\mu} + \sum_{r=1}^{n} \sum_{s_1 < s_2 < \ldots < s_r} \sum_{\sigma_1, \sigma_2, \ldots, \sigma_r} \frac{\partial g^{s_1 s_2 \ldots s_r}_{\sigma_1 \sigma_2 \ldots \sigma_r}}{\partial y_\mu} z_{s_1\sigma_1} z_{s_2\sigma_2} \ldots z_{s_r\sigma_r},$$

and it remains to determine the remaining terms. Write for this

$$\frac{\partial G}{\partial z_{i\mu}} = A_{i\mu} + B_{i\mu},$$

where

$$B_{i\mu} = \sum_{r=1}^{n} \sum_{s_1 < s_2 < \ldots < s_r} \sum_{\sigma_1, \sigma_2, \ldots, \sigma_r} g^{s_1 s_2 \ldots s_r}_{\sigma_1 \sigma_2 \ldots \sigma_r} \frac{\partial}{\partial z_{i\mu}} (z_{s_1\sigma_1} z_{s_2\sigma_2} \ldots z_{s_r\sigma_r}).$$

It suffices to show that

$$\tilde{h}(d\rho) = \left( \left( \frac{\partial G}{\partial y_\nu} - \frac{dB_{i\nu}}{dx_i} \right) dy_\nu + \left( \frac{\partial G}{\partial z_{i\nu}} - B_{i\nu} \right) dz_{i\nu} \right) \wedge \omega_0,$$

i.e., that the remaining terms in $\tilde{h}(d\rho)$ are equal to

$$-\frac{dB_{i\mu}}{dx_i} dy_\mu \wedge \omega_0 = -\left( \frac{\partial B_{i\mu}}{\partial x_i} + \frac{\partial B_{i\mu}}{\partial y_\sigma} z_{i\sigma} + \frac{\partial B_{i\mu}}{\partial z_{k\sigma}} \cdot z_{ik\sigma} \right) dy_\mu \wedge \omega_0.$$

By a rather difficult computation we arrive at the expressions

$$\begin{aligned} \frac{\partial B_{i\mu}}{\partial x_i} = \frac{\partial g^s_\mu}{\partial x_s} &+ \sum_{r=1}^{n-1} \sum_{s_1 < s_2 < \ldots < s_r} \sum_{\sigma_1, \sigma_2, \ldots, \sigma_r} \frac{1}{(r+1)!} \left( \sum_{k < s_1} \frac{\partial g^{k s_1 s_2 \ldots s_r}_{\rho\sigma_1\sigma_2 \ldots \sigma_r}}{\partial x_k} \right. \\ &\left. + \sum_{s_1 < k < s_2} \frac{\partial g^{s_1 k s_2 \ldots s_r}_{\sigma_1 \mu \sigma_2 \ldots \sigma_r}}{\partial x_k} + \ldots + \sum_{k > s_r} \frac{\partial g^{s_1 s_2 \ldots s_r k}_{\sigma_1 \sigma_2 \ldots \sigma_r \mu}}{\partial x_k} \right) \epsilon^{s_1 s_2 \ldots s_r}_{k_1 k_2 \ldots k_r} \\ &\cdot (\delta_{\mu\rho} z_{k_1\sigma_1} z_{k_2\sigma_2} \ldots z_{k_r\sigma_r} - \delta_{\mu\sigma_1} z_{k_1\rho} z_{k_2\sigma_2} \ldots z_{k_r\sigma_r} - \ldots \\ &- \delta_{\mu\sigma_r} z_{k_1\sigma_1} z_{k_2\sigma_2} \ldots z_{k_{r-1}\sigma_{r-1}} z_{k_r\rho}), \end{aligned}$$

$$\begin{aligned} \frac{\partial B_{i\mu}}{\partial y_\rho} z_{i\rho} &= \sum_{r=1}^{n} \sum_{s_1 < s_2 < \ldots < s_r} \sum_{\sigma_1, \sigma_2, \ldots, \sigma_r} \frac{1}{r!} \frac{\partial g^{s_1 s_2 \ldots s_r}_{\sigma_1 \sigma_2 \ldots \sigma_r}}{\partial y_\mu} \epsilon^{s_1 s_2 \ldots s_r}_{k_1 k_2 \ldots k_r} \\ &\cdot (z_{k_1\sigma} z_{k_2\sigma_2} \ldots z_{k_r\sigma_r} \delta_{\mu\sigma_1} + \ldots + z_{k_1\sigma_1} z_{k_2\sigma_2} \ldots z_{k_{r-1}\sigma_{r-1}} z_{k_r\sigma} \delta_{\mu\sigma_r}), \end{aligned}$$

$$\begin{aligned} \frac{\partial B_{i\mu}}{\partial z_{k\sigma}} z_{ik\sigma} &= \sum_{r=1}^{n} \sum_{s_1 < s_2 < \ldots < s_r} \sum_{\sigma_1, \sigma_2, \ldots, \sigma_r} \frac{1}{r!} \frac{\partial g^{s_1 s_2 \ldots s_r}_{\sigma_1 \sigma_2 \ldots \sigma_r}}{\partial z_{k\sigma}} \epsilon^{s_1 s_2 \ldots s_r}_{k_1 k_2 \ldots k_r} \\ &\cdot (z_{k_1 k\sigma} z_{k_2\sigma_2} \ldots z_{k_r\sigma_r} \delta_{\mu\sigma_1} + \ldots + z_{k_1\sigma_1} z_{k_2\sigma_2} \ldots z_{k_{r-1}\sigma_{r-1}} z_{k_r k\sigma} \delta_{\mu\sigma_r}). \end{aligned}$$

Our assertion now follows by comparison of these expressions with (5.7).

We note that the decomposition of the form $\tilde{h}(d\rho)$ into two terms (see (5.6)) is invariant, as can be checked by a direct calculation. Notice that the expression in the first bracket in (5.6), i.e. the coefficient at $dy_\nu \wedge \omega_0$, is just the Euler expression defined by (4.20). This is the main reason why those forms $\tilde{h}(d\rho)$ that are horizontal with respect to $\pi_{2,0}$, are of special interest from the point of view of the variational calculus.

**Lepagian forms.** Using the results of the previous paragraph, we are led to the following definition:

Let $\rho$ be an n-form on $J^r$; $\rho$ is said to be a *Lepagian form*, if the $(n+1)$-form $\tilde{h}(d\rho)$ is horizontal with respect to the projection $\pi_{r+1,0}$.

In this work $\Omega_X^p(J^r)$, $\Omega_Y^p(J^r)$ will denote the spaces of horizontal *p*-forms with respect to $\pi_r$, and horizontal *p*-forms with respect to $\pi_{r,0}$, respectively.

The Lepagian forms on $J^1$ are characterized by the following:

*For an n-form* $\rho \in \Omega_Y^n(J^1)$ *the next four conditions are equivalent:*

1) $\rho$ *is Lepagian.*

2) *In any coordinates defined by a canonical chart on* $J^1$

(5.8) $$A_{k\nu} = 0.$$

3) $\tilde{h}(d\rho)$ *depends only on* $h(\rho)$.

4) *There is one and only one pseudovertical* 1*-form* $E \in \Omega_Y^1(J^2)$ *such that*

(5.9) $$\tilde{h}(d\rho) = E \wedge \omega.$$

The proof is based on the relation (5.6). Conditions 1) and 2) are obviously equivalent. Condition 3) means, in fact, that the mapping $\rho \to \tilde{h}(d\rho)$ is constant on the sets of forms with a given horizontal part. It could be reformulated more exactly by saying that $\rho$ is a Lepagian if and only if for any $\bar{\rho} \in \Omega_Y^n(J^1)$ such that $h(\bar{\rho} - \rho) = 0$ we have $\tilde{h}(d(\bar{\rho} - \rho)) = 0$. But it becomes clear that then 3) is equivalent with 2). Let us examine 4). If $\rho$ is Lepagian, we take

(5.10) $$E = \frac{1}{F}\left(\frac{\partial G}{\partial y_\nu} - \frac{d}{dx_i}\left(\frac{\partial G}{\partial z_{i\nu}}\right)\right) \cdot (dy_\nu - z_{k\nu} dx_k),$$

where $\omega$ is a volume element on *X*, and the function $F > 0$ is defined by the relation $\omega = F\omega_0$. One may prove that this expression is invariant with respect to coordinate transformations in $J^2$. If we admit that there is another 1-form $\bar{E}$ satisfying 4), we arrive at the equality $(\bar{E} - E) \wedge \omega = 0$, and then, by the assumption of the pseudover-ticality of both $\bar{E}$ and *E*, $\bar{E} = E$. Thus 4) follows from 1). Conversely, if $\tilde{h}(d\rho)$ is of the desired form, we apply (5.6) and see at once that $A_{k\nu} = 0$. This completes the proof.

(For the classical approach to the problem of an invariant derivation of the Euler equations, see [27], [4]. Our exposition is, in fact, equivalent with the approach sketched in [4]. A modern treatment has been given in [35]. Notice that the approach given here is based upon the existence of the mappings *h* and $\tilde{h}$ replacing the equiva-lence relations of *Lepage* [27], [4].)

**Lagrangian forms and the Euler form.** According to our previous definition [20] by a *Lagrangian form* of degree *r* on a fibered manifold $(Y, \pi, X)$ we mean each $n$-form on $J^r$ (remember that $n = \dim X$). Real functions on $J^r$ are called *Lagrange functions* of degree *r* on $(Y, \pi, X)$.

Clearly, to each Lagrangian form $\lambda$ of degree *r* there corresponds a Lagrange function, $L_\lambda$, of degree $r+1$, defined by the relation

(5.11) $$h(\lambda) = L_\lambda \cdot \pi_{r+1}^* \omega.$$

In some special cases (compare with the proposition of Section 4, dealing with the properties of horizontal and pseudovertical forms), e.g. iin the case when $\lambda$ is horizontal with respect to $\pi_{r,r-1}$, the Lagrange function $L_\lambda$ can be considered as defined on $J^r$.

Let us now consider the mapping $\rho \to \tilde{h}(d\rho)$ restricted to the set of Lepagian forms on $J^1$. As we know from the above theorem, the mapping is constant on the classes of Lepagian forms with the same horizontal parts. On the other hand, the mapping *h* is surjective (i.e. maps onto $\Omega_X^n(J^1)$, see (4.22)). This means that the formula

*(5.12)* $\tilde{h}(d\rho) = E(h(\rho)) \wedge \omega$

gives rise to the mapping

(5.13) $\Omega_X^n(J^1) \ni \lambda \to E(\lambda) \in \Omega_Y^1(J^2)$,

which will be called the *Euler mapping*. Each $E(\lambda)$, with $\lambda \in \Omega_X^n(J^1)$, will be called the *Euler form* defined by the Lagrangian form $\lambda$. Evidently, the Euler form might be considered as defined by the corresponding Lagrange function $L_\lambda$, which is, according to the horizontality of Lepagian forms with respect to $\pi_{1,0}$, also of degree 2 (see (5.10)).

The function $\lambda \to E(\lambda)$ is linear (over $\mathbf{R}$). In this Section we are going to give a characterization of its kernel, or, which is the same, to find conditions under which the Euler equations of the variational problem defined by a Lagrangian form of degree 1, are identically fulfilled.

**A note on a classical proposition.** Let $U \subset \mathbf{R}^n$ be an open set and consider the fibered manifold $(U \times \mathbf{R}, pr, U)$ and its 1-jet prolongation $(U \times \mathbf{R} \times \mathbf{R}^n, pr, U)$, with the natural projections on *U*, denoted by the same symbol *pr*. Let *L* be a differentiable function on $U \times \mathbf{R} \times \mathbf{R}^n$, i.e. a Lagrange function of degree 1 on $(U \times \mathbf{R}, pr, U)$. The corresponding Euler equation then can be written as

$$E(L) = \frac{\partial L}{\partial y} - \frac{d}{dx_k}\left(\frac{\partial L}{\partial \dot{y}_k}\right) = 0.$$

It is usually considered as a partial differential equation for a function $\varphi: U \to \mathbf{R}$ (see, say [8], [11]). It is known that the equation is identically fulfilled (i.e. for all $\varphi$) if and only if *L* is a "divergence expression" [8]. This means that in the natural coordinates $(x_1, x_2, \ldots, x_n, y, \dot{y}_1, \dot{y}_2, \ldots, \dot{y}_n)$ on $U \times \mathbf{R} \times \mathbf{R}^n$,

$$L = \frac{\partial f_i}{\partial x_i} + \frac{\partial f_i}{\partial y}\dot{y}_i = \frac{df_i}{dx_i}$$

for some functions $f_1, f_2, \ldots, f_n$ of the variables $x_1, x_2, \ldots, x_n, y$.

In the literature there appears an assertion that an analogous theorem is also true for more functional arguments $\varphi_1, \varphi_2, \ldots, \varphi_m$ (i.e. functions with values in $\mathbf{R}$). The next example shows, however, that there are Lagrange functions which are not of the "*divergence type*", but do lead to the vanishing Euler expressions.

Let $U \subset \mathbf{R}^2$ be an open domain, $(U \times \mathbf{R}^2, pr, U)$ the fibered manifold defined by the natural projection, $J^1 = U \times \mathbf{R}^2 \times \mathbf{R}^4$ its 1-jet prolongation, $L$ a differentiable function on $J^1$. We shall write $(x_1, x_2, y_1, y_2, z_{11}, z_{12}, z_{21}, z_{22})$ for the natural coordinates on $J^1$. The Lagrange function gives rise to the following equations (the *Euler equations*)

$$\frac{\partial L}{\partial y_\mu} - \frac{d}{dx_i}\left(\frac{\partial L}{\partial z_{i\mu}}\right) = 0, \quad \mu = 1, 2,$$

where summation over $i = 1, 2$ takes place. Let us consider the case of $L$ defined by means of an arbitrary function $f$ of variables $(x_1, x_2, y_1, y_2)$, by the formula

$$L = \left(\frac{\partial f}{\partial x_2} + \frac{\partial f}{\partial y_2} z_{22}\right) \cdot z_{11} - \left(\frac{\partial f}{\partial x_1} + \frac{\partial f}{\partial y_2} z_{12}\right) \cdot z_{12}.$$

The direct substitution shows that

$$E_1 = \frac{\partial}{\partial y_1} - \frac{d}{dx_1}\left(\frac{\partial L}{\partial z_{11}}\right) - \frac{d}{dx_2}\left(\frac{\partial L}{\partial z_{21}}\right) = 0,$$

$$E_2 = \frac{\partial}{\partial y_2} - \frac{d}{dx_1}\left(\frac{\partial L}{\partial z_{12}}\right) - \frac{d}{dx_2}\left(\frac{\partial L}{\partial z_{22}}\right) = 0,$$

while $L$ is not of the "divergence type" (it depends bilinearly on the $z_{i\mu}$).

It is seen from this example that the case of more functional arguments essentially differs from that one of one functional variable. It is not known to the author whether the problem of identical vanishing of the Euler expressions, or, alternatively, the problem of a characterization of the kernel of the Euler mapping, has been discussed in its full generality (i.e. regardless of the number of the “functional arguments”). In the next two paragraphs we are going to extend the mentioned classical result of *Courant* and *Hilbert* [8] so much used in theoretical physics, to the general case.

**The Euler mapping.** Let us turn to the Euler mapping (5.13). As a consequence of the proved theorem, we have the following proposition:

*Consider the Euler mapping*

$$\Omega_X^n(J^1) \ni \lambda \to E(\lambda) \in \Omega_Y^1(J^2).$$

*A necessary and sufficient condition for* $E(\lambda) = 0$ *is that there exists an n-form* $\rho$ *on Y such that*

1) $h(\rho) = \lambda$,
2) $d\rho = 0$.

*To each* $\lambda$ *with* $E(\lambda) = 0$ *there exists one and only one* $\rho$ *satisfying 1) and 2).*

Suppose $E(\lambda) = 0$, and work with a canonical chart on $J^2$. Write $(x_i, y_\mu, z_{i\mu}, z_{ij\mu})$**,** $i \le j$, for the corresponding coordinates. Put

$$\lambda = \mathscr{L} dx_1 \wedge dx_2 \wedge \ldots \wedge dx_n.$$

The function $\mathscr{L}$ is supposed to satisfy the system of partial differential equations

$$\frac{\partial \mathscr{L}}{\partial y_\mu} - \frac{d}{dx_k}\left(\frac{\partial \mathscr{L}}{\partial z_{k\mu}}\right) = 0$$

equivalent with

$$\frac{\partial^2 \mathscr{L}}{\partial z_{l\sigma}\,\partial z_{k\mu}} + \frac{\partial^2 \mathscr{L}}{\partial z_{k\sigma}\,\partial z_{l\mu}} = 0, \quad \frac{\partial \mathscr{L}}{\partial y_\mu} - \frac{\partial^2 \mathscr{L}}{\partial x_k \partial z_{k\mu}} - \frac{\partial^2 \mathscr{L}}{\partial y_\sigma\,\partial z_{k\mu}} z_{k\sigma} = 0.$$

It follows that $\mathscr{L}$ must be of the form

$$\mathscr{L} = f_0 + \sum_{r=1}^{n} \sum_{s_1<s_2<\ldots<s_r} \sum_{\sigma_1,\sigma_2,\ldots,\sigma_r} f^{s_1 s_2\ldots s_r}_{\sigma_1\sigma_2\ldots\sigma_r} \cdot z_{s_1\sigma} z_{s_2\sigma_2}\ldots z_{s_r\sigma_r},$$

where $f_0$ and $f^{s_1 s_2\ldots s_r}_{\sigma_1\sigma_2\ldots\sigma_r}$ do not depend on $z_{j\nu}$, and $f^{s_1 s_2\ldots s_r}_{\sigma_1\sigma_2\ldots\sigma_r}$ are antisymmetric in $\sigma_1$, $\sigma_2,\ \ldots,\ \sigma_r$. The second condition then can be written as

$$\begin{aligned}
&\frac{\partial f_0}{\partial y_\mu} - \frac{\partial f^k_\mu}{\partial x_k} + \sum_{r=1}^{n-1} \sum_{s_1<s_2<\ldots<s_r} \sum_{\sigma_1,\sigma_2,\ldots,\sigma_r} \left( \frac{\partial f^{s_1 s_2\ldots s_r}_{\sigma_1\sigma_2\ldots\sigma_r}}{\partial y_\mu} - \sum_{k<s_1} \frac{\partial f^{k s_1 s_2\ldots s_r}_{\mu\sigma_1\sigma_2\ldots\sigma_r}}{\partial x_k} \right.\\
&- \sum_{s_1<k<s_2} \frac{\partial f^{s_1 k s_2\ldots s_r}_{\sigma_1\mu\sigma_2\ldots\sigma_r}}{\partial x_k} - \ldots - \sum_{k>s_r} \frac{\partial f^{s_1 s_2\ldots s_r k}_{\sigma_1\sigma_2\ldots\sigma_r\mu}}{\partial x_k} - \frac{\partial f^{s_1 s_2\ldots s_r}_{\mu\sigma_2\ldots\sigma_r}}{\partial y_{\sigma_1}}\\
&\left. - \frac{\partial f^{s_1 s_2 s_3\ldots s_r}_{\sigma_1\mu\sigma_3\ldots\sigma_r}}{\partial y_{\sigma_2}} - \ldots - \frac{\partial f^{s_1 s_2\ldots s_{r-1} s_r}_{\sigma_1\sigma_2\ldots\sigma_{r-1}\mu}}{\partial y_{\sigma_r}} \right) z_{s_1\sigma_1} z_{s_2\sigma_2}\ldots z_{s_r\sigma_r}\\
&+ \sum_{\sigma_1,\sigma_2,\ldots,\sigma_n} \left( \frac{\partial f^{12\ldots n}_{\sigma_1\sigma_2\ldots\sigma_n}}{\partial y_\mu} - \frac{\partial f^{12\ldots n}_{\mu\sigma_2\ldots\sigma_n}}{\partial y_{\sigma_1}} - \frac{\partial f^{123\ldots n}_{\sigma_1\mu\sigma_3\ldots\sigma_n}}{\partial y_{\sigma_2}} - \ldots - \frac{\partial f^{12\ldots n-1 n}_{\sigma_1\sigma_2\ldots\sigma_{n-1}\mu}}{\partial y_{\sigma_n}} \right)\\
&\cdot z_{1\sigma_1} z_{2\sigma_2}\ldots z_{n\sigma_n} = 0.
\end{aligned}$$

Since the coefficients should vanish separately, one gets some conditions for $f_0$, and $f^{s_1 s_2\ldots s_r}_{\sigma_1\sigma_2\ldots\sigma_r}$:

$$\frac{\partial f_0}{\partial y_\mu} - \frac{\partial f^k_\mu}{\partial x_k} = 0,$$

$$\begin{aligned}
&\frac{\partial f^{s_1 s_2\ldots s_r}_{\sigma_1\sigma_2\ldots\sigma_r}}{\partial y_\mu} - \frac{\partial f^{s_1 s_2\ldots s_r}_{\mu\sigma_2\ldots\sigma_r}}{\partial y_{\sigma_1}} - \frac{\partial f^{s_1 s_2 s_3\ldots s_r}_{\sigma_1\mu\sigma_3\ldots\sigma_r}}{\partial y_{\sigma_2}} - \ldots - \frac{\partial f^{s_1 s_2\ldots s_{r-1} s_r}_{\sigma_1\sigma_2\ldots\sigma_{r-1}\mu}}{\partial y_{\sigma_r}}\\
&- \sum_{k<s_1} \frac{\partial f^{k s_1 s_2\ldots s_r}_{\mu\sigma_1\sigma_2\ldots\sigma_r}}{\partial x_k} - \sum_{s_1<k<s_2} \frac{\partial f^{s_1 k s_2\ldots s_r}_{\sigma_1\mu\sigma_2\ldots\sigma_r}}{\partial x_k} - \ldots - \sum_{k>s_r} \frac{\partial f^{s_1 s_2\ldots s_r k}_{\sigma_1\sigma_2\ldots\sigma_r\mu}}{\partial x_k} = 0,
\end{aligned} \tag{5.14}$$

$$\frac{\partial f^{1\ \cdots\ n}_{\sigma_1\cdots\sigma_n}}{\partial y_\mu} - \frac{\partial f^{1\ 2\ \ n}_{\mu\ \mu\cdots\sigma_n}}{\partial y_{\sigma_1}} - \ldots - \frac{\partial f^{1\ \ n-1\ \ n}_{\sigma_1\cdots\sigma_{n-1}\ \mu}}{\partial y_{\sigma_r}} = 0.$$

where $1 \le r \le n-1$.. Suppose that we have some functions $f_0$, $f^{s_1 s_2\ldots s_r}_{\sigma_1\sigma_2\ldots\sigma_r}$ satisfying all the conditions (5.14) and construct a local $n$-form $\rho$, defined as

$$\rho = f_0 dx_1 \wedge dx_2 \wedge \ldots \wedge dx_n + \sum_{r=1}^{n} \sum_{s_1 < s_2 < \ldots < s_r} \sum_{\sigma_1, \sigma_2, \ldots, \sigma_r} \frac{1}{r!} f^{s_1 s_2 \ldots s_r}_{\sigma_1 \sigma_2 \ldots \sigma_r} \tag{5.15}$$
$$\cdot dx_1 \wedge dx_2 \wedge \ldots \wedge dy_{\sigma_1} \wedge \ldots \wedge dy_{\sigma_r} \wedge \ldots \wedge dx_n$$

(compare with (5.1)). An immediate comparison with formula (5.2) shows that $d\rho = 0$. Moreover, $\rho$ is defined on $Y$ and $h(\rho) = \lambda$, by (4.11). Suppose that there is another form, $\bar{\rho}$, such that $h(\bar{\rho}) = \lambda$ and that $\bar{\rho}$ is defined on $Y$. Then for all $\gamma \in \Gamma(\pi)$

$$j^1\gamma^* h(\bar{\rho} - \rho) = 0 = j^1\gamma^* \pi^*_{1,0}(\bar{\rho} - \rho) = \gamma^*(\bar{\rho} - \rho),$$

which implies that $\bar{\rho} - \rho = 0$. This proves uniqueness. It follows that $\rho$ is independent of the choice of particular coordinates. To prove the converse, take (5.15) with $f_0$, $f^{s_1 s_2 \ldots s_r}_{\sigma_1 \sigma_2 \ldots \sigma_r}$ not depending on $z_{i\nu}$, apply the formula (5.2) and then use (5.14). This completely proves the theorem.

**Lagrange functions of degree 1 leading to zero Euler form.** As we noted before, the question on the form of the Lagrange functions leading to zero Euler expressions has been studied in some simple cases in classical literature [8], and is frequently discussed in connection with application of the calculus of variations in theoretical physics (see e.g. [37], [38]). It is mainly of practical interest to have a characterization of such functions, since variational problems are usually defined by means of them and not by means of exterior differential forms.

A coordinate description of such functions can be derived from our theorem on the Euler mapping by means of the *Poincaré lemma* (see e.g. [26]). Consider a form $\rho$ on $Y$ such that $d\rho = 0$. By the Poincaré lemma $\rho$ can be written locally as $d\eta$ for a suitable $(n-1)$-form $\eta$ on $Y$. The only thing we need is just to determine the function $L$ from the condition

$$h(d\eta) = L\pi_1^*\omega.$$

Then it is clear that $L$ will give zero Euler expressions. Conversely, the procedure makes it possible to obtain all Lagrange functions of degree 1 with the property. Let us pass to the formulation of the proposition.

*Let $\lambda$ be a horizontal Lagrangian form of degree 1 on a fibered manifold $(Y, \pi, X)$. Write $(x_i, y_\mu, z_{i\mu})$ for coordinates on $J^1$, corresponding to a canonical chart. Let*

$$\lambda = L dx_1 \wedge dx_2 \wedge \ldots \wedge dx_n$$

*be the coordinate expression for $\lambda$. Then $E(\lambda) = 0$ if and only if there exist functions $f_{s_1, s_2, \ldots, s_r; \sigma_1, \sigma_2, \ldots, \sigma_p}$, $r + p = n - 1$, $1 \le s_1 \le s_2 \le \ldots \le s_r \le n$, $1 \le \sigma_1 \le \sigma_2 \le \ldots \le \sigma_p \le m$, of coordinates $x_i$, $y_\mu$, such that*

$$L = \sum_{s_1 < s_2 < \ldots < s_r} \sum_{\sigma_1, \sigma_2, \ldots, \sigma_p} \left( \frac{\partial f_{s_1, s_2, \ldots, s_r; \sigma_1, \sigma_2, \ldots, \sigma_p}}{\partial x_k} + \frac{\partial f_{s_1, s_2, \ldots, s_r; \sigma_1, \sigma_2, \ldots, \sigma_p}}{\partial y_\mu} z_{k\mu} \right) \tag{5.16}$$
$$\cdot z_{i_1 \sigma_1} z_{i_2 \sigma_2} \ldots z_{i_p \sigma_p} \epsilon^{12 \ldots r\, r+1\, r+2 \ldots n}_{k s_1 s_2 \ldots s_r i_1 i_2 \ldots i_p}.$$

*In this expression summation over* $\mu$, $k$, $i_1$, $i_2$, $\dots$, $i_p$ *takes place.*

For the proof it suffices to compute the coordinate expression for $h(d\eta)$ with a general $(n-1)$-form $\eta$ on $Y$ written as

$$\eta = \sum_{s_1 < s_2 < \dots < s_r} \sum_{\sigma_1, \sigma_2, \dots, \sigma_p} f_{s_1, s_2, \dots, s_r; \sigma_1, \sigma_2, \dots, \sigma_p} dx_{s_1} \wedge dx_{s_2} \wedge \dots \wedge dx_{s_r}$$
$$\wedge dy_{\sigma_1} \wedge dy_{\sigma_2} \wedge \dots \wedge dy_{\sigma_r}, \quad r + p = n - 1.$$

(See [21].)

**Example.** Since formula (5.16) is rather complicated it seems to be more effective to proceed from the beginning in each individual case. If e.g. $\dim X = 2$, the 1-forms on $Y$ are expressed as

$$\eta = f_i dx_i + g_\mu dy_\mu .$$

After some calculation

$$h(d\eta) = \left( \frac{\partial f_i}{\partial x_i} + \left( \frac{\partial g_\mu}{\partial x_i} - \frac{\partial f_i}{\partial y_\mu} \right) z_{j\mu} + \frac{\partial g_\mu}{\partial y_\sigma} z_{i\sigma} z_{j\mu} \right) \epsilon_{ij}^{12} \cdot dx_1 \wedge dx_2 ,$$

which defines $L$ as

$$L = \left( \frac{\partial f_i}{\partial x_i} + \left( \frac{\partial g_\mu}{\partial x_i} - \frac{\partial f_i}{\partial y_\mu} \right) z_{j\mu} + \frac{\partial g_\mu}{\partial y_\sigma} z_{i\sigma} z_{j\mu} \right) \epsilon_{ij}^{12} .$$

If $\dim X = 1$ we start with a 0-form $\eta$, i.e. with a function $F$ on $Y$, and get

$$h(dF) = \frac{dF}{dx} dx, \quad L = \frac{dF}{dx} .$$

Here $x$ denotes a coordinate on $X$ and $d / dx$ is the formal derivative with respect to the coordinate $x$.

## 6. Canonical variational problems

**Variational problems on compact base manifolds.** According to *Hermann* [13], [14], we say that there is given a *variational problem*, if we have the following objects:

1) a compact oriented manifold $X$ with boundary $\partial X$; denote $n = \dim X$ and suppose that $\partial X$ has the induced orientation;

2) a manifold $Y$ with $\dim Y = n + m$, $m \geq 0$;

3) a differential $n$-form $\rho$ on $Y$;

4) a differential ideal $I$ of differential forms on $Y$ (i.e. an ideal with respect to the exterior algebra structure).

(We note that this definition generalizes the approach of *É. Cartan* and *Lepage* to the calculus of variations. *Hermann* [13] calls the defined variational problems "canonical" but we reserve this word for slightly modified ones. The reasons will become clear in the next few paragraphs.)

Suppose we have a variational problem. The data 1) – 4) allow us to consider a real function, also denoted by $\rho$, defined on the set $\Phi$ of all submanifold maps from *X* to *Y*, by the relation

$$(6.1)\qquad \Phi \ni \varphi \to \rho(\varphi) = \int_X \varphi^* \rho \in \mathbf{R}.$$

Here, as usual, $\varphi^*\rho$ denotes the pull-back of $\rho$ by $\varphi$. The variational theory studies the behavior of the function $\rho$ restricted to the set

$$(6.2)\qquad \Phi_I = \{\varphi \in \Phi || \varphi^* \eta = 0, \eta \in I\}.$$

The theory arising is relatively simple since it is based on the systematic use of the *Stokes' formula* for integration of differential forms (see [36], [7]).

In the variational theory, the behavior of the functional (6.1) is studied in such a way that each individual $\varphi_0 \in \Phi_I$ is subject to prescribed variations, and the induced changes of the number (6.1) are observed.

Let $\varphi_0 \in \Phi$; a differentiable mapping $\nu : X \to TY$ with the property that $\nu(\varphi_0(x)) \in T_{\varphi_0(x)}Y$ for all $x \in X$, is called the *vector field along the morphism* $\varphi_0$ (see [12], [13]). It follows from the compactness of *X* that there is a vector field $\bar{\nu}$ on *Y* satisfying, for all $x \in X$,

$$\bar{\nu}(\varphi_0(x)) = \nu(\varphi_0(x))$$

(compare with [12]). If now $\bar{\nu}$ is an arbitrary vector field on *Y*, one can consider its one-parameter group $\alpha_t^{\bar{\nu}}$ of local transformations, and the mapping $t \to \alpha_t^{\bar{\nu}} \circ \varphi_0$ generated by $\bar{\nu}$ on $\varphi_0$ (the mapping could be called a *deformation* of $\varphi_0$ by $\bar{\nu}$). There arises a function

$$(6.3)\qquad (-\epsilon, \epsilon) \ni t \to \int_X (\alpha_t^{\bar{\nu}} \circ \varphi_0)^* \rho \in \mathbf{R},$$

defined for all sufficiently small *t*. According to the well-known formula for the Lie derivative $\vartheta(\bar{\nu})\rho$ of $\rho$ by $\bar{\nu}$,

$$(6.4)\qquad \vartheta(\bar{\nu})\rho = di(\bar{\nu})\rho + i(\bar{\nu})d\rho,$$

we obtain the so-called *first variation formula* in the form

$$(6.5)\qquad \left\{\frac{d}{dt}\rho(\alpha_t^{\bar{\nu}} \circ \varphi_0)\right\} = \int_X \varphi_0{}^* i(\bar{\nu})d\rho + \int_{\partial X} \varphi_0{}^* i(\bar{\nu})\rho;$$

here the mentioned Stokes' formula was applied.

Untill now, the vector field $\bar{\nu}$ was quite general. However, for our variational problem we have to take $\bar{\nu}$ with the property that if $\varphi_0^* \eta = 0$ for all $\eta \in I$, then also

$\varphi_0^* \alpha_t^{\bar{\nu}*} \eta = 0$ for all $\eta \in I$, i.e.

$$(6.6) \qquad \varphi_0^* \vartheta(\bar{\nu}) \eta = 0$$

for all $\eta$. Clearly, the reasoning could be turned and we get condition (6.6) as a sufficient and necessary one for both $\varphi_0^* \eta = 0$, $\varphi_0^* \alpha_t^{\bar{\nu}*} \eta = 0$ to be zero for all $\eta \in I$. In other words we can say that the vector field $\bar{\nu}$ generates variations of the morphism $\varphi_0$ *in the set* $\Phi_I$, if and only if condition (6.6) is fulfilled.

We are now in a position to define the notion of a critical point of our variational problem, or, as we also say, an extremal of our variational problem: A submanifold mapping $\varphi_0 \in \Phi_I$ is said to be a *critical point* (or an *extremal*) of the variational problem *in the set* $\Phi_I$, if for any vector field $\bar{\nu}$ on *Y* satisfying $\varphi_0^* \vartheta(\bar{\nu}) \eta = 0$ for all $\eta \in I$ we have

$$(6.7) \qquad \varphi_0^* i(\bar{\nu}) d\rho = 0.$$

According to [13] we say that $\varphi_0$ is a *critical point of the first kind* (or an *extremal of the first kind*) if it satisfies (6.7) for all vector fields $\nu$ on *Y*.

(For further discussion, e.g., for the definition of transversality of $\bar{\nu}$ to $\varphi$ at the boundary $\partial X$, see [13].)

**A note on the theory of variational problems.** The next remark shows that one must be careful when applying the above theory of variational problems to fibered manifolds. As we shall see, the decomposition (6.4) is not quite satisfactory for the theory of variations, unless the basic form $\rho$ is chosen very carefully. In particular, it is not a priori clear whether the term of the form of exterior differential (i.e. $di(\bar{\nu})\rho$) is uniquely separated. If this is not the case, the second term ($i(\bar{\nu})d\rho$) contains a part which can be transformed, in the first variation formula, into the integral over $\partial X$. The considerations will become more clear in the case of the following variational problem:

Take the following objects:

1) a compact manifold $X \subset U$ with boundary $\partial X$, where $U \subset \mathbf{R}^n$ is an open set; $\dim X = n$;

2) $Y = U \times \mathbf{R}^m \times L(\mathbf{R}^n, \mathbf{R}^m)$, where $U \times \mathbf{R}^m \times L(\mathbf{R}^n, \mathbf{R}^m)$ is considered as the 1-jet prolongation of the fibered manifold $(U \times \mathbf{R}^m, pr, U)$ (pr is the natural projection);

3) a horizontal Lagrangian *n*-form $\lambda$ on *Y* (horizontality with respect to the projection onto *U*); we write in natural coordinates $(x_i, y_\mu, z_{i\mu})$

$$\lambda = \mathscr{L} dx_1 \wedge dx_2 \wedge \ldots \wedge dx_n;$$

4) the ideal *I* generated by 1-forms

$$(6.8) \qquad \omega_\mu = dy_\mu - z_{i\mu} dx_i.$$

As in the general case, let $\Phi$ denote the set of all submanifold maps of *X* into $U \times \mathbf{R}^m \times L(\mathbf{R}^n, \mathbf{R}^m)$ and $\Phi_I$ its subset, annihilating *I*. It is seen at once that $\bar{\varphi} \in \Phi$ is an element of $\Phi_I$ if and only if there exists a morphism $\varphi : X \to \mathbf{R}^m$ satisfying $\bar{\varphi} \in j^1\varphi$; in other words,

$$\Phi_I = \{\bar{\varphi} \in \Phi \,|\, \bar{\varphi} \in j^1\varphi\}.$$

Consider condition (6.6). Write

(6.9) $$\bar{\nu} = \nu_k \frac{\partial}{\partial x_k} + \nu_\mu \frac{\partial}{\partial y_\mu} + \nu_{k\mu} \frac{\partial}{\partial z_{k\mu}}.$$

By a direct computation

$$i(\bar{\nu})\omega_\mu = \nu_\mu - z_{i\mu}\nu_i,$$

$$dj^1\varphi^* i(\bar{\nu})\omega_\mu = \left( \frac{\partial \nu_\mu}{\partial x_k} - \frac{\partial^2 \varphi_\mu}{\partial x_k \partial x_i} \nu_i - \frac{\partial \varphi_\mu}{\partial x_i} \frac{\partial \nu_i}{\partial x_k} \right) dx_k,$$

$$i(\bar{\nu})d\omega_\mu = -\nu_{i\mu} dx_i + \nu_i dz_{i\mu},$$

$$j^1\varphi^* i(\bar{\nu})d\omega_\mu = \left( \frac{\partial^2 \varphi_\mu}{\partial x_i \partial x_k} \cdot \nu_i - \nu_{k\mu} \right) dx_k,$$

where

$$\varphi(x_1, x_2, \ldots, x_n) = (\varphi_1(x_1, x_2, \ldots, x_n), \varphi_2(x_1, x_2, \ldots, x_n), \ldots, \varphi_m(x_1, x_2, \ldots, x_n)).$$

Condition (6.6) thus leads to the relation

$$\nu_{k\mu} = \frac{\partial \nu_\mu}{\partial x_k} - \frac{\partial \varphi_\mu}{\partial x_i} \frac{\partial \nu_i}{\partial x_k}$$

along the morphism $\varphi$.

We are now prepared to discuss the notion of a critical points of the considered variational problem. A mapping $\bar{\varphi} : U \to U \times \mathbf{R}^m \times L(\mathbf{R}^n, \mathbf{R}^m)$ is a critical point of the variational problem if it is of the form $j^1\varphi$ for some morphism $\varphi : U \to U \times \mathbf{R}^m$,, and if

(6.10) $$\varphi^* i(\bar{\nu})d\lambda = 0$$

for all $\bar{\nu}$ of the form

(6.11) $$\bar{\nu} = \nu_k \frac{\partial}{\partial x_k} + \nu_\mu \frac{\partial}{\partial y_\mu} + \left( \frac{\partial \nu_\mu}{\partial x_k} - \frac{\partial \varphi_\mu}{\partial x_i} \frac{\partial \nu_i}{\partial x_k} \right) \frac{\partial}{\partial z_{k\mu}}.$$

Our wish is to show that this definition does *not*, in general, fit in. One gets

$$\varphi^* i(\bar{\nu})d\lambda = \left( \frac{\partial \mathscr{L}}{\partial y_\mu} \left( \nu_\mu - \frac{\partial \varphi_\mu}{\partial x_k} \nu_k \right) + \frac{\partial \mathscr{L}}{\partial z_{i\mu}} \frac{\partial}{\partial x_i} \left( \nu_\mu - \frac{\partial \varphi_\mu}{\partial x_k} \nu_k \right) \right)$$
$$\cdot dx_1 \wedge dx_2 \wedge \ldots \wedge dx_n,$$

and according to the arbitrariness of the functions $\nu_k$ and $\nu_\mu$, the morphism $\bar{\varphi} \in j^1\varphi$ should satisfy the equations

(6.12) $$\frac{\partial \mathscr{L}}{\partial y_\mu} = 0, \quad \frac{\partial \mathscr{L}}{\partial z_{i\mu}} = 0.$$

However, these conditions *differ* from the usual definition of extremals as solutions of the *Euler equations*.

Notice also, that the $(n+1)$-form $d\lambda$ is not, in general, horizontal with respect to the projection on $\mathbf{R}^m$ and therefore cannot give the Euler equations (from (6.10)). In part, this discussion could serve as a motivation for further development of the theory of variations: if one wants to apply the described variational theory to the problems on jet spaces, one should use such differential forms that give their exterior differential independent of $\nu_{k\mu}$ (6.9). In this way we come to the *Lepagian forms*, defined in the previous Section.

The problem briefly sketched here, will be stated more precisely in next paragraphs. We will show how to construct a "suitable" basic form $\rho$ by means of a given Lagrangian horizontal form $\lambda$ (with respect to the projection on the base manifold) so that $\rho$ and $\lambda$ define the same variational problem.

**Variational problems on non-compact base manifolds.** Let $(Y,\pi,X)$ be our fibered manifold, $n = \dim X$. Formulations of other types of variational problems that ones discussed above appears e.g. in theoretical physics where it is not *a priori* known whether the base manifold *X* is compact or not. In these cases we cannot, in general, integrate over all manifold *X* in (6.1). Thus, there are two possibilities, essentially equivalent: 1) if we want to preserve the notion of the basic "functional" (called sometimes the *action function*) we have to integrate over *n*-chains on the manifold *X* (see [20]), and 2) we can develop the theory based on integration over all compact subsets of *X*. Especially because of simplicity, we shall follow here the second way. We note that, of course, our definitions holds good for both compact and non-compact *X*, and are closely related to the basic assumptions on the variational problems in the general relativity theory.

Let us return to vector fields. The *support* of a vector field is the closure of the set where the vector field is different from 0. Let *c* be a compact subset of *X*. A projectable vector field $\Xi$ on *Y* is called *c-admissible* if its support, $supp\Xi$, satisfies

(6.13) $\quad supp\Xi \subset \pi^{-1}(c).$

It is easy to see that a *c*-admissible vector field $\Xi$ has its integral curves starting at $y \in Y$ such that $y \notin supp\Xi$, of the form $t \to \alpha_t^{\Xi}(y) = y$. Let $\gamma \in \Gamma(\pi)$ be an arbitrary cross section. If $x \notin c$ then the vector field $\Xi$ remains unchanged the value $\gamma(x) \in Y$.

We shall say that there is given a *variational problem of order r*, if we have

1) a fibered manifold $(Y,\pi,X)$ with orientable base manifold *X*,
2) a Lagrangian form $\lambda$ of degree *r* on $(Y,\pi,X)$.

If $\gamma \in \Gamma(\pi)$ and $\lambda$ is an *n*-form on $J^r$, then $j^r\gamma^*\lambda$ is an *n*-form on *X*, for any such $\gamma$. To each compact subset $c \subset X$ we can consider the integral

$$\lambda_c(\gamma) = \int_c j^r\gamma^*\lambda.$$

The mapping (real function)

(6.15) $\quad \Gamma(\pi) \ni \gamma \to \lambda_c(\gamma) \in \mathbf{R}$

can be studied in the same manner as in the previous case (6.1), for each compact subset $c \subset X$. Let us proceed to the definitions.

Let $\Xi$ be a projectable vector field on $Y$ and $\alpha_t^{\Xi}$ its local one-parameter group. To each cross section $\gamma \in \Gamma(\pi)$ and sufficiently small $t$, the group induces "*local variations*" $\alpha_t^{\Xi} \gamma \alpha_{-t}^{\xi}$ of the cross section, that are, in fact, local cross sections of $(Y, \pi, X)$. In this expression $\xi$ is defined by the equality $T\pi \circ \Xi = \xi \circ \pi$. Given $\gamma \in \Gamma(\pi)$ and $c \subset X$ we obtain the mapping

$$(-\epsilon, \epsilon) \ni t \to \lambda_{\alpha_t^{\xi}(c)}(\alpha_t^{\Xi} \gamma \alpha_{-t}^{\xi}) = \int_{\alpha_t^{\xi}(c)} (j^r \alpha_t^{\Xi} \gamma \alpha_{-t}^{\xi})^* \lambda \in \mathbf{R},$$

defined for all sufficiently small $t$. Notice that although the family $\alpha_t^{\Xi} \gamma \alpha_{-t}^{\xi}$ consists of local cross sections, the expression on the right is well defined. A *critical point*, or an *extremal*, of the variational problem is then each cross section $\gamma \in \Gamma(\pi)$ satisfying the following condition: The equality

$$\left\{ \frac{d}{dt} \int_{\alpha_t^{\xi}(c)} (j^r \alpha_t^{\Xi} \gamma \alpha_{-t}^{\xi})^* \lambda \right\}_0 = 0$$

takes place for each compact $c \subset X$ and for all c-admissible vector fields $\Xi$ on Y.

According to the transformation rule for integrals [36] we can write

$$\left\{ \frac{d}{dt} \int_{\alpha_t^{\xi}(c)} (j^r \alpha_t^{\Xi} \gamma \alpha_{-t}^{\xi})^* \lambda \right\}_0 = \left\{ \frac{d}{dt} \int_c \alpha_t^{\xi *} (j^r \alpha_t^{\Xi} \gamma \alpha_{-t}^{\xi})^* \lambda \right\}_0$$

$$= \left\{ \frac{d}{dt} \int_c (j^r \alpha_t^{\Xi} \gamma \alpha_{-t}^{\xi} \circ \alpha_t^{\xi})^* \lambda \right\}_0 .$$

Remember that the definition of the local one-parameter group generating the $r$-jet prolongation of $\alpha_t^{\Xi}$ reads

$$\alpha_t^{j^r \Xi}(j_x^r \gamma) = (j^r \alpha_t^{\Xi} \gamma \alpha_{-t}^{\xi} \circ \alpha_t^{\xi})(x).$$

Thus, the following assertion allows to apply the theory of the Lie derivatives to the variational problems:

*A cross section $\gamma \in \Gamma(\pi)$ is a critical point of the variational problem if and only if for any compact domain $c \subset X$ the condition*

$$\text{(6.16)} \qquad \int_c j^r \gamma^* \vartheta(j^r \Xi) \lambda = 0$$

*holds for all c-admissible vector fields $\Xi$ on Y.*

(Compare with [35], [20].)

In this Section we derive a formula for the Lie derivative of a Lepagian form, defined on the 1-jet prolongation of a given fibered manifold. The formula will be re-lated to both types of variational problems we have discussed above (see (6.5) and (6.16)), and is called the *first variation formula*. We start, however, with some local considerations.

**First variation formula: Local considerations.** It is known how can one obtain the first variation formula by means of some coordinate calculation, in principle, for any r, i.e. for Lagrangian forms of any degree (compare with [8], [11]). We give the result for horizontal Lagrangian forms (or, which is the same, for Lagrange functions) of degree 2.

Let $\lambda \in \Omega_X^n(J^2)$; in the coordinates (2.4), let

$$\lambda = \mathscr{L} dx_1 \wedge dx_2 \wedge \ldots \wedge dx_n.$$

The one can verify that for any projectable vector field $\Xi$ on $Y$ expressed as in (4.18) by the equality

$$j^2\Xi = \xi_k \frac{\partial}{\partial x_k} + \Xi_\mu \frac{\partial}{\partial y_\mu} + \Xi_{i\mu} \frac{\partial}{\partial z_{i\mu}} + \sum_{i \le j} \Xi_{ij\mu} \frac{\partial}{\partial z_{ij\mu}},$$

we get

$$\begin{aligned} \pi_{4,2}^* \vartheta(j^2\Xi)\lambda = & \left( \left( \frac{\partial \mathscr{L}}{\partial y_\mu} - \frac{d}{dx_k}\left( \frac{\partial \mathscr{L}}{\partial z_{k\mu}} \right) \right) - \sum_{i \le j} \frac{d}{dx_j}\left( \frac{d}{dx_k}\left( \frac{\partial \mathscr{L}}{\partial z_{jk\mu}} \right) \right) \right) \\ & \cdot (\Xi_\mu - z_{l\mu}\xi_l) + \sum_k \frac{d}{dx_k}\left( \mathscr{L}\xi_k + \frac{\partial \mathscr{L}}{\partial z_{k\mu}} \cdot (\Xi_\mu - z_{i\mu}\xi_i) + \sum_{j \le k} \frac{\partial \mathscr{L}}{\partial z_{kj\mu}} \right. \\ & \left. \cdot (\Xi_{j\mu} - z_{jl\mu}\xi_l) - \sum_{j \le k} \frac{d}{dx_j}\left( \frac{\partial \mathscr{L}}{\partial z_{jk\mu}} \right) \cdot (\Xi_\mu - z_{l\mu}\xi_l) \right) dx_1 \wedge dx_2 \wedge \ldots \wedge dx_n. \end{aligned} \tag{6.17}$$

If we want to obtain a particular case when $\lambda \in \Omega_X^n(J^1)$, we simply substitute $\partial \mathscr{L} / \partial z_{ik\mu} = 0$:

$$\begin{aligned} \pi_{2,1}^* \vartheta(j^1\Xi)\lambda = & \left( \left( \frac{\partial \mathscr{L}}{\partial y_\mu} - \frac{d}{dx_k}\left( \frac{\partial \mathscr{L}}{\partial z_{k\mu}} \right) \right) (\Xi_\mu - z_{l\mu}\xi_l) \right. \\ & \left. + \frac{d}{dx_k}\left( \mathscr{L}\xi_k + \frac{\partial \mathscr{L}}{\partial z_{k\mu}} \cdot (\Xi_\mu - z_{l\mu}\xi_l) \right) \right) dx_1 \wedge dx_2 \wedge \ldots \wedge dx_n. \end{aligned} \tag{6.18}$$

(6.17) and (6.18) are usually called the *first variation formulas*.

**Local first variation formula: Some special variations.** An important modification of the first variation formulas arises when the vector field $\Xi$ on $Y$ is induced by a vector field, $\xi$, defined on the base manifold $X$. As an example we can take any tensor bundle $(T_\sigma X, \tau_\sigma, X)$ on $X$ and consider the vector field $\xi_\sigma$ introduced in Section 3, (3.22). Accordingly, put in (6.17)

$$\Xi_\mu = \sigma_{\mu p}^{\nu q} \cdot \frac{\partial \xi_p}{\partial x_q} \cdot y_\nu, \tag{6.19}$$

(6.20) $$E_\mu = \frac{\partial \mathcal{L}}{\partial y_\mu} - \frac{d}{dx_k}\left(\frac{\partial \mathcal{L}}{\partial z_{k\mu}}\right) + \sum_{i \le j} \frac{d}{dx_j}\left(\frac{d}{dx_k}\left(\frac{\partial \mathcal{L}}{\partial z_{jk\mu}}\right)\right),$$

and slightly modify the first term in (6.17) “pulling it back” on $J^5$. We get

$$\xi_\mu(\Xi_\mu - z_{l\mu}\xi_l) = \frac{d}{dx_q}(\xi_\mu \cdot \sigma^{\nu q}_{\mu p} \cdot y_\nu \cdot \xi_p) - \left(\xi_\mu z_{l\mu} + \frac{d}{dx_q}(\xi_\mu \sigma^{\nu q}_{\mu p} y_\nu)\right)\xi_l$$

(compare with [34], 5.33). Thus, in this special case the first variation formula (6.17) takes the form

(6.21) $$\begin{aligned} \pi^*_{5,2}\vartheta(j^2\xi_\sigma)\lambda = \Bigg( &-\left(E_\mu \cdot z_{l\mu} + \frac{d}{dx_q}(E_\mu \cdot \sigma^{\nu q}_{\mu p} \cdot y_\nu)\right) \cdot \xi_l \\ &+ \sum_k \frac{d}{dx_k}\Big(E_\mu \sigma^{\nu q}_{\mu p} y_\nu \xi_p + \mathcal{L}\xi_k + \frac{\partial \mathcal{L}}{\partial z_{k\mu}} \cdot (\Xi_\mu - z_{i\mu}\xi_i) \\ &+ \sum_{j \le k} \frac{\partial \mathcal{L}}{\partial z_{kj\mu}}(\Xi_{j\mu} - z_{jl\mu}\xi_l) \\ &- \sum_{j \le k} \frac{d}{dx_j}\left(\frac{\partial \mathcal{L}}{\partial z_{jk\mu}}\right) \cdot (\Xi_\mu - z_{l\mu}\xi_l)\Big)\Bigg) dx_1 \wedge dx_2 \wedge \ldots \wedge dx_n \end{aligned}$$

with (6.19), (6.20), (3.12) and (3.14) in view.

Coming out from (6.18) we state the following definition: A cross section $\gamma \in \Gamma(\tau_\sigma)$ is said to be a *weak critical point* of the variational problem defined by $(T_\sigma X, \tau_\sigma, X)$ and $\lambda$, if

$$\int_c j^1\gamma^* \vartheta(j^1\xi_\sigma)\lambda = 0$$

holds for all compact $c \subset X$ and for all vector fields $\xi$ on X satisfying the condition $supp\xi \subset c$.

It is immediately clear that $\gamma \in \Gamma(\tau_\sigma)$ is a weak critical point of the variational problem if and only if it satisfies the *n* equations

$$E_\mu z_{l\mu} + \frac{d}{dx_q}(E_\mu \sigma^{\nu q}_{\mu p} y_\nu) = 0.$$

(Here, as usual, $d/dx_q$ stands for the formal derivative defined in Section 3, (3.14).)

We shall return to the obtained formulas in Section 7 in connection with the so called *generally covariant* variational problems.

**First variational formula.** Let $\rho$ be an arbitrary *n*-form on $J^1$, and suppose that $\rho \in \Omega^n_Y(J^1)$. Let $\Xi$ be a projectable vector field on *Y*. In this paragraph we derive a formula for the Lie derivative $\vartheta(j^1\Xi)h(\rho)$. According to certain relations between forms and vector fields taking place on the infinite jet prolongation $J^\infty$ of our fibered manifold $(Y, \pi, X)$ this formula may be written on $J^\infty$ as directly corresponding with the first variation formula (6.18) in the coordinate form.

According to the relations (4.24), (4.9),

$$
\begin{aligned}
&\pi^*_{\infty,1}\vartheta(j^1\Xi)h(\rho) = \pi^*_{\infty,1}h(\vartheta(j^1\Xi)\rho)\\
(6.22)\qquad &= \pi^*_{\infty,1}h(i(j^1\Xi)d\rho) + \pi^*_{\infty,1}h(di(j^1\Xi)\rho)\\
&= i(v(j^\infty\Xi))\pi^*_{\infty,2}\tilde{h}(d\rho) + \pi^*_{\infty,2}h(di(j^1\Xi)\rho).
\end{aligned}
$$

In fact, this is the well-known formula for the Lie derivative given in Section 1, modified to our situation on a fibered manifold. We note that for vertical vector fields $\Xi$, when $j^\infty\Xi = v(j^\infty\Xi)$, this formula becomes

$$(6.23)\qquad \pi^*_{2,1}\vartheta(j^1\Xi)h(\rho) = i(j^2\Xi)\tilde{h}(d\rho) + h(di(j^1\Xi)\rho).$$

Referring to the note in the second paragraph of this Section, we shall prove two propositions, characterizing the decomposition (6.23) of the Lie derivative $\vartheta(j^1\Xi)h(\rho)$ into the two terms.

*If for all vertical vector fields* $\Xi$ *on Y*

$$(6.24)\qquad di(j^2\Xi)\tilde{h}(d\rho) = 0,$$

then $\tilde{h}(d\rho) = 0$. In other words, the term $i(j^2\Xi)\tilde{h}(d\rho)$ is a closed form for all vertical vector fields $\Xi$ if and only if it identically vanishes.

To prove it, write

$$\omega_0 = dx_1 \wedge dx_2 \wedge \ldots \wedge dx_n,$$

$$j^2\Xi = \Xi_\mu \frac{\partial}{\partial y_\mu} + \Xi_{i\mu} \frac{\partial}{\partial z_{i\mu}} + \sum_{i\le j} \Xi_{ij\mu} \frac{\partial}{\partial z_{ij\mu}}$$

in the coordinates (2.4) on $J^2$. The form $\tilde{h}(d\rho)$ can be written as

$$\tilde{h}(d\rho) = (g_\mu dy_\mu + g_{i\mu} dz_{i\mu}) \wedge \omega_0$$

(5.6), and we get

$$i(j^2\Xi)\tilde{h}(d\rho) = (g_\mu \Xi_\mu + g_{i\mu}\Xi_{i\mu})\omega_0.$$

Differentiating

$$
\begin{aligned}
di(j^2\Xi)\tilde{h}(d\rho) = &\left(\frac{\partial g_\mu}{\partial y_\sigma}\Xi_\mu + \frac{\partial \Xi_\mu}{\partial y_\sigma} g_\mu + \frac{\partial g_{i\mu}}{\partial y_\sigma}\Xi_{i\mu} + \frac{\partial \Xi_{i\mu}}{\partial y_\sigma} g_{i\mu}\right)\\
&\cdot dy_\sigma \wedge \omega_0 + \left(\frac{\partial g_\mu}{\partial z_{k\sigma}}\Xi_\mu + \frac{\partial \Xi_{i\mu}}{\partial z_{k\sigma}} g_{i\mu} + \frac{\partial g_{i\mu}}{\partial z_{k\sigma}}\Xi_{i\mu}\right) dz_{k\sigma} \wedge \omega_0.
\end{aligned}
$$

Since the derivatives of $\Xi_\mu$ are independent, we have

$$g_\mu = 0, \quad g_{i\mu} = 0,$$

proving the proposition.

The next proposition concerns with the uniqueness of the decomposition (6.23).

*Let $\rho$ be a Lepagian form on $J^1$. Then there exist a unique $(n+1)$-form $\eta$ on $J^2$ and an n-form $\lambda$ on $J^1$ satisfying the following conditions:*

1) *For all vertical vector fields $\Xi$ on Y,*

$$\pi_{2,1}^{*}\vartheta(j^1\Xi)h(\rho)=i(j^2\Xi)\eta+h(di(j^1\Xi)\lambda).$$

2) *$\eta$ is horizontal with respect to $\pi_{2,0}$.*

3) *$\lambda$ is horizontal with respect to $\pi_{1,0}$.*

To prove it, suppose that there are two pairs, $(\eta_1,\lambda_1)$, and $(\eta_2,\lambda_2)$, of forms, satisfying 1) and 2) and 3). Put

$$\eta_0=\eta_1-\eta_2,\quad \lambda_0=\lambda_1-\lambda_2.$$

The form $\eta_0$, is horizontal with respect to $\pi_{2,0}$, $\lambda_0$ is horizontal with respect to $\pi_{1,0}$, and the equality

$$i(j^2\Xi)\eta_0+h(di(j^1\Xi)\lambda_0)=0$$

holds for all vertical vector fields $\Xi$ on $Y$. By (6.23) we have for an arbitrary vertical vector field $\Xi$

$$\vartheta(j^2\Xi)h(\lambda_0)-i(j^2\Xi)(h(d\lambda_0)-\eta_0)=0. \tag{6.25}$$

Write

$$h(\lambda_0)=\mathscr{L}dx_1\wedge dx_2\wedge\ldots\wedge dx_n,$$

and, according to (5.6) and the definition of $B_{i\mu}$

$$\tilde{h}(d\lambda_0)=\left(\left(\frac{\partial\mathscr{L}}{\partial y_\mu}-\frac{dB_{i\mu}}{dx_i}\right)dy_\mu+\left(\frac{\partial\mathscr{L}}{\partial z_{i\mu}}-B_{i\mu}\right)dz_{i\mu}\right)\wedge dx_1\wedge dx_2\wedge\ldots\wedge dx_n.$$

Since $\Xi$ is vertical, $\vartheta(j^2\Xi)h(\lambda_0)$ is expressed as

$$\vartheta(j^2\Xi)(\lambda_0)=\left(\frac{\partial\mathscr{L}}{\partial y_\mu}\Xi_\mu+\frac{\partial\mathscr{L}}{\partial z_{i\mu}}\Xi_{i\mu}\right)dx_1\wedge dx_2\wedge\ldots\wedge dx_n;$$

hence, by (6.25),

$$B_{i\mu}=0,$$

and

$$\tilde{h}(d\lambda_0)=dh(\lambda_0).$$

Thus, (6.25) becomes

$$i(j^2\Xi)(\eta_0)=0.$$

By our first assumption, $\eta_0$ must be of the form $\nu\wedge\omega_0$ and, by 2), $\nu=f_\mu dy_\mu$. It follows that $f_\mu=0$ and

$$\eta_0 = \eta_1 - \eta_2 = 0.$$

The existence of the forms $\eta$ and $\lambda$ is proved by (6.23), and we are done.

In accordance with our previous terminology, the expression (6.22) (and, as a special case, (6.23)) is called the *first variation formula* for variational problems defined by *Lepagian* forms $\rho$.

Notice that we have, in fact, developed an appropriate variational theory when starting with a *Lepagian* form on the 1-jet prolongation of the given fibered manifold. Given one such form, we can formulate the variational problems as before, and obtain in a completely invariant way conditions characterizing the critical points of such variational problems. It is necessary, however, to complete the theory by showing how to obtain a Lepagian form from a given horizontal Lagrangian form in such a way that both forms define the same variational problem. We shall be busy with this question in a few next paragraphs.

With these considerations in mind, we define: Each of the two variational problems defined in this work is said to be *canonical*, if it is determined by means of a Lepagian form.

**Lepagian forms associated with a given horizontal Lagrangian form.** In order to complete our considerations concerning the first variation formula, we should give a general rule how to obtain a Lepagian form defining the same variational problem as the given horizontal Lagrangian form. However, it is sufficiently known [4], [27], [35] that this cannot be done in a unique way. In fact, one can put some other assumptions on the Lepagian form, as we noted at the end of Section 4 (see (4.22)). The Lepagian form $\Theta$ given there has been obtained as a very simple, from the analytical point of view, solution of our problem: the higher-order terms in the exterior differentials, $dy_\mu - z_{i\mu} dx_i$, $dz_{i\mu} - z_{ki\mu} dx_k$, appear "linearly" and not "multilinearly".

Consider the general problem. Let $\lambda$ be an arbitrary horizontal Lagrangian form of degree 1 on $(Y, \pi, X)$. The problem is, whether there exists an $n$-form $\rho \in \Omega^n_Y(J^1)$ such that

1) $h(\rho) = \lambda$,
2) $\tilde{h}(d\rho)$ is horizontal with respect to $\pi_{2,0}$ (i.e. $\rho$ is a Lepagian form).

Each $\rho$ satisfying these two conditions, is called a *Lepagian equivalent* of $\lambda$.

It is not so difficult to find a coordinate condition for such a form $\rho$. Define $G$, $A_{i\mu}$, $B_{i\mu}$ in the same way as in Section 5 (see (5.4), (5.5), etc.) If there is $\rho$ satisfying 1) and 2), then 1) means that $G = \mathscr{L}$ (where we write, as usual, $\lambda = \mathscr{L} dx_1 \wedge dx_2 \wedge \ldots \wedge dx_n$), and 2) means that

$$A_{i\mu} = \frac{\partial G}{\partial z_{i\mu}} - B_{i\mu} = 0,$$

i.e.

$$\frac{\partial \mathscr{L}}{\partial z_{i\mu}} = \sum_{r=1}^{n} \sum_{s_1 < s_2 < \ldots < s_r} \sum_{\sigma_1, \sigma_2, \ldots, \sigma_r} g^{s_1 s_2 \ldots s_r}_{\sigma_1 \sigma_2 \ldots \sigma_r} \frac{\partial}{\partial z_{i\mu}} (z_{s_1 \sigma_1} z_{s_2 \sigma_2} \ldots z_{s_r \sigma_r}).$$

This condition can be fulfilled in various ways. For illustration, if we are looking for the form $\rho$ as in the mentioned example of Section 4 but not depending on $z_{ij\mu}$,

$$\rho = g_0 dx_1 \wedge dx_2 \wedge \ldots \wedge dx_n$$
$$+ \sum_{s,\sigma} g_\sigma^s \, dx_1 \wedge dx_2 \wedge \ldots \wedge dx_{s-1} \wedge dy_\sigma \wedge dx_{s+1} \wedge \ldots \wedge dx_n,$$
$$g_{\sigma_1 \sigma_2 \ldots \sigma_r}^{s_1 s_2 \ldots s_r} = 0, \quad r > 1,$$

this condition gives

$$g_\sigma^s = \frac{\partial \mathscr{L}}{\partial z_{s\sigma}},$$

and together with 1), $G = \mathscr{L}$,

$$g_0 = \mathscr{L} - \frac{\partial \mathscr{L}}{\partial z_{s\sigma}} z_{s\sigma}.$$

The *n*-form $\rho$ satisfies conditions 1) and 2). Its invariance with respect to changes of coordinates can be directly proved. In accordance with Section 3 we write

$$\rho = \Theta(\lambda).$$

The mapping $\lambda \to \Theta(\lambda)$ presents an example of the desired rule.

Let us consider an arbitrary *n*-form $\rho$ on $J^0 = Y$. Clearly, the form is Lepagian. Let $\pi_{1,0}^* \rho$ be expressed in coordinates as

$$\pi_{1,0}^* \rho = g_0 dx_1 \wedge dx_2 \wedge \ldots \wedge dx_n + \sum_{r=1}^{n} \sum_{s_1 < s_2 < \ldots < s_r} \sum_{\sigma_1, \sigma_2, \ldots, \sigma_r} \frac{1}{r!} g_{\sigma_1 \sigma_2 \ldots \sigma_r}^{s_1 s_2 \ldots s_r}$$
$$\cdot dx_1 \wedge dx_2 \wedge \ldots \wedge dy_{\sigma_1} \wedge \ldots \wedge dy_{\sigma_r} \wedge \ldots \wedge dx_n;$$

then $h(\rho) = \mathscr{L} \omega_0$ and

$$\mathscr{L} = g_0 + \sum_{r=1}^{n} \sum_{s_1 < s_2 < \ldots < s_r} \sum_{\sigma_1, \sigma_2, \ldots, \sigma_r} g_{\sigma_1 \sigma_2 \ldots \sigma_r}^{s_1 s_2 \ldots s_r} z_{s_1 \sigma_1} z_{s_2 \sigma_2} \ldots z_{s_r \sigma_r}$$

with $g_0$, $g_{\sigma_1 \sigma_2 \ldots \sigma_r}^{s_1 s_2 \ldots s_r}$ not depending on $z_{i\mu}$. We can thus determine the functions $g_0$, $g_{\sigma_1 \sigma_2 \ldots \sigma_r}^{s_1 s_2 \ldots s_r}$ in terms of derivatives of $\mathscr{L}$ with respect to $z_{i\mu}$. One immediately gets

$$g_{\nu_1 \nu_2 \ldots \nu_n}^{12 \ldots n} = \frac{\partial^n \mathscr{L}}{\partial z_{1\nu_1} z_{2\nu_2} \ldots z_{n\nu_n}},$$

$$g_{\nu_1 \nu_2 \ldots \nu_{n-1}}^{s_1 s_2 \ldots s_{n-1}} = \frac{\partial^{n-1} \mathscr{L}}{\partial z_{s_1 \nu_1} z_{s_2 \nu_2} \ldots z_{s_{n-1} \nu_{n-1}}}$$
$$- g_{\sigma_1 \sigma_2 \ldots \sigma_n}^{12 \ldots n} \frac{\partial^{n-1}}{\partial z_{s_1 \nu_1} z_{s_2 \nu_2} \ldots z_{s_{n-1} \nu_{n-1}}} (z_{1\sigma_1} z_{2\sigma_2} \ldots z_{n\sigma_n}),$$

$$\ldots$$

$$g^{s_1 s_2 \ldots s_p}_{\nu_1 \nu_2 \ldots \nu_p} = \frac{\partial^p \mathscr{L}}{\partial z_{s_1 \nu_1} z_{s_2 \nu_2} \ldots z_{s_p \nu_p}} - \sum_{j=p+1}^{n} \sum_{k_1 < k_2 < \ldots < k_j} \sum_{\sigma_1, \sigma_2, \ldots, \sigma_j} g^{k_1 k_2 \ldots k_j}_{\sigma_1 \sigma_2 \ldots \sigma_j}$$

$$\cdot \frac{\partial^p}{\partial z_{s_1 \nu_1} z_{s_2 \nu_2} \ldots z_{s_p \nu_p}} (z_{k_1 \sigma_1} z_{k_2 \sigma_2} \ldots z_{k_j \sigma_j}),$$

$\ldots$

$$g_0 = \mathscr{L} - \sum_{r=1}^{n} \sum_{s_1 < s_2 < \ldots < s_r} \sum_{\sigma_1, \sigma_2, \ldots, \sigma_r} g^{s_1 s_2 \ldots s_r}_{\sigma_1 \sigma_2 \ldots \sigma_r} z_{s_1 \sigma_1} z_{s_2 \sigma_2} \ldots z_{s_r \sigma_r}.$$

Turning our point of view we can take these identities after some necessary changes, as definitions of the local expression (5.1) of an $n$-form $\rho$ when the horizontal form $\lambda = \mathscr{L} dx_1 \wedge dx_2 \wedge \ldots \wedge dx_n$ is given. Thus, having $\lambda$, we set

$$\Delta(\lambda) = f_0 dx_1 \wedge dx_2 \wedge \ldots \wedge dx_n + \sum_{r=1}^{n} \sum_{s_1 < s_2 < \ldots < s_r} \sum_{\sigma_1, \sigma_2, \ldots, \sigma_r} \frac{1}{r!} f^{s_1 s_2 \ldots s_r}_{\sigma_1 \sigma_2 \ldots \sigma_r} \tag{6.27}$$

$$\cdot dx_1 \wedge dx_2 \wedge \ldots \wedge dy_{\sigma_1} \wedge \ldots \wedge dy_{\sigma_r} \wedge \ldots \wedge dx_n,$$

where

$$f^{12\ldots n}_{\sigma_1 \sigma_2 \ldots \sigma_n} = \frac{1}{n!} \epsilon^{12\ldots n}_{k_1 k_2 \ldots k_n} \frac{\partial^n \mathscr{L}}{\partial z_{k_1 \sigma_1} z_{k_2 \sigma_2} \ldots z_{k_n \sigma_n}},$$

$$f^{s_1 s_2 \ldots s_{n-1}}_{\sigma_1 \sigma_2 \ldots \sigma_{n-1}} = \frac{1}{(n-1)!} \epsilon^{s_1 s_2 \ldots s_{n-1}}_{k_1 k_2 \ldots k_{n-1}} \cdot \Bigg( \frac{\partial^{n-1} \mathscr{L}}{\partial z_{k_1 \sigma_1} z_{k_2 \sigma_2} \ldots z_{k_{n-1} \sigma_{n-1}}} - f^{12\ldots n}_{\nu_1 \nu_2 \ldots \nu_n}$$

$$\cdot \frac{\partial^{n-1}}{\partial z_{k_1 \sigma_1} z_{k_2 \sigma_2} \ldots z_{k_{n-1} \sigma_{n-1}}} (z_{1 \nu_1} z_{2 \nu_2} \ldots z_{n \nu_n}),$$

$\ldots$

$$f^{s_1 s_2 \ldots s_p}_{\sigma_1 \sigma_2 \ldots \sigma_p} = \frac{1}{p!} \epsilon^{s_1 s_2 \ldots s_p}_{k_1 k_2 \ldots k_p} \Bigg( \frac{\partial^p \mathscr{L}}{\partial z_{k_1 \sigma_1} z_{k_2 \sigma_2} \ldots z_{k_p \sigma_p}}$$

$$- \sum_{j=p+1}^{n} \sum_{l_1 < l_2 < \ldots < l_j} \sum_{\nu_1, \nu_2, \ldots, \nu_j} f^{l_1 l_2 \ldots l_j}_{\nu_1 \nu_2 \ldots \nu_j} \frac{\partial^j}{\partial z_{k_1 \sigma_1} z_{k_2 \sigma_2} \ldots z_{k_p \sigma_p}} (z_{l_1 \nu_1} z_{l_2 \nu_2} \ldots z_{l_j \nu_j}),$$

$\ldots$

$$f_0 = \mathscr{L} - \sum_{r=1}^{n} \sum_{s_1 < s_2 < \ldots < s_r} \sum_{\sigma_1, \sigma_2, \ldots, \sigma_r} f^{s_1 s_2 \ldots s_r}_{\sigma_1 \sigma_2 \ldots \sigma_r} z_{s_1 \sigma_1} z_{s_2 \sigma_2} \ldots z_{s_r \sigma_r}.$$

A question arises whether the $n$-form $\Delta(\lambda)$ is independent of a particular choice of coordinates. Leaving the question unnoticed we just point out that the functions $f^{12\ldots n}_{\sigma_1 \sigma_2 \ldots \sigma_n}$, being the coefficients at $dy_{\sigma_1} \wedge dy_{\sigma_2} \wedge \ldots \wedge dy_{\sigma_n}$ satisfy the rule

$$f^{12\ldots n}_{\sigma_1 \sigma_2 \ldots \sigma_n} = \bar{f}^{12\ldots n}_{\lambda_1 \lambda_2 \ldots \lambda_n} \frac{\partial \bar{y}_{\lambda_1}}{\partial y_{\sigma_1}} \frac{\partial \bar{y}_{\lambda_2}}{\partial y_{\sigma_2}} \ldots \frac{\partial \bar{y}_{\lambda_n}}{\partial y_{\sigma_n}}.$$

The rule expresses the coefficients in a new canonical chart associated with a fiber chart $(\bar{U},\bar{\psi})$ on $Y$. As for our functions

$$f^{12\dots n}_{\sigma_1\sigma_2\dots\sigma_n} = \frac{1}{n!}\epsilon^{12\dots n}_{k_1k_2\dots k_n}\frac{\partial^n \mathscr{L}}{\partial z_{k_1\sigma_1} z_{k_2\sigma_2}\dots z_{k_n\sigma_n}},$$

according to the general transformation rules

$$\frac{\partial \bar{z}_{k\rho}}{\partial z_{i\mu}} = \frac{\partial \bar{y}_\rho}{\partial y_\mu}\cdot\frac{\partial x_i}{\partial \bar{x}_k}, \quad \mathscr{L} = \bar{\mathscr{L}}\cdot \det\left(\frac{\partial \bar{x}}{\partial x}\right),$$

where

$$\det\left(\frac{\partial \bar{x}}{\partial x}\right) = \epsilon^{12\dots n}_{k_1k_2\dots k_n}\frac{\partial \bar{x}_1}{\partial x_{k_1}}\frac{\partial \bar{x}_2}{\partial x_{k_2}}\dots\frac{\partial \bar{x}_n}{\partial x_{k_n}},$$

we obtain

$$\frac{\partial^n \mathscr{L}}{\partial z_{1\nu_1} z_{2\nu_2}\dots z_{n\nu_n}}$$
$$= \det\left(\frac{\partial \bar{x}}{\partial x}\right)\frac{\partial^n \bar{\mathscr{L}}}{\partial \bar{z}_{k_1\rho_1}\bar{z}_{k_2\rho_2}\dots\bar{z}_{k_n\rho_n}}\frac{\partial \bar{y}_{\rho_1}}{\partial y_{\nu_1}}\frac{\partial \bar{y}_{\rho_2}}{\partial y_{\nu_2}}\dots\frac{\partial \bar{y}_{\rho_n}}{\partial y_{\nu_n}}\cdot\frac{\partial x_1}{\partial \bar{x}_{k_1}}\frac{\partial x_2}{\partial \bar{x}_{k_2}}\dots\frac{\partial x_n}{\partial \bar{x}_{k_n}},$$

and

$$f^{12\dots n}_{\sigma_1\sigma_2\dots\sigma_n} = \frac{1}{n!}\epsilon^{\sigma_1\sigma_2\dots\sigma_n}_{\nu_1\nu_2\dots\nu_n}\frac{\partial^n \mathscr{L}}{\partial z_{1\nu_1} z_{2\nu_2}\dots z_{n\nu_n}}$$
$$= \frac{1}{n!}\epsilon^{\sigma_1\sigma_2\dots\sigma_n}_{\nu_1\nu_2\dots\nu_n}\cdot\frac{\partial \bar{y}_{\rho_1}}{\partial y_{\nu_1}}\cdot\frac{\partial \bar{y}_{\rho_2}}{\partial y_{\nu_2}}\dots\frac{\partial \bar{y}_{\rho_n}}{\partial y_{\nu_n}}\epsilon^{12\dots n}_{k_1k_2\dots k_n}$$
$$\cdot\frac{\partial^n \bar{\mathscr{L}}}{\partial \bar{z}_{1\rho_1}\bar{z}_{2\rho_2}\dots\bar{z}_{n\rho_n}}\det\left(\frac{\partial \bar{x}}{\partial x}\right)\frac{\partial x_1}{\partial \bar{x}_{k_1}}\frac{\partial x_2}{\partial \bar{x}_{k_2}}\dots\frac{\partial x_n}{\partial \bar{x}_{k_n}}$$
$$= \bar{f}^{12\dots n}_{\rho_1\rho_2\dots\rho_n}\frac{\partial \bar{y}_{\rho_1}}{\partial y_{\sigma_1}}\frac{\partial \bar{y}_{\rho_2}}{\partial y_{\sigma_2}}\dots\frac{\partial \bar{y}_{\rho_n}}{\partial y_{\sigma_n}}.$$

We thus can hope that the functions $f_0$ and $f^{s_1s_2\dots s_r}_{\sigma_1\sigma_2\dots\sigma_r}$ being derived from the well-defined functions $f^{s_1s_2\dots s_n}_{\sigma_1\sigma_2\dots\sigma_n}$, satisfy the desired transformation rules. Thus, our next considerations are of local character.

Properties of $\Delta(\lambda)$ we are interested in, are collected in the following proposition:

*The mapping*

$$\Omega^n_X(J^1)\ni\lambda\to\Delta(\lambda)\in\Omega^n_Y(J^1)$$

*has the following properties:*

1) *For any* $\lambda$, $h(\Delta(\lambda))=\lambda$.
2) *For any* $\lambda\in\Omega^n_X(J^1)$, *the n-form* $\Delta(\lambda)$ *is Lepagian.*

3) *If* $\rho = \pi_{1,0}^{*}\rho_0$ *for some n-form* $\rho_0$ *on Y, then* $\Delta(h(\rho)) = \rho$.

It follows from the definition of $\Delta(\lambda)$ that 1) is true. To prove 2), we must determine the expression

$$A_{k\nu} = \frac{\partial f_0}{\partial z_{k\nu}} + \sum_{r=1}^{n} \sum_{s_1<s_2<\ldots<s_r} \sum_{\sigma_1,\sigma_2,\ldots,\sigma_r} \frac{\partial f_{\sigma_1\sigma_2\ldots\sigma_r}^{s_1s_2\ldots s_r}}{\partial z_{k\nu}} z_{s_1\sigma_1} z_{s_2\sigma_2} \ldots z_{s_r\sigma_r}.$$

By the definition of $f_0$,

$$\frac{\partial f_0}{\partial z_{k\nu}} = \frac{\partial \mathscr{L}}{\partial z_{k\nu}} - \sum_{r=1}^{n} \sum_{s_1<s_2<\ldots<s_r} \sum_{\sigma_1,\sigma_2,\ldots,\sigma_r} \left( \frac{\partial f_{\sigma_1\sigma_2\ldots\sigma_r}^{s_1s_2\ldots s_r}}{\partial z_{k\nu}} z_{s_1\sigma_1} z_{s_2\sigma_2} \ldots z_{s_r\sigma_r} \right.$$

$$\left. + f_{\sigma_1\sigma_2\ldots\sigma_r}^{s_1s_2\ldots s_r} \frac{\partial}{\partial z_{k\nu}} (z_{s_1\sigma_1} z_{s_2\sigma_2} \ldots z_{s_r\sigma_r}) \right)$$

$$= \frac{\partial \mathscr{L}}{\partial z_{k\nu}} - f_\nu^k - \sum_{r=2}^{n} \sum_{s_1<s_2<\ldots<s_r} \sum_{\sigma_1,\sigma_2,\ldots,\sigma_r} \frac{\partial}{\partial z_{k\nu}} (z_{s_1\sigma_1} z_{s_2\sigma_2} \ldots z_{s_r\sigma_r}) \cdot f_{\sigma_1\sigma_2\ldots\sigma_r}^{s_1s_2\ldots s_r}$$

$$- \sum_{r=1}^{n} \sum_{s_1<s_2<\ldots<s_r} \sum_{\sigma_1,\sigma_2,\ldots,\sigma_r} \frac{\partial f_{\sigma_1\sigma_2\ldots\sigma_r}^{s_1s_2\ldots s_r}}{\partial z_{k\nu}} z_{s_1\sigma_1} z_{s_2\sigma_2} \ldots z_{s_r\sigma_r},$$

and from the definition of $f_\nu^k$,

$$f_\nu^k = \frac{\partial \mathscr{L}}{\partial z_{k\nu}} - \sum_{r=2}^{n} \sum_{k_1<k_2<\ldots<k_r} \sum_{\sigma_1,\sigma_2,\ldots,\sigma_r} f_{\sigma_1\sigma_2\ldots\sigma_r}^{k_1k_2\ldots k_r} \frac{\partial}{\partial z_{k\nu}} (z_{s_1\sigma_1} z_{s_2\sigma_2} \ldots z_{s_r\sigma_r}).$$

Thus $A_{k\nu} = 0$ and $\Delta(\lambda)$ is Lepagian. 3) follows from the definition, motivated, in fact, by 3) as an assumption.

**Critical points of canonical variational problems.** The canonical variational theory is based on the assumption that the basic Lagrangian form is Lepagian. If it is not, one can find a Lepagian equivalent to the given form and build up all the theory with the Lepagian equivalent. Certainly, if we consider a fibered manifold $(Y, \pi, X)$ and a Lagrangian form $\lambda \in \Omega_Y^n(J^1)$ then choosing a compact domain $c \subset X$ we have for all $\gamma \in \Gamma(\pi)$ and any Lepagian equivalent $\rho$

$$\int_c j^1\gamma^*\lambda = \int_c j^1\gamma^* h(\lambda) = \int_c j^1\gamma^* h(\rho) = \int_c j^1\gamma^*\rho.$$

Thus the first variation formula of the canonical variation theory (6.22) or (6.23) can be used.

Let us discuss the case when $X$ is not necessarily compact. The considerations can be directly extended to the case of extremals of the first kind of our variational problem on compact base manifold (see the first paragraph of this Section). First we state the following result, more strong than (6.16):

*Let us consider a variational problem defined by a fibered manifold* $(Y, \pi, X)$ *and a Lagrangian form* $\lambda \in \Omega_Y^n(J^1)$ *which is supposed to be Lepagian. Then* $\gamma \in \Gamma(\pi)$ *is a* *critical point of the variational problem if and only if to each compact domain $c \subset X$ the relation*

$$\int_c j^1\gamma^*\vartheta(j^1\Xi)\lambda = 0$$

*holds for all vertical c-admissible vector fields $\Xi$ on $Y$.*

We need to show that if the assertion is true for all vertical vector fields, then it is true for all projectable vector fields. It can be shown that this follows from the pseudoverticality of the Euler form $E(h(\lambda))$ (see (5.9)).

We note that the definition of critical points by means of the only vertical vector fields has been given by Sniatycki [35].

The well-known proposition on critical points can now be stated:

*A cross section $\gamma \in \Gamma(\pi)$ is a critical point of the canonical variational problem defined by a fibered manifold $(Y, \pi, X)$ and a Lepagian form $\lambda \in \Omega^n_Y(J^1)$ if and only if the Euler form $E(h(\lambda))$ vanishes along $\gamma$,*

$$E(h(\lambda)) \circ j^2\gamma = 0.$$

This is a direct consequence of the definition and of the first variation formula (6.23).

## 7. Invariant variational problems

**Classes of symmetry transformations.** Let $(Y, \pi, X)$ be a fibered manifold, $\dim X = n$, $\dim Y = n + m$, and assume that $X$ is orientable. Throughout this Section we suppose that we have a Lepagian form $\lambda$ on the 1-jet prolongation $J^1$ of the fibered manifold. We shall study the behavior of the function

$$(7.1) \qquad \Gamma(\pi) \ni \gamma \to \eta_\iota(\gamma) = \int_c j^1\gamma^*\lambda \in \mathbf{R}$$

for any compact domain $c \subset X$, i.e. the behavior of the canonical variational problem, given by $(Y, \pi, X)$ and $\lambda$, under some classes of transformations acting on $(Y, \pi, X)$. In particular, we wish to characterize those one-parameter groups (maybe local) of automorphisms of $(Y, \pi, X)$ that are in a close relation to the notion of the critical point of the canonical variational problem.

The classification of the transformations we are going to speak about, has been given by *Trautman* [38] and discussed, from different viewpoints, by many authors (see e.g. [18], [19], [20], [22], [28], [31], [35], [37]). We shall be busy with the geometric aspects of the theory, and refer for papers containing applications and other considerations to [15]. We note that the work [15] contains a list of papers dealing with applications to the general relativity theory.

In general, the notion of an *invariant transformation* is not necessarily connected with fibered manifolds, or with the calculus of variations, and is basic in differential geometry. If $M$ denotes a differentiable manifold and $\alpha$ a local automorphism of $M$, then $\alpha$ is said to be an *invariant transformation* of a differential form $\eta$ on *M*, if

$$\alpha^* \eta = \eta$$

(compare with [30], etc.).

Accordingly, a*n n-form $\lambda$ defined on $J^1$ is said to be invariant with respect to a local automorphism $(\alpha_1, \alpha_0)$ of $(J^1, \pi_1, X)$, if*

(7.2) $$\alpha_1^* \lambda = \lambda.$$

We alternatively say that $\alpha_1$ is an invariant transformation of $\lambda$.

In the calculus of variations in fibered manifolds we often use local automorphisms of $(Y, \pi, X)$, transforming cross sections again into cross sections. If $\alpha_1$ is an invariant transformation of $\lambda$ such that $\alpha_1 = j^1\alpha$ for some local automorphism $(\alpha, \alpha_0)$ of $(Y, \pi, X)$, we shall also say that $(\alpha, \alpha_0)$, or just $\alpha$, is an invariant transformation of $\lambda$.

Denote by $E(h(\lambda))$ the Euler form defined by $h(\lambda)$ (5.12). Let $(\alpha, \alpha_0)$ be a local automorphism of the fibered manifold $(Y, \pi, X)$ and denote $(j^2\alpha, \alpha_0)$ its 2-jet prolongation (3.4). *The automorphism $j^2\alpha$ is said to be a generalized invariant transformation of the horizontal n-form $h(\lambda)$, if it preserves the Euler form associated with* $h(\lambda)$, i.e.

(7.3) $$j^2\alpha^* E(h(\lambda)) = E(h(\lambda))$$

(compare with the local definition given in [38]).

Notice that according to the property of the mapping *h* we have at the same time defined the notion of the generalized invariant transformations for an *arbitrary* horizontal Lagrangian form on $(Y, \pi, X)$ (compare with the theorem of Section 5 dealing with the properties of the Lepagian forms (5.9)).

We pass to the definition of the symmetry transformations, or just symmetries, of the form $h(\lambda)$. An important feature of these transformations is that they are associated with the notion of the critical point from the calculus of variations.

A local automorphism $(\alpha, \alpha_0)$ of the fibered manifold $(Y, \pi, X)$ is called a *symmetry transformation* of the *n*-form $\lambda$ and a cross section $\gamma \in \Gamma(\pi)$ of the fibered manifold, if

1) $\gamma$ is a critical point of the canonical variational problem defined by $(Y, \pi, X)$ and $\lambda$, i.e.

$$E(h(\lambda)) \circ j^2\gamma = 0,$$

2) $\alpha \circ \gamma \circ \alpha_0^{-1}$ is a critical point of the same canonical variational problem, i.e.

(7.4) $$E(h(\lambda)) \circ j^2\alpha \circ j^2\gamma = 0.$$

As above, the definition of the symmetry transformations for the case when $h(\lambda) = \lambda$ is evident.

The following proposition characterizes classes of invariant, generalized invariant, and symmetry transformations.

*Let $\lambda \in \Omega_Y^n(J^1)$ be a Lepagian form on a fibered manifold $(Y, \pi, X)$, and suppose*

*that a cross section* $\gamma \in \Gamma(\pi)$ *satisfies* $E(h(\lambda)) \circ j^2\gamma = 0$. *Let* $(\alpha, \alpha_0)$ *be a local automorphism of* $(Y, \pi, X)$. *Then:*

1) *If* $\alpha$ *is an invariant transformation of the form* $h(\lambda)$, *then it is a generalized invariant transformation of the form* $h(\lambda)$.

2) *If* $\lambda$ *is a generalized invariant transformation of the form* $h(\lambda)$, *then it is a symmetry transformation of the form* $h(\lambda)$ *and the cross section* $\gamma$.

To prove it, consider a point $j_x^2\gamma \in J^2$, arbitrary tangent vectors $\xi_0$, $\xi_1$, ,…, $\xi_n$ at the point, and an arbitrary $(n+1)$-form $\eta$ on $J^1$. By definition of the pull-back (Section 1) and *h* (4.14)

$$\left\langle j^2\alpha^* \tilde{h}(\eta)(j_x^2\gamma), \xi_0 \times \xi_1 \times \ldots \times \xi_n \right\rangle = \left\langle \tilde{h}(\eta)(j^2_{\alpha_0(x)}\alpha\gamma\alpha_0^{-1}),\right.$$
$$\left. T_{j_x^2\gamma} j^2\alpha \cdot \xi_0 \times T_{j_x^2\gamma} j^2\alpha \cdot \xi_1 \times \ldots \times T_{j_x^2\gamma} j^2\alpha \cdot \xi_n \right\rangle$$
$$= \sum_{k=0}^{n} \left\langle \eta(j^1_{\alpha_0(x)}\alpha\gamma\alpha_0^{-1}), T_{\alpha_0(x)} j^1\alpha\gamma\alpha_0^{-1} \cdot T\pi_2 \cdot T_{j_x^2\gamma} j^2\alpha \cdot \xi_0 \right.$$
$$\times T_{\alpha_0(x)} j^1\alpha\gamma\alpha_0^{-1} \cdot T\pi_2 \cdot T_{j_x^2\gamma} j^2\alpha \cdot \xi_1 \times \ldots \times T\pi_{2,1} \cdot T_{j_x^2\gamma} j^2\alpha \cdot \xi_k$$
$$\left. \times \ldots \times T_{\alpha_0(x)} j^1\alpha\gamma\alpha_0^{-1} \cdot T\pi_2 \cdot T_{j_x^2\gamma} j^2\alpha \cdot \xi_n \right\rangle$$
$$= \sum_{k=0}^{n} \left\langle \eta(j_x^1\alpha), T_x(j^1\alpha\gamma\alpha_0^{-1} \circ \alpha_0) \cdot T\pi_2 \cdot \xi_0 \times T_x(j^1\alpha\gamma\alpha_0^{-1} \circ \alpha_0) \cdot T\pi_2 \cdot \xi_1 \right.$$
$$\left. \times \ldots \times T_{j_x^1\gamma} j^1\alpha \cdot T\pi_{2,1} \cdot \xi_k \times \ldots \times T_x(j^1\alpha\gamma\alpha_0^{-1} \circ \alpha_0) \cdot T\pi_2 \cdot \xi_n \right\rangle,$$

where we have used properties (3.6) and (3.7) of $j^r\alpha$. Remembering that by definition (3.4),

$$j^1\alpha\gamma\alpha_0^{-1} \circ \alpha_0 = j^1\alpha \circ j^1\gamma,$$

we obtain

$$\left\langle j^2\alpha^* \tilde{h}(\eta)(j_x^2\gamma), \xi_0 \times \xi_1 \times \ldots \times \xi_n \right\rangle = \sum_{k=0}^{n} \left\langle \eta(j^1\alpha(j_x^1\gamma)),\right.$$
$$T_{j_x^1\gamma} j^1\alpha \circ T_x j^1\gamma \circ T\pi_2 \cdot \xi_0 \times T_{j_x^1\gamma} j^1\alpha \circ T_x j^1\gamma \circ T\pi_2 \cdot \xi_1$$
$$\left. \times \ldots \times T_{j_x^1\gamma} j^1\alpha \circ T\pi_{2,1} \cdot \xi_k \times \ldots \times T_{j_x^1\gamma} j^1\alpha \circ T_x j^1\gamma \circ T\pi_2 \cdot \xi_n \right\rangle$$
$$= \sum_{k=0}^{n} \left\langle j^1\alpha^*\eta(j_x^1\gamma), T_x j^1\gamma \circ T\pi_2 \cdot \xi_0 \times T_x j^1\gamma \circ T\pi_2 \cdot \xi_1 \times \ldots \times T\pi_{2,1} \cdot \xi_k \right.$$
$$\left. \times \ldots \times T_x j^1\gamma \circ T\pi_2 \cdot \xi_n \right\rangle = \left\langle \tilde{h}(j^1\alpha^*\eta)(j_x^2\gamma), \xi_0 \times \xi_1 \times \ldots \times \xi_n \right\rangle,$$

and see that the following useful relation

(7.5) $$j^2\alpha^* \tilde{h}(\eta) = \tilde{h}(j^1\alpha^*\eta)$$

takes place. We shall use this relation in the case when $\eta = d\lambda$. Notice that then as a partial result we get that if $\lambda$ is Lepagian then also $j^1\alpha^*\lambda$ is: If the left-hand side of (7.5) is horizontal with respect to $\pi_{2,0}$, then also the right-hand side must be. Now we can write the identity

$$j^2\alpha^*\tilde{h}(d\lambda) = \tilde{h}(dj^1\alpha^*\lambda),$$

and use the definition (5.12) of the Euler form. This implies

$$j^2\alpha^* E(h(\lambda)) \wedge \alpha_0^*\omega = E(h(j^1\alpha^*\lambda)) \wedge \alpha_0^*\omega.$$

It is seen at once that this implies the identity

(7.6) $$j^2\alpha^* E(h(\lambda)) - E(h(j^1\alpha^*\lambda) = 0.$$

equivalent with (7.5).

Assume now that $\alpha$ is an invariant transformation of the $n$-form $h(\lambda)$. Then (7.2) and (7.3) together with the identity (6.6) imply that $\alpha$ is a generalized invariant transformation of $h(\lambda)$. Assume that $\alpha$ is a generalized invariant transformation of the $n$-form $h(\lambda)$; then

$$\langle j^2\alpha^* E(h(\lambda))(j_x^2\gamma), \xi_0\rangle = \langle E(h(\lambda))(j^2\alpha(j_x^2\gamma)), Tj^2\alpha \cdot \xi_0\rangle$$
$$= \langle E(h(\lambda))(j^2\gamma), \xi_0\rangle.$$

If $E(h(\lambda))(j_x^2\gamma) = 0$ then the relation

$$E(h(\lambda)) \circ j^2\alpha(j_x^2\gamma) = 0$$

must be fulfilled, and this finishes the proof.

(Compare with [38] and [20], where similar questions are explained in a somewhat different manner.)

**Invariant transformations.** Let $\Xi$ be a projectable vector field on $Y$, and let $\alpha_t^\Xi$ be the local one-parameter group of $\Xi$. It is easy to find conditions under which the transformations $\alpha_t^\Xi$ are invariant transformations of the horizontal $n$-form $h(\lambda)$. We get the following proposition:

*Let $\lambda$ be a Lepagian form on $J^1$ and define the corresponding Lagrange function L of degree* 1 *on* $(Y, \pi, X)$ *by the relation*

(7.7) $$h(\lambda) = L\pi_1^*\omega.$$

*Let $\Xi$ be a projectable vector field on Y with $T\pi \circ \Xi = \xi \circ \pi$. Then $\Xi$ generates invariant transformations $\alpha_t^\Xi$ of $h(\lambda)$ if and only if it satisfies the condition*

(7.8) $$\langle dL, j^1\Xi\rangle \pi_1^*\omega + L\vartheta(\xi)\omega = 0.$$

Equation (7.8) for $\Xi$ is known as the *Noether equation* [38]. If we define the function $\operatorname{div}\xi$ on $X$ by the relation

$$\operatorname{div}\xi \cdot \omega = \vartheta(\xi)\omega,$$

then the Noether equation takes the form

(7.9) $\langle dL, j^1\Xi\rangle + L\cdot div\xi = 0.$

**Generalized invariant transformations.** With the same notation as in the previous paragraph we state the following proposition characterizing the generalized invariant transformations:

*The following conditions are equivalent:*

1) $\Xi$ *generates generalized invariant transformations of* $h(\lambda)$.
2) *There exists an n-form* $\eta$ *on Y such that* $d\eta = 0$ *and*

(7.10) $h(\vartheta(j^1\Xi)\lambda - \eta) = 0.$

3) *The relation*

$$\vartheta(j^2\Xi)E(h(\lambda)) = E(h(\vartheta(j^1\Xi)\lambda)) = 0$$

*holds.*

Conditions 1) and 3) are obviously equivalent. Suppose that 2) holds; using (7.6) one gets

$$E(h(\vartheta(j^1\Xi)\lambda)) = 0,$$

and using our theorem concerning the kernel of the Euler mapping (Section 5) we get 3). 2) follows from 3) in the same way.

Condition (7.10) determining the vector fields $\Xi$ whose local one-parameter groups are generalized invariant transformations, is called the *Noether-Bessel-Hagen equation* [38]. We note, however, that, in general, the term $h(\eta)$ in the equation is not of the "divergence type" (compare with our note in Section 4). Thus, our result (7.10) is more precise than that one with the "divergence term".

(See [22].)

**Symmetry transformations.** In connection with the notion of asymmetry transformation we pose the following problem:

Assume that we have the following objects:

1) a fibered manifold $(Y, \pi, X)$ with orientable base space $X$,
2) a Lagrangian form $\lambda$ of degree 1 on $(Y, \pi, X)$, which is Lepagian,
3) a vector space $\mathcal{V}$ (over $\mathbf{R}$) of projectable vector fields.

In other words, we have a canonical variational problem with a prescribed vector space of projectable vector fields. We shall say that these objects define a $\mathcal{V}$*-symmetric variational problem on* $(Y, \pi, X)$. The question is to find those critical points of the canonical variational problem, for which the elements of $\mathcal{V}$ generate symmetry transformations. Thus, a cross section $\gamma \in \Gamma(\pi)$ will be called *a solution* of the $\mathcal{V}$-symmetric variational problem, i*f*

1) $\gamma$ satisfies the Euler equations

$$E(h(\lambda)) \circ j^2\gamma = 0,$$

2) for each $\Xi \in \mathcal{V}$

(7.11) $$E(h(\lambda)) \circ \alpha_t^{j^2\Xi} \circ j^2\gamma = 0.$$

Notice that given $\gamma \in \Gamma(\pi)$, the same conditions 1) and 2) define just what we mean by a symmetry transformation $\alpha_t^{\Xi}$. Thus, at the same time the condition (7.11) can be considered as determining $\Xi$ up to certain extent.

The solutions of the $\mathcal{V}$-symmetric variational problem are characterized by the following proposition:

*A cross section* $\gamma \in \Gamma(\pi)$ *is a solution of the* $\mathcal{V}$*-symmetric variational problem if and only if*

1) $\gamma$ *satisfies the Euler equations*

$$E(h(\lambda)) \circ j^2\gamma = 0,$$

2) *for each* $\Xi \in \mathcal{V}$

(7.12) $$E(h(\vartheta(j^1\Xi)\lambda) \circ j^2\gamma = 0.$$

In other words a necessary and sufficient condition for $\gamma$ to be a solution is that $\gamma$ satisfies the Euler equations defined by $h(\lambda)$ and the Euler equations defined by $h(\vartheta(j^1\Xi)\lambda)$, for all $\Xi \in \mathcal{V}$.

We shall prove it. Suppose that for each sufficiently small $t$, the cross section $\alpha^{\Xi}\gamma\alpha_{-t}^{\xi}$ is an extremal. That is, for all vertical vector fields $\Theta$ on $Y$ the relation

(7.13) $$(j^1\alpha_t^{\Xi}\gamma\alpha_{-t}^{\xi})^* i(j^2\Theta)\tilde{h}(d\lambda) = 0$$

holds. Then, according to the general rules for computation with the Lie derivative and the pull-back of differential forms (Section 1)

$$\begin{aligned} j^1\gamma^*\alpha_t^{j^1\Xi *}\vartheta(j^1\Theta)h(\lambda) &= j^2\gamma^*\alpha^{j^2\Xi *} i(j^2\Theta)\tilde{h}(d\lambda) + j^2\gamma^*\alpha_t^{j^2\Xi *} h(di(j^1\Theta)\lambda) \\ &= \alpha_t^{\xi *}(j^2\alpha_t^{\Xi}\gamma\alpha_{-t}^{\xi})^* i(j^2\Theta)\tilde{h}(d\lambda) + j^2\gamma^*\alpha_t^{j^2\Xi *} h(di(j^1\Theta)\lambda), \end{aligned}$$

where we have used (6.23), (3.4), and write $\xi$ for the vector field on $X$ defined as the projection of $\Xi$: $T\pi \circ \Xi = \xi \circ \pi$. According to (7.13)

$$j^2\gamma^*\alpha_t^{j^2\Xi *}(\pi_{2,1}^*\vartheta(j^1\Theta)h(\lambda) - h(di(j^1\Theta)\lambda) = 0$$

and we get from the definition of $h$

$$j^1\gamma^*\alpha_t^{j^1\Xi *} i(j^1\Theta) d\lambda = 0.$$

Since the Lie derivative commutes with the contraction by a vector field and with exterior differentiation, we have

$$j^1\gamma^* i(j^1\Theta)\vartheta(j^1\Xi)d\lambda = j^1\gamma^* i(j^1\Theta)d\vartheta(j^1\Xi)\lambda$$
$$= j^2\gamma^* h(i(j^1\Theta)d\vartheta(j^1\Xi)\lambda) = j^2\gamma^* i(j^2\Theta)\tilde{h}(d\vartheta(j^1\Xi)\lambda) = 0.$$

According to our assumption, the last formula is satisfied by all vertical vector fields $\Theta$ on $Y$. As $\lambda$ is Lepagian, so is $\vartheta(j^1\Xi)\lambda$ for any $\Xi$. Thus we see that $\gamma\in\Gamma(\pi)$ is an extremal related to the Lagrangian form $\vartheta(j^1\Xi)\lambda$. This means that it also satisfies (7.12). To prove the converse it suffices to apply the identity (7.6) in the form

$$\vartheta(j^2\Xi)E(h(\lambda)) - E(h(\vartheta(j^1\Xi)\lambda)) = 0.$$

This completes the proof.

Thus, considering a $\mathcal{V}$-symmetric variational problem, we have to deal with the Lagrangian forms $\lambda$ and $\vartheta(j^1\Xi)\lambda$, $\Xi\in\mathcal{V}$. If $L$ is the Lagrange function of degree 1 defined by $\lambda$ (7.7), then it is clear how to find the Lagrange functions defined by $\vartheta(j^1\Xi)\lambda$. If we define $L_\Xi$ by the condition

$$L_\Xi \pi_1^*\omega = h(\vartheta(j^1\Xi)\lambda),$$

then

$$L_\Xi = \langle dL, j^1\Xi\rangle + L\,\mathrm{div}\xi. \quad (7.14)$$

The last formula may be used for an explicit construction of the corresponding Euler equations (7.12).

**Example.** As an illustration we shall formulate a special $\mathcal{V}$-symmetric variational problem in terms of partial differential equations.

Let us consider the fibered manifold $(T_\sigma X, \tau_\sigma, X)$ introduced in Section 3 (example). The fibered manifold has the property that each vector field on $X$ can be "lifted" to $T_\sigma X$ so that the obtained vector field is projectable and its projection is just the given vector field on $X$. Fibered manifolds $(Y,\pi,X)$ where such construction of vector fields on $Y$ is possible, admit some interesting $\mathcal{V}$-symmetric variational problems. In this paragraph we wish to study such problems in the case of the tensor bundle $(T_\sigma X, \tau_\sigma, X)$.

Let us consider a $\mathcal{V}$-symmetric variational problem given by the following:

1) the fibered manifold $(T_\sigma X, \tau_\sigma, X)$ of tensors of type $\sigma$ on $X$,

2) a Lagrangian form $\lambda$ of degree 1 on $(Y,\pi,X)$; assume for simplicity that $\lambda$ is Lepagian,

3) the vector space $\mathcal{V}_X$ of projectable vector fields on $T_\sigma X$ of the form (3.22), i.e., generated by vector fields defined on $X$.

We shall discuss equations for solutions of this problem in terms of a canonical chart on $J^r T_\sigma X$ associated with a fibered chart $(V,\psi)$ on $T_\sigma X$. The corresponding coordinate functions will be denoted by $(x_i, y_A, z_{iA}, \ldots z_{i_1 i_2 \ldots i_r A})$ (2.4).

Let $\xi$ be a vector field on $X$; we write

$$\xi = \xi_k \frac{\partial}{\partial x_k}.$$

The corresponding vector field $\xi_\sigma \in \mathcal{V}_X$ is then given as

$$\xi_\sigma = \xi_k \frac{\partial}{\partial x_k} + \xi^A \frac{\partial}{\partial y_A},$$

where

$$\xi^A = \sigma_{Ap}^{Bq} \frac{\partial \xi_p}{\partial x_q} y_B.$$

Further write

$$j^1\xi_\sigma = \xi_k \frac{\partial}{\partial x_k} + \xi^A \frac{\partial}{\partial y_A} + \xi_k^A \frac{\partial}{\partial z_{kA}}$$

with

$$\xi_k^A = \sigma_{Ap}^{Bq}\left(\frac{\partial^2 \xi_p}{\partial x_k \partial x_q} \cdot y_B + \frac{\partial \xi_p}{\partial x_q} \cdot z_{kB}\right) - z_{lA} \frac{\partial \xi_l}{\partial x_k},$$

and

$$\omega_0 = dx_1 \wedge dx_2 \wedge \ldots \wedge dx_n,$$

$$h(\lambda) = \mathcal{L}\omega_0,$$

(7.15) $$h(\vartheta(j^1\xi_\sigma)\lambda) = \mathcal{L}_\xi \omega_0,$$

$$\mathcal{L}_\xi = \frac{\partial \mathcal{L}}{\partial x_i}\xi_i + \frac{\partial \mathcal{L}}{\partial y_A}\xi^A + \frac{\partial \mathcal{L}}{\partial z_{iA}}\xi_i^A + \mathcal{L}\frac{\partial \xi_i}{\partial x_i}.$$

With this notation, necessary and sufficient conditions for a cross section $\gamma \in \Gamma(\tau_\sigma)$ to be a solution of this $\mathcal{V}_X$-symmetric variational problem are expressed as the system

(7.16) $$\frac{\partial \mathcal{L}}{\partial y_A} - \frac{d}{dx_i}\left(\frac{\partial \mathcal{L}}{\partial z_{iA}}\right) = 0,$$

(7.17) $$\frac{\partial \mathcal{L}_\xi}{\partial y_A} - \frac{d}{dx_i}\left(\frac{\partial \mathcal{L}_\xi}{\partial z_{iA}}\right) = 0$$

of partial differential equations for the cross section $\gamma$ (clearly, here $\xi$ runs over the whole space of vector fields on $X$).

Since in the second equation all derivatives of $\xi$ can be considered as independent variables, we expect that this equation will lead to some other ones for the coefficients. After some calculations we get

$$\frac{\partial \mathcal{L}_\xi}{\partial y_C} - \frac{d}{dx_s}\left(\frac{\partial \mathcal{L}_\xi}{\partial z_{sC}}\right) = a_i^C \xi_i + a_{pq}^C \frac{\partial \xi_p}{\partial x_q} + a_{p,kq}^C \cdot \frac{\partial^2 \xi_p}{\partial x_k \partial x_q} + a_{p,sqk}^C \cdot \frac{\partial^3 \xi_p}{\partial x_s \partial x_q \partial x_k}.$$

With the abbreviation

$$E_A = \frac{\partial \mathcal{L}}{\partial y_A} - \frac{d}{dx_i}\left(\frac{\partial \mathcal{L}}{\partial z_{iA}}\right),$$

the functions $a_i^C, a_{pi}^C, a_{p,kq}^C, a_{p,skq}^C$ are given as

$$a_i^C = \frac{\partial E_C}{\partial x_i},$$

$$a_{pq}^C = E_A(\sigma_{Ap}^{Cq} + \delta_{CA}\delta_{pq}) + \sigma_{Ap}^{Bq} y_B \frac{\partial E_C}{\partial y_A} + \frac{\partial E_C}{\partial z_{kA}}(\sigma_{Ap}^{Bq} + \delta_{AB}\delta_{pq}) z_{kB}$$

$$+ \left( \frac{\partial^2 \mathscr{L}}{\partial z_{sC} \partial z_{qA}} + \frac{\partial^2 \mathscr{L}}{\partial z_{sA} \partial z_{qC}} \right) z_{spA} - \sigma_{Ap}^{Bq} z_{skB} \frac{\partial^2 L}{\partial z_{sC} \partial z_{kA}},$$

$$a_{p,kq}^C = \frac{1}{2}\sigma_{Ap}^{Br} y_B \left( \frac{\partial E_C}{\partial z_{kA}} \delta_{qr} + \frac{\partial E_C}{\partial z_{qA}} \delta_{kr} \right) - \frac{1}{2}\sigma_{Ap}^{Br} z_{sB}$$

$$\cdot \left( \frac{\partial^2 \mathscr{L}}{\partial z_{sC} \partial z_{kA}} \delta_{qr} + \frac{\partial^2 \mathscr{L}}{\partial z_{sC} \partial z_{qA}} \delta_{kr} + \frac{\partial^2 \mathscr{L}}{\partial z_{sA} \partial z_{kC}} \delta_{qr} + \frac{\partial^2 \mathscr{L}}{\partial z_{sA} \partial z_{qC}} \delta_{kr} \right)$$

(7.18)

$$+ \frac{1}{2} z_{pA} \left( \frac{\partial^2 \mathscr{L}}{\partial z_{qA} \partial z_{kC}} + \frac{\partial^2 \mathscr{L}}{\partial z_{kA} \partial z_{qC}} \right),$$

$$a_{p,skq}^C = \frac{1}{2}\sigma_{Ap}^{Br} y_B \left( \frac{\partial^2 \mathscr{L}}{\partial z_{kC} \partial z_{sA}} \delta_{qr} + \frac{\partial^2 \mathscr{L}}{\partial z_{kC} \partial z_{qA}} \delta_{sr} + \frac{\partial^2 \mathscr{L}}{\partial z_{qC} \partial z_{kA}} \delta_{sr} + \frac{\partial^2 \mathscr{L}}{\partial z_{qC} \partial z_{sA}} \delta_{kr} \right).$$

If we let them vanish we obtain, together with $E_C = 0$ (7.16), necessary and sufficient conditions for the solutions of our $\mathscr{V}_X$-symmetric variational problem.

We shall not discuss the arising conditions. Just notice that they contain the second derivatives of the Lagrange function with respect to $z_{iA}$. This means that the second derivatives should be in high degree symmetric along the solution $\gamma \in \Gamma(\tau_\sigma)$ of the $\mathscr{V}_X$-symmetric variational problem.

The described example serves as a motivation for the study of a type of variational problems on tensor bundles with high degree of symmetry. By the way described here, laying stress upon the symmetries of the variational problem in the large rather than upon symmetry properties of each individual solution, we are led to the notion of generally covariant variational problems. These are defined in the next paragraph.

**General covariance.** It is of great interest in the general relativity theory to study the so called *generally covariant* (or *generally invariant*) variational problems (see [15], [25], [31], [37], etc.). The definition of such problems has been given by *Trautman* in his fundamental work on the structure of canonical variational problems in fibered manifolds [38]. Let us study the notion of the generally covariant theory from the point of view of the formalism developed in this paper.

The notion of the generally covariant variational problems can be introduced on each fibered manifold $(Y, \pi, X)$ where there is given a mapping, *u*, from the space of all vector fields on *X* into the space of all projectable vector fields on *Y*, commuting with the projection $T\pi$:

$$T\pi \circ u(\xi) = \xi$$

for each vector field $\xi$ on *X*. As discussed in the previous paragraph, this is so in the case of all tensor bundles $(T_\sigma X, \tau_\sigma, X)$ on the given manifold $X$. The mapping $u$ is given by

$$u(\xi) = \xi_\sigma. \tag{7.19}$$

Let us precise what we are going to deal with. Assume that we have a canonical variational problem, given by the fibered manifold $(T_\sigma X, \tau_\sigma, X)$ and by a Lepagian form $\lambda$ on $J^1 T_\sigma X$. This canonical variational problem is called *generally covariant*, if for each vector field $\xi$ on $X$, the vector field $\xi_\sigma$ on $T_\sigma X$ generates generalized invariant transformations of the form $h(\lambda)$.

Thus, generally covariant problems behave rather simply with respect to some special variations, especially those generated by vector fields on the base manifold $X$.

We note that the assumption concerning the Lagrangian form (i.e. that $\lambda$ is Lepagian) is not essential; the definition of generally covariant variational problems (not necessarily canonical) can be given without changes taking instead of any $\rho \in \Omega_Y^n(J^1)$ a Lepagian equivalent of $\rho$.

Using the theorem, characterizing the generalized invariant transformations, we can say:

*The canonical variational problem is generally covariant if and only if*

$$E(h(\vartheta(j^1\xi_\sigma)\lambda) = 0 \tag{7.20}$$

*for all vector fields $\xi$ on X.*

This theorem shows how to proceed if one wants to *find* some Lagrangian forms (or Lagrange functions) defining generally covariant variational problems, or to check whether a Lagrangian form defines a generally covariant variational problem. It suffices for this to find a solution of the system

$$a_i^C = 0, \quad a_{pq}^C = 0, \quad a_{p,kq}^C = 0, \quad a_{p,sqk}^C = 0$$

of partial differential equations with the left-hand sides given by (7.18). Of course this system is now considered as the system for the Lagrange function $\mathscr{L}$.

Analogous definitions can be given in the case of Lagrangian forms on higher jet spaces. The only difference one can expect is that the calculations be more difficult.

**Differential conservation laws.** Assume that a symmetric variational problem is given.

*Let $(Y, \pi, X)$ be a fibered manifold, $\lambda$ a Lepagian form. Suppose that $\gamma \in \Gamma(\pi)$ is a critical point of the canonical variational problem defined by $(Y, \pi, X)$ and $\lambda$, and that we have a projectable vector field $\Xi$ on Y generating generalized invariant transformations of $h(\lambda)$. Then there is a local $(n-1)$-form $\eta$ on Y such that*

$$dj^1\gamma^*(i(j^1\Xi)\lambda - \eta) = 0. \tag{7.21}$$

The assertion follows from (6.23), (6.16).

The situation described by this theorem well refers to the situation in physical applications of symmetric variational problems (compare with [11], [15], [20], [25], [31], [34], [35], [37], [38]). The expression on the left-hand side of (7.21) can be rewritten as the divergence of a vector field on the base manifold $X$, constructed from the Lagrangian form $\lambda$, the form $\eta$ and the vector field $\Xi$ (see [38], [20]). We often say that it express a *conservation law in the differential form*. This is motivated by physical reasons (see e.g. [15]).

Rather similar situation, at least from the technical point of view, arises with the weak critical points of variational problems defined in tensor bundles (see Section 6); each weak critical point and each vector field on the base manifold, generating generalized invariant transformations, give rise to a differential conservation law. In the special case of generally covariant theory we obtain in this way infinite number of relations, everyone being connected with a vector field on the base manifold. We shall not go into the details in this direction, and pass to another problem. Roughly speaking, one can prescribe "conservation laws" and look for the critical fields (*critical points*) along which these conservation laws will hold.

Suppose that we have a closed $n$-form $\eta$ on $Y$ (i.e. $d\eta = 0$). Remember that according to the general formula (5.2) for $d\eta$ the only forms $\rho \in \Omega_Y^n(J^1)$ satisfying $d\rho = 0$ are pull-backs $\pi_{1,0}^* \eta$, where $d\eta = 0$. Given $\eta$, a projectable vector field $\Xi$ on $Y$ and a Lepagian $n$-form $\lambda$ on $J^1$ one can request that the conditions

$$j^1\gamma^*(di(j^1\Xi)\lambda - \eta) = 0, \quad E(h(\lambda) \circ j^2\gamma = 0$$

should be satisfied by a cross section $\gamma \in \Gamma(\pi)$ (notice that locally, $\eta$ in is, by the *Poincaré lemma*, of the form $d\overline{\eta}$). $\eta$ may be given, without loss of generality, by its horizontal part, $h(\eta)$ (see the theorem of Section 4 concerning the kernel of the Euler mapping). In this way we obtain some partial differential equations for the critical point $\gamma \in \Gamma(\pi)$, and a conservation law at the same time. The reader will find that this is, in fact, a modification of the *Noether-Bessel-Hagen equation*, applied to critical points instead of generators of groups of transformations.

## 8. Summary

This work contains an exposition of foundations of the variational calculus in fibered manifolds. The emphasis is laid on the geometrical aspects of the theory. Especially functionals defined by means of real functions (*Lagrange functions*) or differential forms (*Lagrangian forms*) on the first jet prolongation of a given fibered manifold are studied. *Critical points* (critical cross sections) of the functionals are examined and the *Euler equations* for them are derived in a completely invariant manner. The *first variation formula* is derived by means of the so-called *Lepagian forms*. All variations appearing in the theory are generated by vector fields. Jet prolongations of projectable vector fields are defined. The *Euler form*, associated with a given Lagrange function (or Lagrangian form) is introduced by means of the Euler equations of the calculus of variations. Necessary and sufficient conditions for the vanishing of the Euler form are stated in terms of differential forms and their exterior differential. The corresponding conditions for a Lagrange function leading to identically vanishing Euler equations are given. Some special Lepagian forms are studied. *Classes of symmetries* of a variational problem are defined. *Invariant*, *generalized invariant*, and *symmetry transformations* are characterized in terms of the Lie derivatives. The variational problem with prescribed symmetry transformations is formulated, and necessary and sufficient conditions for its solutions are given. The geometrical aspects of the so-called *generally covariant* variational theories are studied. Definitions and theorems are well adapted to the situation in physical field theories.

## Acknowledgments

The author wishes to thank Dr. I. Kolár from Dept. of Mathematics, CSAV, for his helpful criticism and interest in this work. He also wishes to thank Professor A. Trautman from the University of Warsaw for his constructive suggestions and encouragement. He would like to thank *APhEN* for interesting discussions and continuing support.

## References

[1] Abraham, R., Lectures of Smale on differential topology (mimeographed notes).
[2] Bishop, R.L., R.J. Crittenden, *Geometry of Manifolds*, New York and London, 1964.
[3] Boardman, J.M., Singularities of differentiable maps, Inst. Hautes Études Sci. Publ. Math. No. 33 (1967), 21-57.
[4] Boerner, H., Über die Legendresche Bedingung und die Feldtheorien in der Variationsrechnung der mehrfachen Integrale, Math. Z. 46 (1940), 720-742.
[5] Bourbaki, N., *Topological Vector Spaces*, Moscow, 1959 (Russian).
[6] Cartan, É., *Lecons sur les Invariants Intégraux*, Paris (1922).
[7] Choquet-Bruhat, Y., *Géométrie Différentielle et Systémes Extérieurs*, Paris, 1968.
[8] Courant, R., D. Hilbert, *Methods of Mathematical Physics*, Vol. I, New York and London, 1953.
[9] Dieudonné, J., *Foundations of Modern Analysis*, New York and London, 1960.
[10] Eells, J. Jr., J.H. Sampson, Variational theory in fibre bundles, Proc. U.S.-Japan Seminar in Differential Geometry, Tokyo, 1965.
[11] Gelfand, I.M., S. Fomin, *Calculus of Variations*, New Jersey, 1967.
[12] Gromoll, D.,W. Klingenberg, W. Meyer, *Riemannsche Geometrie im Grossen, Berlin*, Heidelberg, New York, 1968.
[13] Hermann, R., *Differential Geometry and the Calculus of Variations*, New York and London, 1968.
[14] Hermann, R., Second variation for variational problems in canonical form, Bull. Am. Mat. Soc. 71 (1965), 145-149.
[15] Horsky, J., J. Novotny, Conservation laws in general relativity, Czech. J. Phys. B 19 (1969), 419-442.
[16] Kelley, J.L., *General Topology*, Princeton, N. J., 1957.
[17] Kolár, I., Introduction to the theory of jets (mimeographed notes in Czech), CSAV Brno, 1972.
[18] Komorowski, J., A modern version of the E. Noether’s theorems in the calculus of variations, I, Studia Math. 29 (1968), 261-273.
[19] Komorowski, J., A modern version of the E. Noether's theorems in the calculus of variations, II, Studia Math. 32 (1969), 181-190.
[20] Krupka, D., Lagrange theory in fibred manifolds, Rep. Math. Phys. 2 (1971), 121-133.
[21] Krupka, D., On the structure of the Euler mapping, to be published in: Arch. Math., Brno.
[22] Krupka, D., On generalized invariant transformations, to be published in: Rep. Math. Phys., Torun.
[23] Kuranishi, M., *Lectures on Involutive Systems of Partial Differential Equations*, Sao Paulo, 1967.

[24] Kurosh, A.G., *Lectures on Higher Algebra*, Moscow, 1959 (Russian).
[25] Landau, L.D., E.M. Lipshitz, *The Classical Theory of Fields*, Pergamon Press, 1971.
[26] Lang, S., *Introduction to Differentiable Manifolds*, New York and London, 1962.
[27] Lepage, Th.H.J., Sur les champs géodésiques du Calcul des Variations, Bull. Acad. Roy. Belg., Cl. Sci. V, Sér. 22 (1936), 716, 1036.
[28] Logan, J.D., Generalized invariant variational problems, J. Math. Anal. Appl. 38 (1972), 175-186.
[29] Noether, E., Invariante Variationsprobleme, Göttingen Nachr. 1918, 235.
[30] Nomizu, K., *Lie Groups and Differential Geometry*, Publ. Math. Soc. Jap. 2, 1956.
[31] Novotny, J., *Principles of symmetry, conservation of energy and momentum and chronometrical invariance in general relativity*, RNDr. Dissertation, J.E. Purkyne University, Brno, Czechoslovakia, 1971.
[32] Palais, R. S., Manifolds of sections of fiber bundles and the calculus of variations, Proc. Symp. Pure Math. XVIII, Part 1, Chicago, 1970, 195-205.
[33] Robertson, A.P., W. Robertson, *Topological Vector Spaces*, Cambridge Univ. Press, 1964.
[34] Schild, A., Lectures on general relativity theory, in: *Relativity Theory and Astrophysics*, Am. Math. Soc., Providence, Rhode Island, 1967.
[35] Sniatycki, J., On the geometric structure of classical field theory in Lagrangian formulation, Proc. Camb. Phil. Soc. (1970), 68, 475-484.
[36] Sternberg, S., *Lectures on Differential geometry*, N. J., 1965.
[37] Trautman, A., Conservation laws in general relativity, in: *Gravitation*, ed. by L. Witten, New York, 1962.
[38] Trautman, A., Noether equations and conservation laws, Commun. math. Phys. 6 (1967), 248-261.
[39] Ver Eecke, Connections d'ordre infini, Cahiers de Topol. et Géom. Diff. XI, (1969), 281-321.

Added in proofs: The paper [21] will appear in Arch. Math. 10 (1974), the paper [22] in Rep. Math. Phys. 5 (1974).